\documentclass[twocolumn,
    aps,
    prb,
    superscriptaddress,
    longbibliography]{revtex4-2}
\usepackage{graphicx}
\usepackage{amsmath}
\usepackage[export]{adjustbox}
\usepackage{enumitem}
\usepackage{amssymb}
\usepackage{hyperref}
\usepackage[normalem]{ulem}
\usepackage{adjustbox}
\usepackage{physics2}
\usephysicsmodule{ab,braket}
\usepackage{makecell}
\usepackage[dvipsnames]{xcolor}
\newcommand{\bigzero}{\mbox{\normalfont\Large\bfseries 0}}
\newcommand{\rvline}{\hspace*{-\arraycolsep}\vline\hspace*{-\arraycolsep}}

\usepackage{lineno}
\usepackage{bbm}
\usepackage{bm}
\usepackage{comment}
\usepackage{setspace}
\usepackage{float}
\usepackage{blkarray}
\usepackage{relsize}
\usepackage{comment}

\begin{document}
\title{Every Benchmark All at Once}

\author{Ana Silva and Eliska Greplova}
\affiliation{QuTech and Kavli Institute of Nanoscience,
Delft University of Technology}
\thanks{A.C.OliveiraSilva@tudelft.nl}

\begin{abstract}
As quantum technology matures, the efficient benchmarking of quantum devices remains a key challenge. Although sample-efficient benchmarking techniques have recently been proposed, there is still a gap in adapting these techniques to contemporary experiments. In this work, we reformulate standard, interleaved, simultaneous, correlated and leakage randomized benchmarking techniques in the modern language of the gate-set shadow protocol, and discuss the relation between this framework and character randomized benchmarking. This reformulation provides a common language in which familiar randomized benchmarking fitting models arise from different choices of gate set, sampling scheme, and probe operator. We further analyze the role of multiplicity in the gate-set shadow protocol. As a particular example, we show how reconstructing the matrix governing the evolution of average sequence functions can provide additional information about the underlying noise on the idle qubit. We also examine how multiplicity constraints which average sequence functions can be constructed in postprocessing, particularly when the input and measurement settings cannot be varied.
Our work supports the use of the gate-set shadow protocol as a flexible framework for estimating target quantities across different gate sets, while illustrating how multiplicity can both obstruct and enable the extraction of physically meaningful error metrics.
\end{abstract}

\maketitle

\section{Introduction}\label{sec:intro}

The ability to build high performance quantum processors relies on acquiring precise control over quantum systems. To this end, the capacity of performing targeted operations that may only address parts of the system is crucial. However, near-term quantum devices are prone to errors of multi-factorial origin. These may stem, for example, from imperfect manufacturing, inaccuracies in control fields, unforeseen interactions that couple different parts of the system, or undesired interactions with the environment. While quantum error correcting codes have been developed that allow for fault-tolerant quantum computing~\cite{DevMunNem2013,RyaBohLee2021,Goo2025}, their efficacy requires gate errors to be bounded below a certain threshold~\cite{Pres1997,kniLafZur1998,DevMunNem2013}. Additionally, error correcting schemes typically assume errors to be uncorrelated~\cite{DevMunNem2013,YanChuGuo2024}. Given these constraints, and the myriad of effects leading to noise, diagnostic tools that provide reliable noise characterization are essential to the development of high performance quantum processors.

Quantum process tomography is a well-established technique that provides full gate error reconstruction~\cite{ChuNiel1997,AriPre2001}. Its implementation, however, is challenging, with two issues arising as limitations: firstly, the measurement overhead required for tomographic reconstruction scales exponentially in the number of qubits~\cite{TorWooAcha2023}, and secondly, its gate noise estimations are vulnerable to inaccuracies arising from state preparation and measurement errors (SPAM)~\cite{BalKalDeu2014}. A commonly employed characterization technique that addresses both these difficulties is randomized benchmarking (RB)~\cite{MagGamEme2011,MagGamEme2012,HelRotOno2022}. By forgoing full error reconstruction, RB allows for efficiently probing average noise metrics, while still guaranteeing that these estimates are independent of SPAM errors. To gain information on specific average noise properties, different RB protocols have been developed that tailor to efficiently extract a targeted noise metric. This can be, for example, learning the noise affecting a specific gate in the gate set (interleaved RB)~\cite{MagGamJoh2012,HarFla2017}, estimating the coherence of the noise (purity RB)~\cite{WalGraHar2015,DirHelWeh2019}, or estimating crosstalk (simultaneous RB and crosstalk RB)~\cite{GamCorMer2013,MckCroWoo2020}.

Recently, further RB variants have been developed with the aim of improving the scalability and efficiency of error characterization of multi-qubit gate operations. Randomized mirror circuits~\cite{ProSerRud2022} and binary randomized benchmarking~\cite{HinHotBlu2024} provide scalable approaches for estimating error rates in multi-qubit Clifford gate sets, with the latter protocol removing the need for motion reversal circuits. Cycle benchmarking~\cite{ErhWallPos2019} uses randomized compiling to estimate the average process fidelity of a dressed cycle, formed by composing a target cycle of multi-qubit gates with a random Pauli cycle. This method can also be extended to characterize non-Clifford gates~\cite{ErhWallPos2019}. In addition, universal randomized benchmarking~\cite{CheDinHua2022} provides a formalism to extend RB-like protocols to gate sets that do not form a group.

Complementary to these scalability oriented developments is the question of efficiently exploiting existing data~\cite{HelIoaKit2023}: once randomized benchmarking data has been collected, what properties may be estimated in postprocessing? This data reuse perspective is central to classical shadow tomography~\cite{HuaKuePres2020}. Developed for quantum state characterization, classical shadows provide a sample efficient approach that, albeit not aiming for full state reconstruction, still allows for estimating multiple properties of a quantum state from the same data in post-processing~\cite{HuaKuePres2020}. Crucially, the required number of samples to attain a certain estimation precision is not directly dependent on the number of qubits, but rather on the type of observables we wish to learn~\cite{HuaKuePres2020}. The classical shadow protocol relies on data collected from several repetitions of state evolution through randomly selected unitary gates. Unlike RB, however, the end result in classical shadow tomography is a dataset that allows for the estimation of multiple expectation values, instead of a single figure of merit.

Given that knowledge of many noise facets are actually desirable to render a more comprehensive picture of the noise corrupting quantum operations in a real device, it would be desirable to import the multiple observable estimation capabilities of classical shadow to the realm of RB techniques. The gate-set shadow protocol, developed in Ref.~\cite{HelIoaKit2023}, is a recent noise characterization protocol that allows to re-frame RB into a similar framework as the classical shadow protocol. In this approach, the same dataset can be used in postprocessing to construct different average sequence functions, tailored to estimate different RB-like decay rates that encode information about the underlying gate noise. The flexibility afforded by this postprocessing step is not unlimited: the quantities that can be extracted are determined by the chosen gate set, sampling scheme, input state and measurement operators. In this work, we focus on benchmarking gate sets that form a group, 
as originally considered in the foundational work introducing the gate-set shadow formalism~\cite{HelIoaKit2023}.

While the protocol establishes an important development in the field of RB techniques, adopting this new approach can be hindered by a lack of clarity on how it relates to established RB protocols and allows for the extraction of comparable information. This naturally motivates the main question of this work: What does the gate-set shadow framework add to existing RB protocols? We distinguish three cases: cases where the framework facilitates extracting additional noise diagnostic information, cases where it provides a more convenient SPAM robust route for extracting the same information, and cases where it exposes identifiability limitations already implicitly present in randomized benchmarking data. We examine these different roles of the gate-set shadow framework through two complementary contributions.

Our first contribution is to demonstrate how the gate-set shadow formalism provides a unified framework for describing established RB protocols. In doing so, we show how familiar RB fitting models arise from the combined choice of benchmarking gate set, input state, measurement operators, and probe operator. In particular, we illustrate how the flexibility of postprocessing provides a convenient alternative to character RB, demonstrating  how the SPAM robustness inherent to the gate-set shadow framework can be exploited for concrete benchmarking tasks. Our goal here is to help practitioners view the method of Ref.~\cite{HelIoaKit2023} not as a fixed protocol, but as a flexible language for constructing RB sequence functions that are adapted to the practitioner's specific needs. We complement our use cases with open source code that can serve as basis for practical implementations of the protocol~\cite{SilGrep2025_git,SilGrep_zenodo}.

This first contribution is covered in Sections~\ref{rel_charRB},~\ref{sta_and_inter_RB_sec} and~\ref{correlated_crosstalk_sec}, and provide concrete examples of RB as a gate-set shadow protocol. In Sec.~\ref{rel_charRB}, we discuss the relation between the gate-set shadow protocol and character RB. In Sec.~\ref{sta_and_inter_RB_sec}, we formulate standard RB and interleaved RB within the gate-set shadow framework, using the benchmarking of a CNOT gate as a concrete example. In Sec.~\ref{correlated_crosstalk_sec}, we further show how the marginal channels obtained from the crosstalk tomography method of Ref.~\cite{HelIoaKit2023} can be postprocessed into effective Pauli error rates, thereby recasting the same information in a more interpretable error-rate language.

Our second contribution addresses multiplicity in the gate-set shadow formalism. From a pragmatic point of view, multiplicity means that even after applying strategies like character RB, the RB decay signal can still contain several decay modes~\cite{ClaRieWan2021}. This means we are not estimating a single scalar parameter, but have a problem more akin to reconstructing a matrix. The resulting reconstructed matrix is only unique up to a similarity transformation (i.e., a change of basis)~\cite{GamCorMer2013,HelRotOno2022,ClaRieWan2021,BluGamNie2013,ProRudYou2017,NieGamRud2021}. We show that, in the presence of multiplicity, the usual multi-decay fitting task can be viewed as a system identification problem, provided that multiple input states and measurement settings are accessible. The resulting average sequence functions then define an effective input-output dynamics. This perspective provides transparency to how gauge ambiguities can arise in RB sequence data, even under the common simplifying assumption of a gate-independent noise process and negligible SPAM noise. Although gauge ambiguities in RB have been discussed before~\cite{ProRudYou2017}, the mapping to a system identification problem provides a concrete illustration of how they enter in the gate-set shadow framework, whenever multiplicity is present.

The consequences of multiplicity are illustrated in Sections~\ref{op_crosstalk_sec} and~\ref{leakage_RB_sec}. In Sec.~\ref{op_crosstalk_sec}, we apply the system-identification perspective to estimate the fidelity of operations on the first qubit, while the second qubit remains idle. This approach makes explicit the ambiguity in extracting the desired subsystem fidelity from the estimated decay modes. We then explore whether additional assumptions, such as an uncorrelated noise ansatz, can partially alleviate this ambiguity, and what properties can be reliably extracted. In Sec.~\ref{leakage_RB_sec}, we revisit the leakage RB example of Ref.~\cite{ClaRieWan2021} to examine the practical limitations that arise when multiplicity is present, but access to different input states is restricted. In particular, we show that even for the same choice of probe operator, distinct input states can lead to different fitting models with different practical advantages in the finite sampling regime. This observation emphasizes how varying the probe operator alone in postprocessing has restricted effectiveness in the presence of multiplicity.

The remainder of the paper is organized as follows. We begin in Sec.~\ref{gss_review_cons_sec} by reviewing the gate-set shadow protocol. Sections~\ref{sta_and_inter_RB_sec} to \ref{leakage_RB_sec} then develop the examples described above. We conclude in Sec.~\ref{conclusions_sec} with a summary of the main results. Appendices~\ref{fitting_models_app}, \ref{app_num_det}, and \ref{unital_rec_app} provide further details on the mathematical reformulation of RB using the gate-set shadow formalism, as well as selected aspects of our numerical simulations.

\section{Gate-set shadow protocol: Background and Connections} \label{gss_review_cons_sec}
In this section, we start by reviewing the gate-set shadow protocol developed in Ref.~\cite{HelIoaKit2023}. We then compare it with character randomized benchmarking (character RB)~\cite{HelXueVan2019,ClaRieWan2021}, highlighting their similarities and key differences. Finally, we end this section by establishing a connection between the gate-set shadow protocol and system identification methods for linear time-invariant systems in control engineering.

\subsection{Overview of the Gate-Set Shadow Protocol}

\begin{figure*}[t]
\includegraphics[width=0.97\linewidth]{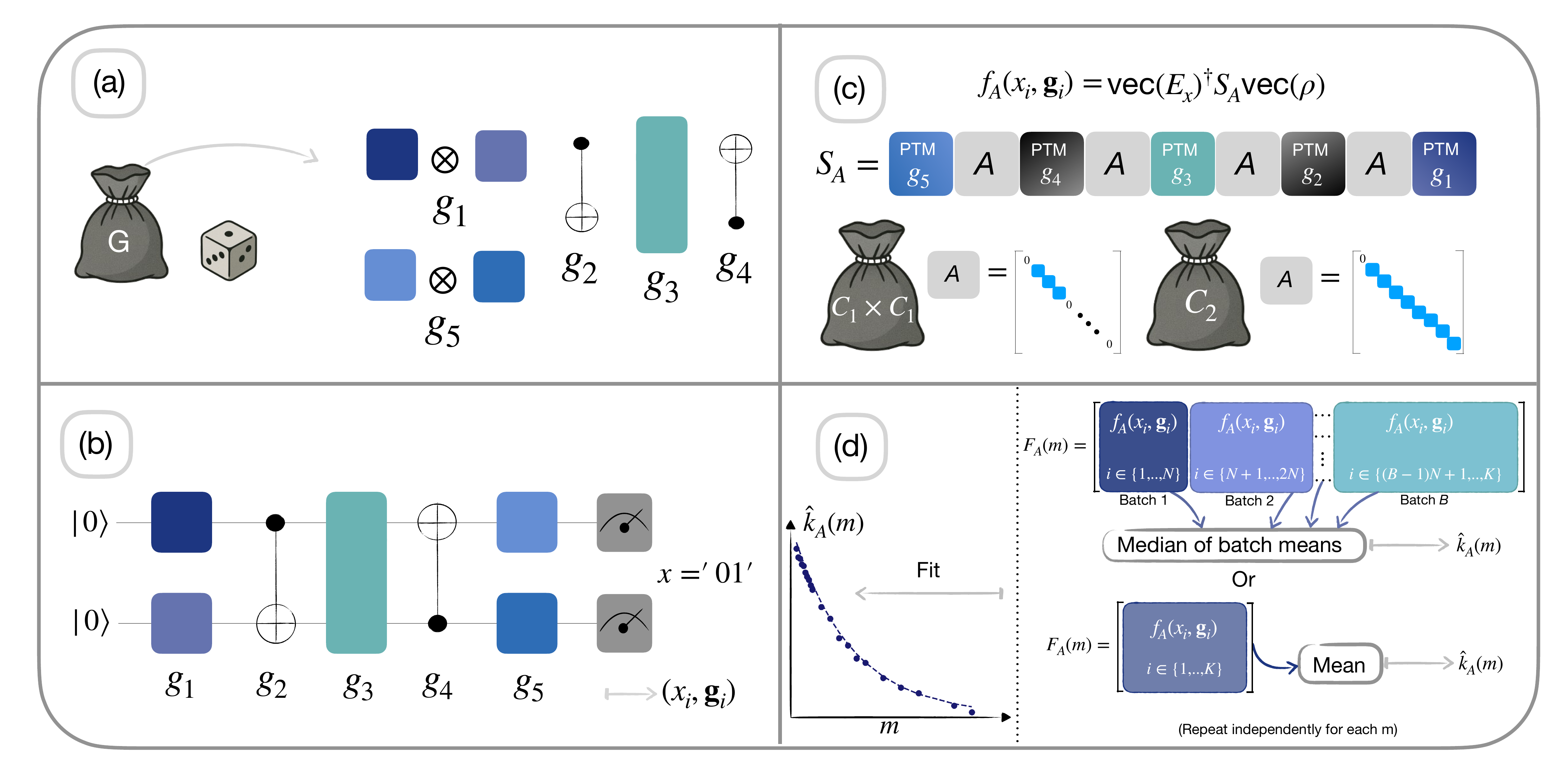}
\caption{Overview of the gate-set shadow protocol. (a) We start by selecting gates uniformly at random from a given gate set $G$. (b) This creates a sequence of random gates that are then applied to the input state. The resulting outcome is measured in a computational basis. Iterating over this process allows us to create a dataset formed by the pairing of random sequences with the corresponding outcomes, $(x_i,\mathbf{g}_i)$. (c) For each dataset entry $(x_i,\mathbf{g}_i)$, we construct the sequence correlation function $f_A$. A typical choice for $A$ is to fix it as a subspace projector. Specific choices of subspace projectors (in the PTM representation) are depicted in the figure for the case of the two-qubit Clifford  group ($\mathcal{C}_2$) and the local Clifford group ($\mathcal{C}_1\times \mathcal{C}_1$) as gate sets. The blue blocks visually indicate the non-zero matrix elements of the corresponding projectors. (d) Taking the mean (or median-of-means, MoM) of all the sequence correlation functions gives an estimate of the average sequence function, $\hat{k}_A(m)$, which can then be fitted to a decay model to extract targeted fidelities.  \label{fig_GSS_overview} }
\end{figure*}
The gate set shadow protocol developed in Ref.~\cite{HelIoaKit2023} shares many similarities with RB-like methods, but adds to it the flexibility and data driven power of classical shadow tomography, which can simultaneously estimate multiple quantities from the same data.~\cite{HuaKuePres2020}. The classical shadow aspect of the protocol lies in the separation between data acquisition and the choice of quantity to be estimated. In standard RB, the experimental data collection usually leads to a particular survival probability. In the gate-set shadow protocol, by contrast, the data consists of the measurement outcomes together with the random gate sequences that produced them. Only in postprocessing does one choose a probe operator $A$, which converts the dataset into a particular average sequence function. Different choices of $A$ therefore allow different average sequence functions to be estimated from the same experimental data.

During the data acquisition phase, similar to RB, we apply a random sequence of $m$ gates. This is followed by a measurement in a computational basis to retrieve a bit-string outcome, $x_i$. Thus, the fundamental units of information are composed by the pairs $(x_i, \mathbf{g}_i)$, i.e. the bit-string outcome $x_i$ resulting from applying the random sequence of gates $\mathbf{g}_i=\{g_1,g_2,...,g_m\}$. A complete dataset is then made of $S=LK$ samples of pairs $(x_i, \mathbf{g}_i)$ where $K$ is the number of randomizations for each sequence length and $L$ is the number of different sequence lengths. Experimentally, the protocol is therefore almost identical to RB, with the difference that the end recovery gate inverting the sequences is removed (see Fig.~\ref{fig_GSS_overview} (a)-(b) for an overview of the experimental phase of the protocol). We note that this last feature is not unique to the gate-set shadow protocol: some RB variants also avoid the end recovery gate~\cite{ErhWallPos2019,HinHotBlu2024}. Here, we emphasize this point to contrast the data collection stage with more traditional RB protocols.

To describe the measurement process in the postprocessing step~\cite{HelIoaKit2023,HuaKuePres2020}, the language of positive-operator-valued measures (POVMs) is employed. POVM measurements are characterized by the POVM elements, $\{E_x\}_x$, each corresponding to a measurement outcome $x$. These are positive operators whose sum produces the identity, i.e. $\sum_x E_x=\mathbbm{1}$.

In the gate-set shadow formalism, the postprocessing stage starts with building a set of sequence functions, each relating a recorded outcome $x_i$ to a particular sequence of operations. This specific sequence of operations does not need to be directly the gate sequence $\mathbf{g}_i$, but can be a modified sequence. In this modified sequence, each gate operation in $\mathbf{g}_i$ is interleaved with a user-defined operator $A$ (referred to as \emph{probe operator}). Following Ref.~\cite{HelIoaKit2023}, these sequence functions are termed sequence correlation functions. In practice, we evaluate them using the Pauli transfer matrix (PTM) formalism~\cite{Chow2012,Green2015} (see Appendix~\ref{PTM_Rep_intro} for a more detailed description of the PTM formalism). In this formalism, we can write the sequence correlation function as
\begin{align}
    f_A(x_i,\mathbf{g}_i) = c^{(0)}_{f_A} \; \text{vec}(E_{x_i})^{\dagger} \; S_A \; \text{vec}(\rho) \; , \label{seqcorrfun_def}
\end{align}
where $c^{(0)}_{f_A}$ is a normalization constant, $E_{x_i}$ is the POVM element corresponding to the bit-string $x_i$, and $\rho$ is the initial state. Here $\text{vec}(\cdot)$ denotes vectorization. The matrix $S_A$, written in the PTM basis, describes the modified sequence used in postprocessing. Concretely,
\begin{align}
    S_A = U_m A U_{m-1} A ... A U_{2}A U_{1}, 
\label{S_Adef}
\end{align}
 
where $U_i$ is the PTM of the ideal noise-free gate $g_i$, and $A$ is the probe operator also expressed in the PTM representation. See App.~\ref{fitting_models_app} for further details of how $f_A(x_i,\mathbf{g}_i)$ is evaluated.

The value of $f_A(x_i,\mathbf{g}_i)$ reflects how compatible the observed outcome $x_i$ is with the action of the modified sequence $S_A$ on the input state. The sequence $S_A$ is constructed from the ideal representation of the sampled gates and from the chosen probe operator $A$. Nevertheless, the outcomes $x_i$ are sampled from the noisy experimental implementation of the corresponding gate sequence $\mathbf{g}_i$. Therefore, the average behaviour of $f_A(x_i,\mathbf{g}_i)$ over many sampled sequences and outcomes still reflects the noise affecting the implemented gates. Different choices of $A$ define different sequences $S_A$, and hence different values of $f_A(x_i,\mathbf{g}_i)$. In this sense, the choice of $A$ helps determine which aspects of the noise information contained in the data are emphasized in the average behaviour of $f_A(x_i,\mathbf{g}_i)$. A particularly convenient choice for the probe operator $A$ is to set it as a subspace projector. In this way, the sequence correlation functions only retain targeted information on the gate noise. Concrete examples of this mechanism are presented in Sec.~\ref{sta_and_inter_RB_sec} and~\ref{correlated_crosstalk_sec}.

Averaging the sequence correlation functions for each sequence length $m$ leads to an estimate of the average sequence function, $\hat{k}_{A}(m)$, typically a decaying function~\cite{HelIoaKit2023} with increasing sequence length. In this way, $\hat{k}_{A}(m)$ plays the role of the average survival probability in RB, and can be used to estimate target figures of merit. An overview of the postprocessing phase of the protocol is give in Fig.~\ref{fig_GSS_overview} (c)-(d).

Computing the average sequence function estimate $\hat{k}_{A}(m)$, for each value of $m$, can be achieved either by using the empirical mean, $\hat{k}_{A}(m) = \frac{1}{K} \sum^K_{i=1} f_A(x_i,\mathbf{g}_i)$, or the median-of-means (MoM). When using the MoM estimator, the $K$ data points obtained for each sequence length are partitioned into $B$ batches. Assuming that all batches contain the same number of points, $B$ is chosen such that the batch size $N=K/B$ is an integer. The MoM estimator is then given by
\begin{align}
\hat{k}_{A}(m) \,&= \text{median} \bigg[ \frac{1}{N} \sum^{b+N-1}_{i=b} f_A(x_i,\mathbf{g}_i) \; ,\nonumber\\ &b  \in \{1, N+1, 2N+1,...,(B-1)N+1 \}\bigg] \;,
\end{align}
where $b$ denotes the starting index of each batch. The mean is computed independently within each batch, and the estimator returns the median of the resulting $B$ batch means. This procedure is repeated separately for each sequence length $m$. Compared to the empirical mean, the MoM estimator is known for its improved robustness against outliers and favorable performance under heavy-tailed distributions~\cite{LugMen2019,PauHog2025}.

The MoM estimator plays an additional role in the gate-set shadow protocol. Its sub-Gaussian performance~\cite{LugMen2019}, combined with bounds on the dynamic shadow norm, enabled Ref.~\cite{HelIoaKit2023} to derive rigorous sample complexity bounds for simultaneously estimating many average sequence functions $k_{A}(m)$, with only logarithmic overhead in the number of probe operators. The dynamic shadow norm provides an upper bound on the variance associated with estimating $k_{A}(m)$. It further depends on the gate set, the POVM set, input state, as well as on the sequence length $m$. If the dynamic shadow norms remain bounded independently of the number of qubits and sequence length, then estimating a fixed number of average sequence functions requires a number of samples independent of the system size and sequence length~\cite{HelIoaKit2023}.

While we have chosen to keep the present overview as accessible as possible, we note that Ref.~\cite{HelIoaKit2023} formally evaluates the sequence correlation functions $f_A(x_i,\mathbf{g}_i)$ in Eq.(\ref{seqcorrfun_def}) by selecting an irreducible representation of the sampled gates and a probe operator acting within the corresponding subspace (see App.~\ref{fitting_models_app} for further details). The matrix products entering Eq.(\ref{seqcorrfun_def}) can then be evaluated using smaller irreducible representation matrices, rather than the full $4^n\times 4^n$ PTM matrices. In all the RB examples considered here, it is sufficient to take a probe operator with support on a single irreducible subspace, i.e. $A=\mathbbm{P}_i A \mathbbm{P}_i$. In that case, even if we do not explicitly restrict the PTM matrices to a particular irreducible representation, Eq.(\ref{seqcorrfun_def}) reduces to matrix products within the target irreducible subspace. Thus, the classical postprocessing cost depends on the number and dimensions of the irreducible representations required to estimate the target figures of merit, as well as on the efficiency with which the corresponding gate and probe matrices can be stored and manipulated.

In the remainder of this work, we focus on the practical implications of the gate-set shadow framework for randomized benchmarking. We refer to Appendix~\ref{fitting_models_app} for a more technical discussion on the construction of the sequence correlation function and its expectation value. As a final remark, throughout this work, unless stated otherwise, we adopt the standard RB assumptions regarding gate noise: gate noise is Markovian, both time and gate independent, and represented by a completely positive and trace preserving (CPTP) channel~\cite{MagGamEme2011,Wall2018,HelRotOno2022}.

\subsection{Relation to Character Randomized Benchmarking}\label{rel_charRB}
When the benchmarking gate set forms a finite group other than the multiqubit Clifford group, the resulting RB model will generally contain multiple decay factors~\cite{GamCorMer2013, HelXueVan2019, MckCroWoo2020, ClaRieWan2021, HelRotOno2022}. This leads to a multi-parameter fitting problem, which is generally harder to perform in the finite sampling regime, often resulting in large confidence intervals for the estimated parameters~\cite{HelXueVan2019}.

Character RB circumvents this issue by allowing the construction of weighted survival probabilities that isolate the contribution of specific decay factors in the original RB signal, while retaining the SPAM robustness of standard RB~\cite{HelXueVan2019,ClaRieWan2021}. Experimentally, this is achieved by uniformly sampling an additional gate from a smaller set contained within the original gate set (formally corresponding to a subgroup of the original gate set group)~\cite{HelXueVan2019,ClaRieWan2021}. The final step of the protocol, prior to measurement, is the implementation of the inverse gate. Importantly, this inverse is constructed solely from the original RB sequence and excludes the contribution of the additional sampled gate. The measured survival probabilities are subsequently combined through an additional weighted average over the sampled subgroup gates (see Refs.~\cite{HelXueVan2019,ClaRieWan2021} for further details). This weighted averaging procedure selectively filters the original RB signal, yielding survival probabilities with fewer decay parameters. The choice of subgroup and weighting factors is therefore crucial, as it determines which decay contributions remain in the final signal.

In the gate-set shadow protocol, the sequence correlation function defined in Eq.(\ref{seqcorrfun_def}) also acts as a filter through the choice of probe operator $A$, selectively retaining specific decay contributions. Unlike character RB, however, this filtering is performed entirely in post-processing, without requiring the experimental sampling of an additional subgroup during data acquisition. As in character RB, suitable choices of $A$ can therefore reduce the number of decay parameters entering the fitting model for the average sequence function. From this perspective, character RB can be viewed as a particular realization of the gate-set shadow framework corresponding to specific choices of probe operators. In Appendix~\ref{char_and_gss_app} we show how particular choices of $A$ reproduce the same type of fitting model appearing in character RB.

\subsection{State-Space Perspective on the Gate-Set Shadow Protocol} \label{state_space_gss_sec}
In control engineering, an important problem is how to identify a dynamical model that best describes the observed relationship between a system's inputs and outputs~\cite{GueSinMaj2021,kees2011}. System identification addresses this type of problem, aiming to reconstruct dynamical models directly from observed input-output data~\cite{GueSinMaj2021,kees2011}. In this setting, a widely used class of models is provided by discrete-time linear time-invariant (LTI) systems, whose dynamics can be expressed as~\cite{kees2011}:
\begin{equation}
    \begin{split}
        &\mathbf{x}(k+1) = A\mathbf{x}(k) + B \mathbf{u}(k) \; ,   \\
        & \mathbf{y}(k) =  C \mathbf{x}(k) + D \mathbf{u}(k) \; ,
    \end{split}
\end{equation}
where $\mathbf{u}(k) \in \mathbb{R}^m$ is the input signal vector, $\mathbf{x}(k)$ is the system's internal state and $\mathbf{y}(k) \in \mathbb{R}^l$ denotes the output vector. The constant matrices $A \in \mathbb{R}^{n\times n}$, $B \in \mathbb{R}^{n\times m}$, $C \in \mathbb{R}^{l\times n}$ and $D \in \mathbb{R}^{l\times m}$ define the system's internal dynamics, and its couplings to the input and output signals. The goal of system identification is then to determine a realization of the matrices $A, B, C$ and $D$ from the observed input-output data. Assuming zero initial conditions ($\mathbf{x}(0)=\mathbf{0}$), and applying unit impulses at $k=0$, the corresponding output responses can be collected into a matrix $h(k)$, expressed for $k \geq1$ as~\cite{GueSinMaj2021,kees2011}:
\begin{equation}
    \begin{split}
        &  h(k) = C A^{k-1} B \; .
    \end{split}
    \label{state_space_model}
\end{equation}

A similar structure is also present in the gate-set shadow protocol. For fixed input states and POVMs, the average sequence function can be written as
\begin{equation}
    k^{(i,j)}_A(m) = c^{(0)}_{i,j} \; \text{vec} \big( M^{(i)} \big)^{\dagger} \; \Phi^{m-1} \; \text{vec} \big( W^{(j)} \big) \; ,
    \label{single_seq_corr_model}
\end{equation}
where $c^{(0)}_{i,j}$ is the normalization constant, $M^{(i)}$ is built from the POVM elements and $W^{(j)}$ from the input state (see Eqs.(\ref{mapping_to_state_space_model_app1})-(\ref{mapping_to_state_space_model_app2}) in Appendix~\ref{fitting_models_app} for further details). Considering now a collection of input states and POVMs, if we define the stacked measurement and input matrices
\begin{equation}
    C =  \begin{pmatrix}
        &\text{vec} \big( M^{(1)} \big)^{\dagger} \\
        &\vdots \\
        & \text{vec} \big( M^{(p)} \big)^{\dagger} 
    \end{pmatrix} \;, \;
    B = \begin{pmatrix}
    \text{vec} \big(W^{(1)} \big) \; , \cdots \; , \text{vec} \big(W^{(q)} \big)\; 
    \end{pmatrix}
    \;,
\end{equation}
we can construct a matrix of average sequence functions, $Y(m)$, as follows:
\begin{equation}
    Y(m) = C \; \Phi^{m-1} \; B \; .
    \label{sim_LTI}
\end{equation}
The matrix $Y(m)$ has a similar structure to the output response of a multi-input multi-output LTI system (see Eq.(\ref{state_space_model})). From this perspective, the sequence length plays the role of a discrete time coordinate, while $\Phi$ acts as the effective internal dynamics. Thus, using this interpretation, $\Phi$ can be seen as encoding how information propagates as the noisy gate sequence grows in length. This similarity motivates the use of standard reconstruction techniques, such as the Eigensystem Realization Algorithm (ERA)~\cite{GueSinMaj2021,JuaPap1985}, to obtain the effective dynamics encoded by $\Phi$.

This class of methods has previously been employed for the characterization of open quantum systems from observable time-series data~\cite{ZhaSar2015,BenRyaLou2019}. Here, we apply similar ideas to the average sequence functions generated by the gate-set shadow protocol. Rather than trying to reconstruct the full quantum process matrix~\cite{BluGamNie2013,NieGamRud2021,MillOstCor2026}, the dynamical matrix $\Phi$ provides an effective description, encoding information on the average noise over the gate set.

The reconstructed $\Phi$ nevertheless inherits the usual ambiguity of state-space realizations: the matrix $\Phi$ is not unique~\cite{GueSinMaj2021,kees2011}. This non-uniqueness is unavoidable, as any transformation of the form  
\begin{equation}
  C \to CS^{-1} \; , \;  \Phi \to S \Phi S^{-1} \; , \; B \to S B \;,
\end{equation}
leaves the matrix $Y(m)$ in Eq.(\ref{sim_LTI}) invariant. Consequently, the reconstructed model is only defined up to similarity transformations, or equivalently, up to a gauge choice. This introduces limitations on the information that can be extracted from the average sequence functions. We note that a similar issue also arises in GST~\cite{BluGamNie2013,NieGamRud2021,MillOstCor2026}. In Sec~\ref{op_crosstalk_sec}, we provide an explicit example illustrating both the benefits and limitations of reconstructing $\Phi$.

\section{Standard and Interleaved RB}~\label{sta_and_inter_RB_sec}
The main goal of randomized benchmarking (RB) techniques is to characterize the quality of the experimentally implemented quantum operations. Conventionally, we quantify this quality through the average gate fidelity $\bar{F}(\mathcal{\tilde{U}}, \mathcal{U})$, where $\mathcal{U}$ is the ideal unitary channel and $\mathcal{\tilde{U}}$ is its experimental implementation. In the Pauli Transfer Matrix (PTM) notation, we can define the average fidelity as~\cite{KimSilRya2014}
\begin{equation}
   \bar{F}(\mathcal{\tilde{U}}, \mathcal{U}) = \frac{\text{Tr} \big(U^{\dagger} \tilde{U} \big) +d}{d(d+1)} \;,
\label{average_fid}
\end{equation}
where $U$ and $\tilde{U}$ are the PTM representations of the ideal unitary channel $\mathcal{U}$ and the experimentally implemented channel $\mathcal{\tilde{U}}$, respectively, and $d$ is the Hilbert space dimension. Both standard RB~\cite{MagGamEme2011,MagGamEme2012} and interleaved RB~\cite{MagGamJoh2012} aim to estimate average gate fidelities. However, their goals are different: standard RB characterizes the average fidelity of a gate set, while interleaved RB aims to characterize the average fidelity of a \emph{single gate}. Mathematically, standard RB is designed to estimate a decay parameter related to $\bar{F}(\Lambda, \mathcal{I})$, where $\mathcal{I}$ is the identity channel and $\Lambda$ is the noise associated with the gate set~\cite{MagGamEme2012,CarWalEm2019}. Interleaved RB instead strives to find an estimate for $\bar{F}(\mathcal{\tilde{U}}, \mathcal{U})$~\cite{MagGamJoh2012}.

In standard RB, the fidelity is estimated by measuring the average sequence fidelity as a function of the number of applied gates (i.e, the sequence length)~\cite{MagGamEme2011,MagGamEme2012,SilGrep2025}. The experimentally obtained data is then fitted to an exponential decay model as a function of the sequence length $m$, i.e. $b_0+a_0 \; \lambda^m$. When using the Clifford gate set, fully random sequences of gates give rise to $\lambda = \frac{\text{Tr}(\Lambda^{(\text{PTM})})-1}{d^2-1}$. On the other hand, the interleaved sequences employed in interleaved RB produce a fitting model where $\lambda$ is directly related to $\text{Tr}(\Lambda^{(\text{PTM})}_U \Lambda^{(\text{PTM})})$~\cite{MagGamEme2012}. Here, $\Lambda^{(\text{PTM})}_U$ denotes the PTM of the noise channel associated with the experimental implementation of the unitary channel $\mathcal{U}$. While $\text{Tr}(\Lambda^{(\text{PTM})}_U \Lambda^{(\text{PTM})})$ is not directly related to the target fidelity we would like to estimate, we can obtain bounds on $\bar{F}(\mathcal{\tilde{U}}, \mathcal{U})$ by combining standard and interleaved RB~\cite{MagGamJoh2012,CarWalEm2019}.

The usefulness of the gate-set shadow protocol becomes especially apparent when the benchmarking gate set is not the standard multi-qubit Clifford group. In such cases, the average sequence function may contain several decay contributions, and estimating a desired fidelity can require isolating specific decay modes. In the gate-set shadow framework, different choices of $A$ can be used to construct different average sequence functions, each retaining different decay contributions. We next illustrate this postprocessing flexibility using the benchmarking of a CNOT gate as a concrete example. We refer the reader to Apps.~\ref{fitting_models_staRB_app}--\ref{fitting_models_intRB_app} for the derivation of the standard and interleaved fitting models presented in this section.

Although our focus is on demonstrating the flexibility of the gate-set shadow formalism, we note that specialized protocols such as cycle benchmarking~\cite{ErhWallPos2019} can offer more scalable alternatives for estimating the fidelity of multi-qubit gate cycles.

\subsection{Example: Interleaved RB beyond the full Clifford group}
\label{inter_G4_main}
While the Clifford group is commonly used as a gate set in RB like protocols, this might not be the most convenient choice. In particular, if compiling Clifford operations requires long sequences of native gates, the resulting experiments can be demanding in terms of the required coherence time~\cite{XueWatHel2019}. In these settings, using smaller gate sets that more naturally align with the available native gates may provide useful alternatives. This motivates considering RB protocols based on gate stes other than the full multi-qubit Clifford group.

Here, we use the task of benchmarking a CNOT gate as a concrete example. Specifically, we take $G$ to be the CNOT-Dihedral group $G(4)$~\cite{GarCros2020}, which belongs to a family of gate sets that has been employed previously in several RB settings~\cite{CrosMagBis2016,HelXueVan2019,GarCros2020,GarKanLan2021}.

All the gates in the CNOT-Dihedral $G(4)$ gate set can be built from the elementary gates listed in $\mathcal{S}$ (see App.~\ref{G4groupRB} for more details):
\begin{equation}
\begin{split}
    \mathcal{S} = \big\{& X \otimes \mathbbm{1}, \mathbbm{1} \otimes X, S \otimes \mathbbm{1}, \mathbbm{1} \otimes S, \\ 
    &\text{CNOT}_{0,1}, \text{CNOT}_{1,0} \big\}.
\end{split}    
    \label{gate_set_cdi_def}
\end{equation}
In total, $G$ is formed by 768 distinct gates. This corresponds to $1/15$ of the total elements in the two-qubit Clifford group and is $4/3$ the size of the local Clifford group $\mathcal{C}_1 \times \mathcal{C}_1$.

Assuming the noise is gate independent~\cite{MagGamEme2011,MagGamEme2012,GamCorMer2013,HelRotOno2022,SilGrep2025}, the average over random sequences gives rise to an effective noise channel that governs the RB signal. Formally, this effective channel is the average of the original noise channel with respect to the operations in the chosen gate set~\cite{GamCorMer2013,HelRotOno2022,ClaRieWan2021,SilGrep2025}. In the PTM representation, this average can be written as~\cite{ClaRieWan2021}
\begin{equation}
    \Lambda^{\text{(PTM)}}_G = \frac{1}{|G|} \sum_{U \in G} U \Lambda^{\text{(PTM)}} U^{\dagger} \; ,
    \label{twirl_G}
\end{equation}
where $|G|$ is the total number of distinct gate operations in the gate set $G$.

The form of $\Lambda^{\text{(PTM)}}_G$ is determined by how the space generated by the Pauli operators decomposes under the action of the gates in $G$. For the Clifford group, conjugation by Clifford gates permutes, up to a possible phase, the non-identity Pauli operators among themselves. Consequently, the traceless Pauli subspace forms a single invariant subspace under the Clifford action and the average in Eq.(\ref{twirl_G}) produces a single non-trivial decay parameter $\lambda=\frac{\text{Tr}(\Lambda^{\text{(PTM)}})-1}{d^2-1}$~\cite{GamCorMer2013}. This provides the origin of the single exponential fit in standard RB.

For other benchmarking groups, the traceless Pauli subspace can split into smaller invariant subspaces that are not mixed with one another by the action of the gates. When the corresponding representation is multiplicity-free (see definition in App.~\ref{RepGroups_intro}), each resulting subspace is associated with a single scalar decay parameter $\lambda_i$. This is the case for the CNOT-Dihedral gate set considered here. In particular, this gate set contains two non-trivial decay parameters, $\lambda_1$ and $\lambda_2$.

The resulting RB decay signal can therefore depend on both decay factors, unless one constructs an average sequence function that isolates a chosen contribution. Denoting the projector onto each invariant subspace by $\mathbb{P}_i$, and defining $\lambda_i(\mathcal{E})=\frac{\text{Tr}(\mathbb{P}_i\mathcal{E}^{(\text{PTM})})}{\text{Tr}(\mathbb{P}_i)}$, then we can express
\begin{equation}
     \text{Tr}(\mathcal{E}^{(\text{PTM})}) = 1 + \sum^2_{i=1} \text{Tr}(\mathbb{P}_i) \; \lambda_i(\mathcal{E})  \;.
    \label{cdi_trace_formula}  
\end{equation}
Since an estimate for $\bar{F}(\mathcal{\tilde{U}}, \mathcal{U})$ requires us to estimate both $\text{Tr}(\Lambda^{\text{(PTM)}})$ and $\text{Tr}(\Lambda^{(\text{PTM})}_U\Lambda^{\text{(PTM)}})$, we need to estimate $\lambda_i(\mathcal{E})$ for both $\mathcal{E}=\Lambda$ (associated to purely random gate sequences) and $\mathcal{E}=\Lambda_{\mathcal{U}}\circ \Lambda$ (linked with interleaved gate sequences)~\cite{MagGamJoh2012}. In the gate-set shadow protocol, these quantities are isolated in postprocessing using $A=\mathbb{P}_i$ for purely random gate sequences and  $A=\mathbb{P}_iU\mathbb{P}_i$ for interleaved sequences, with $U$ the PTM representation of the ideal interleaved gate (see Apps.~\ref{G4groupRB} and~\ref{G4groupRB_int} for further details on the derivation of $\lambda_i$).

\begin{figure}[t]
\centering
\includegraphics[width=0.99\linewidth]{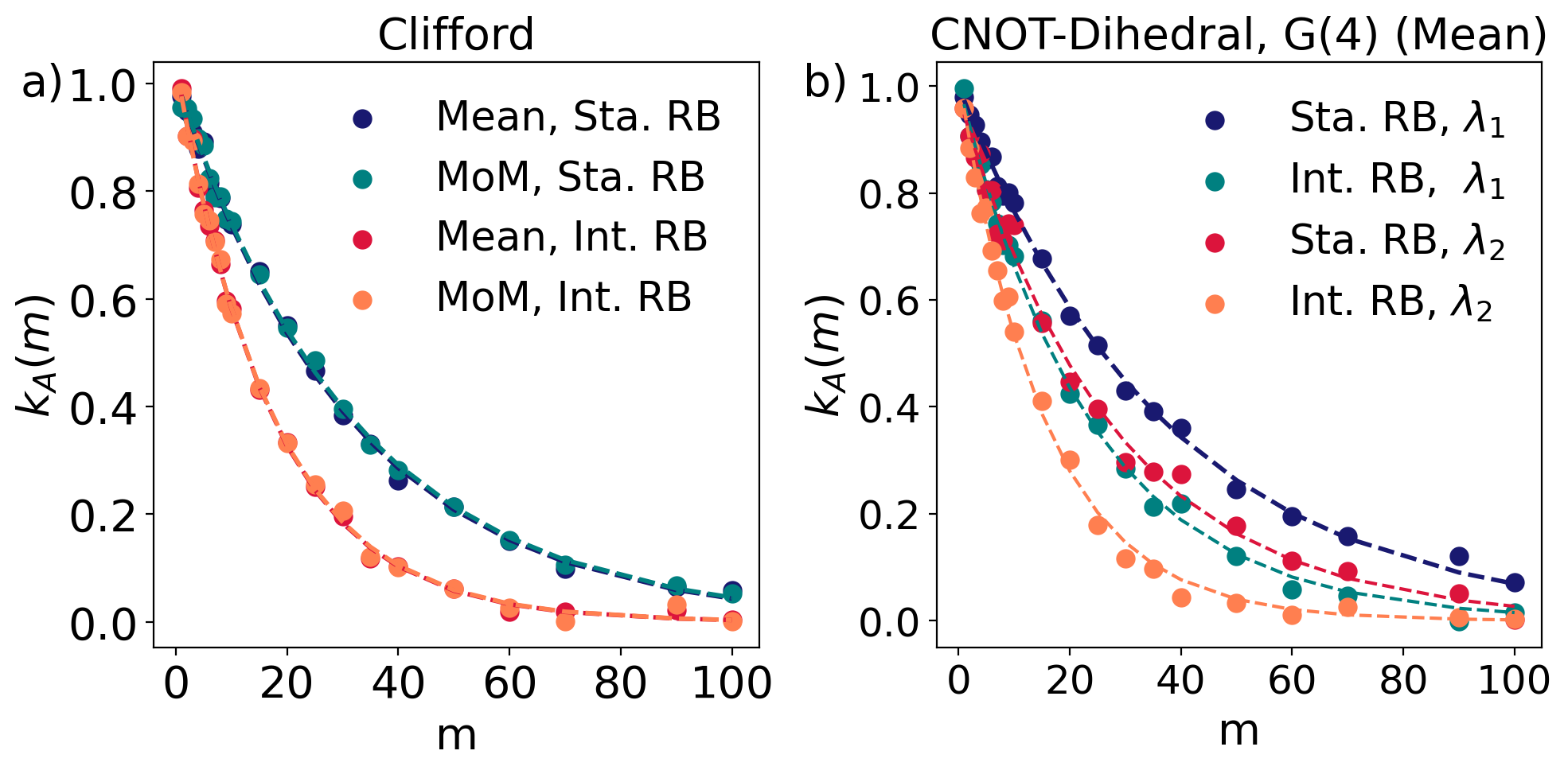}
\caption{\label{fig_RB_cuves_interleaved} Average sequence functions, $k_A(m)$, emulating standard (sta.) and interleaved (int.) RB, with $m$ denoting the sequence length. Panel a) uses the Clifford gate set, while b) uses the $G(4)$ CNOT-Dihedral gate set. For $G(4)$, each average sequence function isolates one decay rate, $\lambda_1$ or $\lambda_2$. The fitted decay rates lead to the single-point estimates of the average gate fidelities in Tab.~\ref{tab_fid_cnot}. For panel b), only the $k_A(m)$ obtained using the sample mean are shown, since both estimators yield similar results. In both panels, the number of randomizations per sequence length was $K=10000$.}
\end{figure}

In Fig.~\ref{fig_RB_cuves_interleaved}, we show the average sequence functions, $k_A(m)$, obtained using either the two-qubit Clifford group or the CNOT-Dihedral $G(4)$ group as gate sets, and with either fully random or interleaved sequences. We perform the gate-set shadow postprocessing step to estimate each decaying factor separately. We note that choosing the Clifford gate set yields essentially the same postprocessing step except that we only need to estimate one decay factor. In the numerical simulations, the noisy CNOT gate is modeled by composing the ideal gate with a coherent two-qubit $zz$ entangling gate, $\mathcal{R}_{zz}(\theta)$, with $\theta=0.1$. Further details of the noise model are provided in App.~\ref{inter_G4_main_app}, including the noise applied to the remaining gates in the gate set.

As discussed, the estimated decay rates $\lambda_i$ provide all the information needed to bound the target gate fidelity~\cite{CarWalEm2019} and obtain the single point estimate of the average gate fidelity proposed in Ref.~\cite{MagGamJoh2012}:

\begin{equation}
    \bar{F}^{\text{(est)}}(\tilde{\mathcal{U}}, \mathcal{U})= 1- \frac{(d-1)}{d} \bigg[1-\frac{ \big(d \;\bar{F}_{\text{int}}-1 \big)}{\big(d \; \bar{F}_{\text{sta}}-1 \big)} \bigg] \; .
\end{equation}
Here, $\bar{F}_{\text{sta}}$ is the average gate set fidelity estimated from averaging over fully random sequences, and $\bar{F}_{\text{int}}$ is the equivalent quantity estimated from averaging over interleaved sequences.
 \begin{table*}[t]\centering
\centering
\begin{tabular}{ccccc}
\hline
 Gate-set & \thead{Average Estimated \\ Gate Fidelity} & \thead{Ground Truth \\ Gate Fidelity} & \thead{Interleaved\\ Fidelity Bounds}  \\ 
 \hline & \\[-1.8ex]
 $\mathcal{C}_2$    & $0.98 \;\;\; (+0.023/-0.023)$    &  $0.986$ & $[0.877, 0.997]$ \\ [2mm] 
 $G(4)$   & $0.98 \; (+0.038/-0.038)$    &  $0.986$ & $[0.870, 0.997]$  \\[1.5mm]
\hline
\hline
\end{tabular}
\caption{\label{tab_fid_cnot} Estimated average fidelity of the CNOT gate, using the single-point estimator of Ref.~\cite{MagGamJoh2012}. Estimates were produced by simulating the gate-set shadow protocol using the two-qubit Clifford gate set, $\mathcal{C}_2$, and the CNOT-Dihedral gate set, $G(4)$. Results were obtained using the sample mean for both groups, since in this case no substantial difference was found between the mean and the median-of-means. Uncertainties are obtained by varying each decay factor by $\pm2$ times its estimated standard deviation (retrieved from the covariance matrix of the fit) and propagating the resulting extrema through the fidelity expressions. Interleaved fidelity bounds are calculated using Eq.\eqref{bound_interleaved_gate} in App.~\ref{C2groupRB_int}.}
\end{table*}

In Tab.~\ref{tab_fid_cnot}, the estimated bound and single-point estimate for the average fidelity of the CNOT gate are provided for both gate sets. Both single-point estimates are comparable in this example, showing that the protocol can reproduce consistent interleaved RB estimates for different choices of gate sets.

As a final remark, we note that choosing between these two particular gate sets also requires adapting the choice of input state and computational basis used for measurements. Indeed, the choice of the Clifford gate set employs the computational $Z-$basis, while for the CNOT-Dihedral gate set the input and POVM elements need to span the relevant irreducible subspaces. Here, this is achieved using a mixed basis, where the first qubit is measured in the $Z-$basis and the second qubit in the $X-$basis (see Eq.(\ref{POVM_and_input_G4}) in App.~\ref{G4groupRB}). While this may at first seem like a disadvantage, it is rather a consequence of the protocol's flexibility: on the one hand, it forces more conscious initial choices of input states and POVM elements; on the other hand, once a suitable dataset has been collected, it provides the postprocessing freedom to extract different decay factors associated with the chosen gate set.

\section{Simultaneous and Correlated RB}
The ability to selectively target and perform operations over a sub-set of qubits in the system is crucial for any successful scalable quantum processor. Such goal requires, for instance, very precise control fields that are able to carry out target operations without unintentionally affecting the state of the idle qubits. Yet, control fields can sometimes spillover to other qubits outside of the target set and induce unintended changes. These errors are termed operation or classical crosstalk~\cite{SarProRud2020, GamCorMer2013}. Additionally, residual interactions between qubit subsystems may persist, coupling their dynamics and preventing selective operational control~\cite{SarProRud2020}.

Within the framework of RB techniques, simultaneous RB offers a route to detect signatures of crosstalk~\cite{GamCorMer2013}. The procedure compares error rates obtained from RB experiments performed separately on individual qubit partitions with those obtained when RB is applied simultaneously to all partitions~\cite{GamCorMer2013}.

In addition to operational crosstalk, it is important to know whether noise sources induce correlations between qubit subsystems, since correlated errors can be particularly damaging for quantum error correction~\cite{KleFran2005,Fow2013,Goo2025}. Correlated RB is a step towards distilling information on the locality and weight of crosstalk errors from simultaneous RB experiments~\cite{MckCroWoo2020}. In this section, we explore how the gate-set shadow protocol may be used to extract similar information to simultaneous and correlated RB. Additionally, we provide examples on how the protocol may be used to gain further insights into the noise affecting an idle qubit and how the crosstalk tomography method proposed in Ref.~\cite{HelIoaKit2023} can be further postprocessed into effective Pauli error rates. For the sake of concreteness, we will focus on a two-qubit system.

\subsection{Benchmarking with an idle qubit}\label{op_crosstalk_sec}
To illustrate the protocol, we consider the task of characterizing the average fidelity of operations acting on a target qubit, while the second qubit remains idle. Hence, we use the gate set $\mathcal{C}_1 \times \mathcal{I}$, where only target operations are applied to the first qubit.

For the gate set $\mathcal{C}_1 \times \mathcal{I}$, we would like to characterize the average fidelity of the effective noise channel acting on the target qubit, namely $\Lambda^{(eff)}_1(\rho_1)= \text{Tr}_2 \Big[ \Lambda(\rho_1 \otimes I/2) \Big]$~\cite{GamCorMer2013}, where $\Lambda$ is the noise channel acting on the full two-qubit system and the idle qubit is taken to be maximally mixed. The resulting average fidelity is given by
\begin{equation}
    \bar{F}_1
    =
    \frac{1+\text{Tr}(\mathbb{P}_1\Lambda)+d_1}
    {d_1(d_1+1)}
    =\lambda_1+\frac{1-\lambda_1}{d_1} \; ,
    \label{c1xI_fid}
\end{equation}
with 
\begin{equation}
    \lambda_1=\frac{\text{Tr}(\mathbb{P}_1\Lambda)}
    {d_1^2-1},
    \label{lambda1_decay}
\end{equation}
In the equation above $d_1=2$ and $\mathbb{P}_1$ denotes the projector onto the traceless operator subspace of the first qubit (see Eq.(\ref{P1_proj_C1xI}) in App.~\ref{group_c1xI_sec} for a precise definition of the projector). In the absence of correlated noise, $\bar{F}_1$ corresponds to the average gate fidelity of the operations acting solely on the first qubit. However, in the presence of correlations, the effective noise experienced by the target qubit may depend on the state of the second qubit. Hence, $\bar{F}_1$ should be interpreted as the average fidelity of the effective channel $\Lambda^{(eff)}_1$~\cite{GamCorMer2013,GirCirJen2022}.

Inspecting Eq.(\ref{lambda1_decay}), a natural choice of probe operator is $A=\mathbb{P}_1$. However, because the gate set $\mathcal{C}_1 \times \mathcal{I}$ gives rise to multiplicities, even this choice does not, in general, guarantee an average sequence function with a single decay factor, and extracting $\lambda_1$ from the resulting signal becomes potentially non-trivial (see App.~\ref{group_c1xI_sec} for further details). An important simplification occurs in the absence of SPAM noise and for uncorrelated gate noise. In this case, using the input state and POVM elements defined in Eq.(\ref{POVM_input_C1xI_app}) of App.~\ref{group_c1xI_sec}, the average sequence function associated with $A=\mathbb{P}_1$ reduces to a single decay model. The resulting signal takes the simple form $k(m)\propto\lambda_1^{m-1}$, allowing $\lambda_1$ to be extracted directly from a single exponential fit. This simplification, however, relies on the simultaneous absence of both gate correlations and SPAM errors. While it is possible that other noise processes (including some correlated noise channels) may reduce to $k(m)\propto\lambda_1^{m-1}$, the uncorrelated limit is special in the sense that this simplification does not require any further assumptions on the individual noise channels beyond them being CPTP.

Rather than seeking conditions under which the average sequence function reduces to a single exponential fit, one may instead embrace the fact that the observed signal is, in general, governed by multiple decay modes. From this perspective, the state-space framework introduced in Sec.~\ref{state_space_gss_sec} can be a useful approach. Instead of focusing on a single average sequence function, it aims to reconstruct the effective dynamics underlying the observed sequence data. This reconstruction does not rely on the assumptions required for the emergence of the single decay model introduced previously. We therefore investigate whether it can be used both to estimate $\lambda_1$ and to extract additional information about the noise process acting on the idle qubit.

We then proceed by considering the following set of input states and POVM elements:
\begin{equation}
\begin{split}
   & \rho^{(\nu)}= \frac{1}{4} \big( \mathbbm{1} + Z \big) \otimes \big(\mathbbm{1} + \mathbf{r}^{\intercal}_{\nu} \boldsymbol{P} \big) \;, \\
   & E^{(\mu)}_0 =  \frac{1}{2} \big(\mathbbm{1}\otimes \mathbbm{1} \big) + \frac{1}{2} \big( Z \otimes P_{\mu} \big) \; , \\
   & E^{(\mu)}_1 =  \frac{1}{2} \big(\mathbbm{1}\otimes \mathbbm{1} \big) - \frac{1}{2} \big( Z \otimes P_{\mu} \big) \;,
\end{split} 
\label{States_and_POVMs_C1xI}
\end{equation}
where $\boldsymbol{P}=(X,Y,Z)$, $\mathbf{r}_1=\frac{1}{\sqrt{3}}(1,1,1)$, $\mathbf{r}_2=\frac{1}{\sqrt{3}}(1,-1,-1)$, $\mathbf{r}_3=\frac{1}{\sqrt{3}}(-1,1,-1)$ and $\mathbf{r}_4=\frac{1}{\sqrt{3}}(-1,-1,1)$ are vectors on the Bloch sphere and $P_{\mu} \in \{\mathbbm{1}, X,Y,Z\}$ is a Pauli operator. The vectors $\mathbf{r_\nu}$ correspond to the vertices of a regular tetrahedron inscribed in the Bloch sphere. This choice gives rise to four linearly independent single-qubit states, $\frac{1}{2}\big(\mathbbm{1} + \mathbf{r}^{\intercal}_{\nu} \boldsymbol{P}\big)$~\cite{JarEngBert2004}. Together with the four POVM elements defined above, these settings provide sixteen distinct state-preparation and measurement configurations. For each of these configurations, we perform an RB experiment on the first qubit using the gate set $\mathcal{C}_1 \times \mathcal{I}$. Therefore, the protocol combines RB on the first qubit with a tomographic characterization of the idle qubit.

Since the states $\rho^{(\nu)}$ and the POVM elements $E^{(\mu)}_x$ are the state preparation and measurement settings of the protocol, imperfections in their implementation give rise to SPAM errors. In the state-space analogy, they affect only the matrices $B$ and $C$ (see Eq.(\ref{sim_LTI})). Hence, while the effective dynamics $\Phi$ remains unchanged, SPAM errors can suppress the amplitudes of certain decay modes, making them difficult to resolve under finite sampling. This is analogous to conventional RB, where SPAM errors can similarly obscure the contribution of certain decay modes~\cite{HelRotOno2022}. For simplicity, the simulations reported in this section include no SPAM errors.

Each configuration yields an average sequence function $k^{(\mu,\nu)}_A(m)$, where the probe operator $A$ remains fixed and is shared by all sequence functions (see Appendix~\ref{group_c1xI_sec} for further details). Collecting all resulting sequence functions into a matrix allows the problem to be expressed in the state-space like form. In this case, the effective dynamics is described by a $4\times 4$ matrix $\Phi$, whose matrix elements are given by
\begin{equation}
    \Phi_{\mu,\nu}=\frac{1}{d^2_1-1} \sum^3_{n=1} \text{Tr}\Big( (\sigma_n\otimes \sigma_{\mu}) \; \Lambda \big(\sigma_n\otimes \sigma_{\nu} \big)  \Big) \; ,
\end{equation} 
where $\sigma_{i}$ denotes a normalized Pauli operator. For uncorrelated noise, the matrix elements simplify to
\begin{equation}
    \Phi_{\mu,\nu}= \lambda_1 \; \text{Tr}\Big(\sigma_{\mu} \; \Lambda_2(\sigma_{\nu})\Big) \;,
    \label{Phi_uncorrelated}
\end{equation} 
with $\Lambda_2$ the noise channel acting on the second qubit. Thus, in the uncorrelated limit, the effective dynamics separates naturally into a scalar contribution set by $\lambda_1$ and a matrix describing the second qubit noise process.

As described in Sec.~\ref{state_space_gss_sec}, techniques such as ERA can be used to estimate $\Phi$ from the average sequence data. However, any realization obtained in this manner is only unique up to a matrix similarity transformation (or, equivalently, up to a gauge choice). Furthermore, while $\Lambda_2$ is expected to be CPTP, this property is not generally preserved under arbitrary similarity transformations. Consequently, even when the underlying noise is uncorrelated and a reconstructed realization of the form $\Phi=pR$ is obtained, the matrix $R$ does not need to correspond to a physical CPTP noise process. We therefore focus on properties that remain invariant under gauge transformations, such as the eigenspectrum of $\Phi$.

For a single-qubit channel, the trace-preserving condition fixes one eigenvalue to be one, while the remaining three eigenvalues are either all real or consist of one real eigenvalue together with a complex conjugate pair~\cite{HelBatTer2019,WonZenLi2026}. Pauli and amplitude-damping noise possess a purely real eigenspectrum, while generic unitary rotations generate a complex-conjugate pair of eigenvalues. Hence, the appearance of a complex-conjugate pair provides a spectral signature of a rotational component in the effective noise process acting on the idle qubit, consistent with the presence of coherent noise. However, finite sampling can introduce small spurious imaginary components in the reconstructed eigenvalues, even when the underlying noise has a purely real spectrum. In practice, these spurious imaginary components are expected to decrease as the amount of sampling increases.

The numerical examples in the remainder of this subsection all draw from the same one-parameter family of two-qubit noise channels,
\begin{equation}
    \Lambda(\rho)=(1-\varepsilon) \; \Lambda_{\text{unc}}(\rho) + \; \varepsilon \; \Lambda_{\text{corr}}(\rho) \; ,
\label{noise_model_mix_uncor_cor}
\end{equation}
where $\Lambda_{\text{unc}}=\Lambda_1\otimes\Lambda_2$ is a product of single-qubit channels acting on the operated and the idle qubit respectively, $\Lambda_{\text{corr}}=R_{xx}(\theta_{\text{corr}})$ is a coherent two-qubit $XX$ rotation with angle $\theta_{\text{corr}}$, and $\varepsilon\in[0,1]$ sets the relative weight of the correlated component. At $\varepsilon=0$, the noise is uncorrelated, so that $\Phi$ factorizes as in Eq.(\ref{Phi_uncorrelated}). As $\varepsilon$ increases, deviations from this form become increasingly significant, until the factorized ansatz is no longer a valid approximation. Unless stated otherwise, the channel on the operated qubit is $\Lambda_1=R_z(\theta_1)$ (with $\theta_1=0.2$), where $R_z(\theta)$ denotes a rotation about the $z$-axis by an angle $\theta$.

\begin{figure*}[t]
\centering
\includegraphics[width=\linewidth]{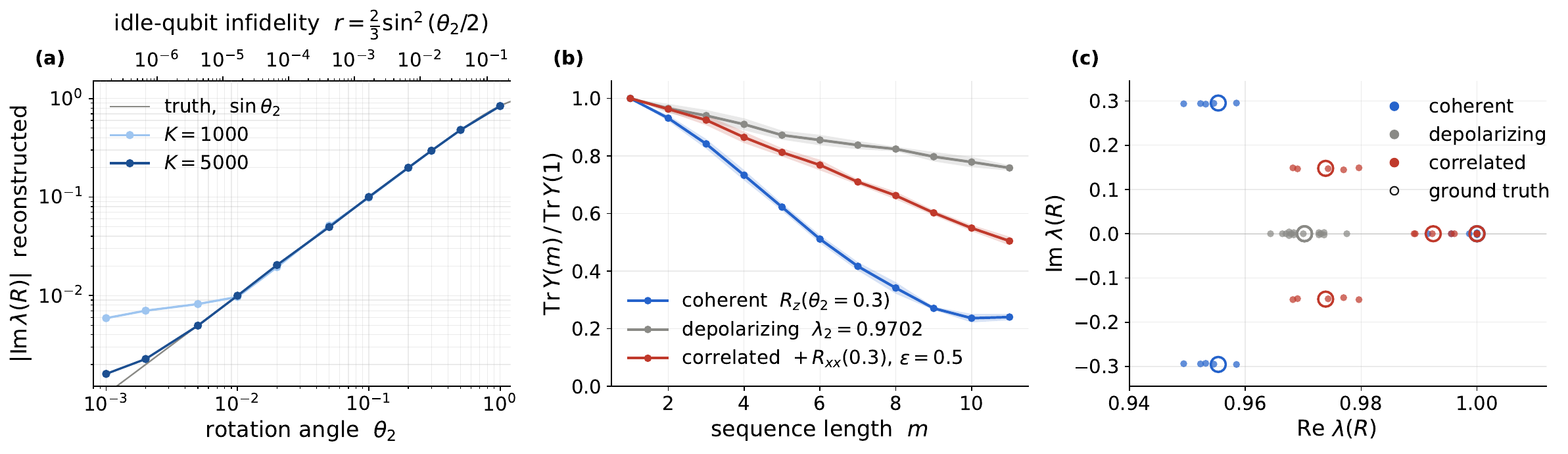}
\caption{Distinguishing noise channels with identical average gate fidelities through their reconstructed eigenspectra. (a) Median $|\mathrm{Im}\,\lambda(R)|$ of the reconstructed idle-qubit channel against the true rotation angle, for $\Lambda_2=R_z(\theta_2)$ at $\varepsilon=0$ with $\theta_1=0.2$ fixed. The median is evaluated from five independent repetitions. The grey line represents the exact value $\sin\theta_2$, while the upper axis shows the corresponding idle-qubit infidelity, $r=\frac{2}{3}\sin^2(\theta_2/2)$. At $K=5000$, the reconstruction tracks the ground-truth down to $\theta_2\approx 5\times 10^{-3}$, corresponding to an infidelity of order $10^{-6}$, before finite sampling errors dominate. Figs. (b) and (c) compare three noise models, all sharing the same effective idle-qubit average fidelity $\bar{F}=0.985112$: a coherent rotation $ R_z(\theta_1=0.2)\otimes R_z(\theta_2=0.3)$, a product of two depolarizing channels with with depolarizing parameters $\lambda_1=0.986711$ and $\lambda_2=0.970224$, and an equal mixture of a coherent rotation $R_z(\theta_2=0.3)$ with correlated $R_{xx}(0.3)$, corresponding to setting $\varepsilon=0.5$ in Eq.(\ref{noise_model_mix_uncor_cor}). The correlated rotation is chosen to have the same idle-qubit marginal fidelity as $R_z(0.3)$. For the first two noise models, the average gate fidelity of the operated qubit is also the same. This makes their resulting $\Phi$ differ only through $\Lambda_2$ (the noise on the idle qubit). (b) Trace of the matrix of average sequence functions, normalized to its value at $m=1$. The data points correspond to the mean over five independent repetitions, with shaded bands corresponding to $\pm$ one standard deviation. Since the protocol does not twirl the idle subsystem, the signals retain information not captured by conventional RB decay rates. (c) Eigenvalues of $R=\Phi/p$: the filled markers correspond to estimates arising from the five independent reconstruction repetitions, while open markers are the exact eigenvalues. The coherent rotation channel has a doubly degenerate real eigenvalue ($\lambda(R)=1$), while the depolarizing channel has a triply degenerate eigenvalue at $\lambda(R)=0.970224$. The median values of $|\mathrm{Im}\,\lambda(R)|$ are $0.2943$, $0.1473$ and $0.0024$ for coherent, correlated and depolarizing, respectively. Sampling noise causes the estimated degenerate eigenvalues to split up for both coherent and depolarizing models. In the correlated noise model, the two real eigenvalues are separated by $0.0076$. However, this difference is too small to be resolved at the present sampling precision. All simulations in (b) and (c) employed $K=5000$.}
\label{fig_coh_vs_stoch}
\end{figure*}

To investigate whether a small rotation angle can produce a detectable signature in $\Phi$, we simulated the protocol at $\varepsilon=0$, with the idle-qubit noise channel  given by $\Lambda_2=R_z(\theta_2)$. We varied the angle $\theta_2$ over twelve values from $0$ to $1$, performing five independent repetitions for each value. For the coherent $R_z$ noise considered here, the protocol effectively yields a Ramsey-like experiment on the idle qubit, performed simultaneously with a standard RB experiment on the first qubit. In this case, $R=\Phi/\lambda_1$ corresponds to the PTM of the idle-qubit rotation $R_z(\theta_2)$, whose eigenvalues are $\lambda(R)=\{1,1,e^{\pm i \theta_2}\}$. For the reconstructed matrix $R$, we let $|\mathrm{Im}\,\lambda(R)|$ denote the largest absolute imaginary part of its eigenvalues. As shown in Fig.~\ref{fig_coh_vs_stoch}~(a), $|\mathrm{Im}\,\lambda(R)|$, follows the ground-truth value $\sin\theta_2$ until it saturates on a floor set by finite sampling that decreases as $K$ grows.

The usefulness of this spectral information becomes particularly clear when two channels cannot be told apart by a single decay rate. Consider two noise models acting on the idle qubit: a coherent rotation $R_z(\theta_2=0.3)$ and a depolarizing channel with depolarizing parameter $\lambda=0.970224$. Both channels give rise to the exact same (single-qubit) RB decay rate, $\lambda_{RB}=0.9702243$. Their spectra, however, differ qualitatively: the coherent channel carries a conjugate pair $e^{\pm i\theta_2}$ together with a doubly degenerate eigenvalue at one, whereas the depolarizing channel carries a triply degenerate real eigenvalue. Fig.~\ref{fig_coh_vs_stoch}~(b) and (c) show the measured sequence functions and the reconstructed spectra for both noise models, simulated at $K=5000$ and with the single-qubit fidelity of the first qubit matched as well. Hence, both their corresponding $\Phi$ matrices differ only through $\Lambda_2$. The reconstruction reproduces this difference: the coherent case returns a conjugate pair at $0.9532\pm 0.2930i$, with imaginary parts two orders of magnitude larger than the values of $\pm 0.0025i$ induced by finite sampling in the depolarizing case. Hence, the two noise models yield distinct, detectable spectral features.

We can also consider a third noise channel, this time with a correlated component. Concretely, we consider the channel produced by setting $\varepsilon=0.5$ and $\Lambda_1 \otimes\Lambda_2=R_z(\theta_1)\otimes R_z(\theta_2)$, with $\theta_{\text{corr}}=\theta_2$, in Eq.(\ref{noise_model_mix_uncor_cor}). In order to compare the average gate fidelity of the noise affecting the idle qubit with that of the two previous uncorrelated models, we consider the effective channel: $\Lambda^{\text{(eff)}}_2=\text{Tr}_1 \big( \Lambda(\mathbbm{1}/2 \otimes \rho_2) \big)$. Tracing out the maximally mixed qubit reduces the $XX$ rotation to a bit-flip channel on the idle qubit, with error probability $\sin^2(\theta_2/2)$. Since this channel has the same average gate fidelity as $R_z(\theta_2)$, their mixture retains the same fidelity, $\bar{F}=0.9851122$. Thus, all three cases in Fig.~\ref{fig_coh_vs_stoch} share the same average gate fidelity for the effective noise channel acting on the idle qubit. Nevertheless, the eigenspectrum reconstruction distinguishes all three cases: the median $|\mathrm{Im}\,\lambda(R)|$ over five repetitions is $0.2943(9)$ for the coherent case, $0.1473(17)$ for the correlated one and $0.0024(13)$ for the depolarizing case, against the exact values of $0.2955$, $0.1478$ and zero, respectively.

\begin{figure*}[t]
\centering
\includegraphics[width=\linewidth]{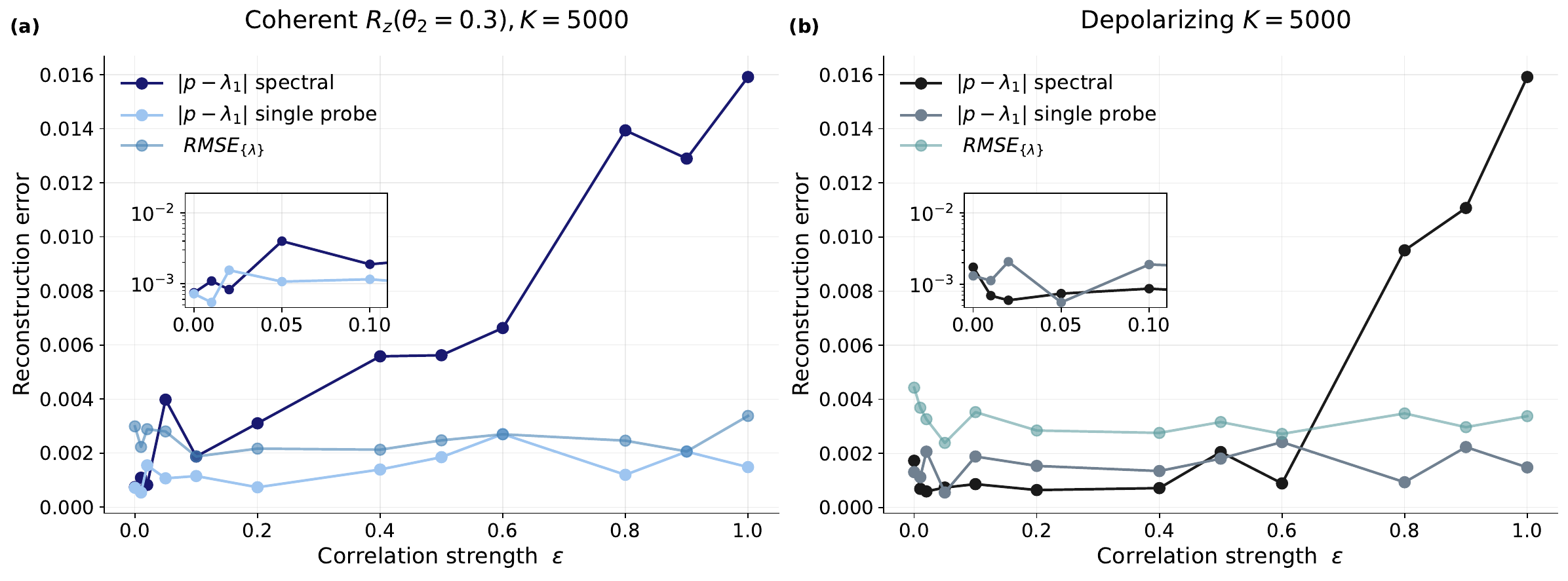}
\caption{With $K=5000$, we repeat the reconstruction of $\Phi$ five independent times for each value of the correlation strength $\varepsilon$. For each reconstruction, we compute the absolute error in the estimate of $\lambda_1$ and the spectral root mean square error ($RMSE_{\{\lambda\}}$). The $RMSE_{\{\lambda\}}$ is obtained by optimally pairing the estimated eigenvalues with those of the true eigenspectrum. Separately, we apply the gate-set shadow protocol with probe operator $A=\mathbb{P}_1$ to estimate $\lambda_1$ and, likewise, independently repeat each simulation five times, for each value of correlation strength. For both methods, $p$ denotes the estimate of $\lambda_1$, obtained either from the single-probe fit or from the largest real eigenvalue of the reconstructed $\Phi$. The curves show the median of these errors over the five repetitions, at each value of $\varepsilon$. The insets to the figures show a close-up of the absolute errors in the estimates of $\lambda_1$ in the low-correlation regime. The term "Reconstruction error" is used on the vertical axis as a collective label for the three error metrics. While the $RMSE_{\{\lambda\}}$ and the single probe estimates remain approximately constant over the full $\varepsilon$ range, the error in the spectral estimate of $\lambda_1$ increases with correlation strength. This reflects the fact that identifying $\lambda_1$ as the largest real eigenvalue is only guaranteed to be accurate in the uncorrelated limit.}
\label{fig_increasingcorr}
\end{figure*}

Although the reconstructed eigenspectra in Fig.~\ref{fig_coh_vs_stoch}~(c) distinguish all three noise models, their physical interpretation requires care. In the presence of noise correlations, the factorization in Eq.(\ref{Phi_uncorrelated}) no longer holds in general. This not only prevents the spectrum of $\Phi$ to be interpreted (up to a scaling factor) as that of the noise channel acting on the idle-qubit, but can also complicate the estimation of $\lambda_1$ from the reconstructed eigenspectrum. In the uncorrelated limit, $\lambda_1$ coincides with the largest real eigenvalue of $\Phi$. However, in the presence of correlations, this correspondence does not need to hold. This is illustrated in Fig.~\ref{fig_increasingcorr} (a) and (b), where estimating $\lambda_1$ from the largest real eigenvalue yields larger errors than the single exponential fit with $A=\mathbb{P}_1$, particularly at higher $\varepsilon$. This difference in performance arises because, for the noise models considered here, $\lambda_1$ remains an eigenvalue of $\Phi$ that can be directly isolated by choosing the probe $A=\mathbb{P}_1$. For the depolarizing model, $\lambda_1$ remains the largest real eigenvalue of $\Phi$ up to higher values of $\varepsilon$, explaining why the spectral estimate remains competitive over a wider range. In contrast, for the coherent model, this correspondence breaks down at smaller values of $\varepsilon$. Nevertheless, the errors in the reconstructed eigenvalues of $\Phi$ remain relatively stable across the entire $\varepsilon$ range. Thus, an accurate reconstruction of the eigenspectrum does not necessarily grantee and accurate estimate of $\lambda_1$.

We emphasize that, for arbitrary correlated noise, $\lambda_1$, as defined in Eq.(\ref{lambda1_decay}), is not necessarily an eigenvalue of $\Phi$. Consequently, the single-probe fit is not guaranteed to recover $\lambda_1$. More generally, RB decay rates are determined by the eigenvalues of the effective dynamics, so quantities that cannot be related to these eigenvalues may not be accessible as SPAM robust decay rates. It is therefore important to assess whether noise correlations are present before assigning a physical interpretation to the reconstructed spectrum. One approach is to perform simultaneous RB and evaluate the correlation coefficient $\delta \alpha =\alpha_{12}- \alpha_{1|2}\alpha_{2|1}$~\cite{GamCorMer2013}. Although this test does not detect all forms of correlated noise~\cite{GamCorMer2013}, obtaining a statistically significant nonzero value for $\delta \alpha$ provides evidence for the presence of correlated noise.

We may wonder whether the eigenspectrum of $\Phi$ can itself reveal the presence of noise correlations. For the protocol considered here, uncorrelated noise implies the factorization in Eq.(\ref{Phi_uncorrelated}), where $\Phi/\lambda_1$ is the PTM of a single-qubit CPTP channel. Hence, its eigenvalues must satisfy the spectral compatibility conditions for single-qubit CPTP channels, which can be expressed as a set of tetrahedron inequalities~\cite{WolGar2010,RudPucZyc2018}. Violating these inequalities provides then evidence of noise correlations. Yet, the converse is not true, as illustrated by the correlated $XX$ noise model considered here. In App.~\ref{group_c1xI_sec}, we examine two additional correlated noise models. For the first, both $\delta \alpha$ and the eigenspectrum compatibility criterion detect correlations, while for the second, only the latter does so. These results suggest that the eigenspectrum compatibility criterion can complement simultaneous RB in detecting correlations in small qubit subsystems.

In the uncorrelated noise limit, the gauge ambiguity inherent to the reconstruction of $\Phi$ prevents us from directly estimating quantities such as the unitarity or the strength of the non-unital component of $\Lambda_2$, since these are not invariant under gauge transformations~\cite{RudPucZyc2018}. Nevertheless, the eigenvalues of a CPTP channel can provide bounds on its unitarity and unitality. For $\Lambda_2$, we can therefore use the reconstructed eigenspectrum of $\Phi/p$, where $p$ is an estimate of $\lambda_1$, to obtain the following bounds~\cite{RudPucZyc2018,HelBatTer2019}:
\begin{equation}
    u(\Lambda_2)  \geq \frac{1+\sum^3_{k=1} |\phi_k|^2 - d}{d(d-1)} \; ,
    \label{unitarity_bound}
\end{equation} 
\begin{equation}
    ||\vec{w}||^2 \leq 1 - |\phi_1|^2 - |\phi_2|^2 - |\phi_3|^2 + 2\phi_1 \phi_2 \phi_3 \; .
    \label{non_unital_bound}
\end{equation}
Here, $\phi_1$, $\phi_2$ and $\phi_3$ denote the three nontrivial eigenvalues of $\Phi/p$, $u(\Lambda_2)$ denotes the unitarity of $\Lambda_2$, and $\vec{w}$ is its non-unital translation vector in the PTM representation (see Eq.(\ref{PTM_matrix_gen}) in App.~\ref{PTM_Rep_intro}). Evaluating Eqs.(\ref{unitarity_bound}) and (\ref{non_unital_bound}) for the two uncorrelated noise channels of Fig.~\ref{fig_coh_vs_stoch} gives $u \geq 0.9931 \pm 0.0086$ and $||\vec{w}||^2 \leq 0 \pm 0.0001$ for the coherent rotation, and $u \geq 0.9115 \pm 0.0058$ and $||\vec{w}||^2 \leq 0.0026 \pm 0.0004$ for the depolarizing channel. These agree with the corresponding exact bounds of $u \geq 1$ and $||\vec{w}||^2 \leq 0$ for the coherent rotation, and $u \geq 0.9120$ and $||\vec{w}||^2 \leq 0.0026$ for the depolarizing channel. For comparison, the exact unitarity of each channel is $u_{\text{coherent}}=1$ and $u_{\text{depolarizing}}=0.9413$, while both channels are unital ($||\vec{w}||^2=0$). The estimates were obtained from the reconstructed eigenvalues of five independent realizations of the protocol. The reported values correspond to their means, with uncertainties given by their standard deviations. Finite sampling can yield unphysical estimates of these bounds, such as a negative upper bound on $||\vec{w}||^2$. Nevertheless, the estimated bounds for the coherent rotation indicate a nearly unitary and unital channel, while those for the depolarizing channel provide weaker constraints. Taken together with the reconstructed eigenspectra, these results illustrate how gauge invariant information can provide a more comprehensive picture of the underlying gate noise, even when gauge ambiguity complicates its complete characterization.

The use of eigenspectra as a diagnostic tool for quantum noise forms the basis of protocols such as spectral quantum tomography and channel spectrum benchmarking~\cite{HelBatTer2019,YaZhuChai2023}. In those approaches, the eigenvalues of a noisy quantum gate are extracted directly from experimentally observed signals and used to infer properties of the underlying noise process affecting the target gate. Here, the spectrum instead arises from the reconstructed effective dynamics governing the gate-set shadow average sequence functions. In the uncorrelated limit, its reconstructed eigenspectrum is directly related to the noise process acting on the idle qubit. However, extending this reconstruction to larger idle subsystems comes at an increasing experimental cost. If $q_2$ is the number of idle qubits, and the gate set has the form $\mathcal{C}_{q_1}\times \mathcal{I}_{q_2}$, then the number of combined input and measurement configurations would scale as $16^{q_2}$. The present approach is then most useful for the characterization of small idle subsystems.

From a practical perspective, the single exponential fit and state-space approaches provide complementary information. When the absence of noise correlations is statically plausible, the single exponential fit provides a simpler root to estimate $\lambda_1$. Independently of this estimate, the ERA reconstruction gives access to gauge invariant quantities such as the eigenspectrum of the effective dynamics, offering additional insight into the nature of the underlying noise process.

\subsection{Detecting correlations with simultaneous benchmarking} \label{correlated_crosstalk_sec}
Having analyzed how correlations can complicate the physical interpretation of target figures of merit, we now focus on a more detailed characterization of correlated noise itself. The use of the gate-set shadow approach for crosstalk tomography was already proposed in the original work introducing the protocol~\cite{HelIoaKit2023}. Here, we build on the reconstruction of unital marginals presented in Ref.~\cite{HelIoaKit2023} to obtain more interpretable diagnostics of correlated noise.

It is instructive to compare the present reconstruction scheme with the approach of Sec.~\ref{op_crosstalk_sec}. There, the gate set leaves one qubit idle, and the reconstructed matrix is related to the noise channel acting on that qubit in the uncorrelated noise limit. Here, both qubits are twirled, and the reconstruction targets the unital marginals of the joint two-qubit channel. These correspond to the diagonal blocks of the channel's PTM associated with the irreducible subspaces of the local Clifford group. These blocks are reconstructed from individually gauge invariant decay rates. The resulting description is necessarily incomplete, as it discards contributions coupling inequivalent irreducible subspaces, and thus may fail to capture some forms of correlated noise. Nevertheless, it provides access to the weight-two PTM block, allowing correlations to be detected by comparing this block with the tensor product of the weight-one unital marginals.

We will first review how the same gate-set shadow data can be postprocessed to obtain both scalar and matrix valued crosstalk metrics, as well as how the corresponding unital marginal channels can be reconstructed. We then map these unital marginals to effective Pauli error rates, allowing us to identify which Pauli components dominate the reconstructed crosstalk signal. Pauli based diagnostics have also been used in other contexts, such as correlated randomized benchmarking~\cite{MckCroWoo2020} and the quantum noise learning protocol proposed in Ref.~\cite{HarFlaWall2020}. Here, we apply them directly to the unital marginals reconstructed from the gate-set shadow sequences.

One immediate advantage of employing the gate-set shadow protocol is that the correlated metric $\delta \alpha=\lambda_{12}-\lambda_{1|2}\lambda_{2|1}$, introduced in Ref.~\cite{GamCorMer2013}, can be estimated while retaining robustness to SPAM errors. Standard application of simultaneous RB requires taking additional measurements and combining them appropriately to extract individual estimates for each decay rate~\cite{GamCorMer2013}. This strategy, however, is sensitive to SPAM noise. Character RB applied to the gate set $\mathcal{C}_1\times\mathcal{C}_1$ provides a SPAM robust route to estimate the individual decay rates~\cite{HelXueVan2019,XueWatHel2019}. The gate-set shadow protocol achieves the same goal directly in postprocessing: different decay rates can be estimated from the same dataset by choosing appropriate probe operators. Further details on the corresponding fitting models for the local Clifford gate set are provided in App.~\ref{fitting_models_simRB_app}.

Crosstalk tomography relies on the reconstruction of the unital marginals of the gate noise channel~\cite{HelIoaKit2023}. The unital marginals are defined in the PTM formalism as $\Lambda^{\text{(PTM)}}_i = \mathbb{P}_i \Lambda^{\text{(PTM)}} \mathbb{P}_i$~\cite{HelIoaKit2023}. They can be estimated from a single gate-set shadow dataset by repeating the postprocessing step with $A$ equal to each local Clifford element $C$. For each probe choice $(A=C)$, the corresponding average sequence function provides a decay rate $\lambda_{i,C}$. This allows us to estimate the unital marginals as

\begin{align}
\Lambda^{\text{(PTM)}}_i = c_i \; \sum_{C \in \mathbbm{C}_i} \lambda_{i,C} \; \mathbb{P}_i C \mathbb{P}_i,
\label{unital_marginal_c1xc1_sec}
\end{align}
where $c_i$ is a normalization constant, and $\mathbbm{C}_i$ denotes the set of local Clifford elements required to reconstruct the unital marginal $\Lambda^{\text{(PTM)}}_i$ (see App.~\ref{unital_rec_app}, Eq.(\ref{marginals_recformula}), for further details).

The crosstalk metric $\delta \alpha$ can be generalized to the level of the unital marginals, as proposed in Ref.~\cite{HelIoaKit2023}. This leads to the following metric:
\begin{equation}
    \delta^{(2)}_{CT}= |\big( \Lambda^{\text{(PTM)}}_1 \otimes \Lambda^{\text{(PTM)}}_2\big)^{-1}\Lambda^{\text{(PTM)}}_3| \; .
    \label{crosstalk_metrics}
\end{equation}
The presence of correlations is reflected in $\delta^{(2)}_{CT}$ by its deviation from the identity matrix. For the coherent $xx$, $yy$, and $zz$ entangling errors considered here, these deviations are most noticeable in the diagonal elements, as illustrated in App.~\ref{unital_rec_app}.

To characterize weight and locality of these errors, as in crosstalk RB~\cite{MckCroWoo2020}, we assemble the reconstructed unital marginals into a block diagonal matrix, which we denote by \emph{full} unital marginal:

\begin{equation}
    \Lambda^{\text{(PTM)}}_F = \text{diag} \big(1, \Lambda^{\text{(PTM)}}_1, \Lambda^{\text{(PTM)}}_2, \Lambda^{\text{(PTM)}}_3  \big)\; .
    \label{full_marginal}
\end{equation}
We then transform it into an effective Pauli noise channel (see Eq.(\ref{pauli_twirl}) in App.~\ref{unital_rec_app}). The resulting Pauli error rates allow us to identify the dominant error components according to their weight and locality.

\begin{figure}[t]
\centering
\includegraphics[width=1\linewidth]{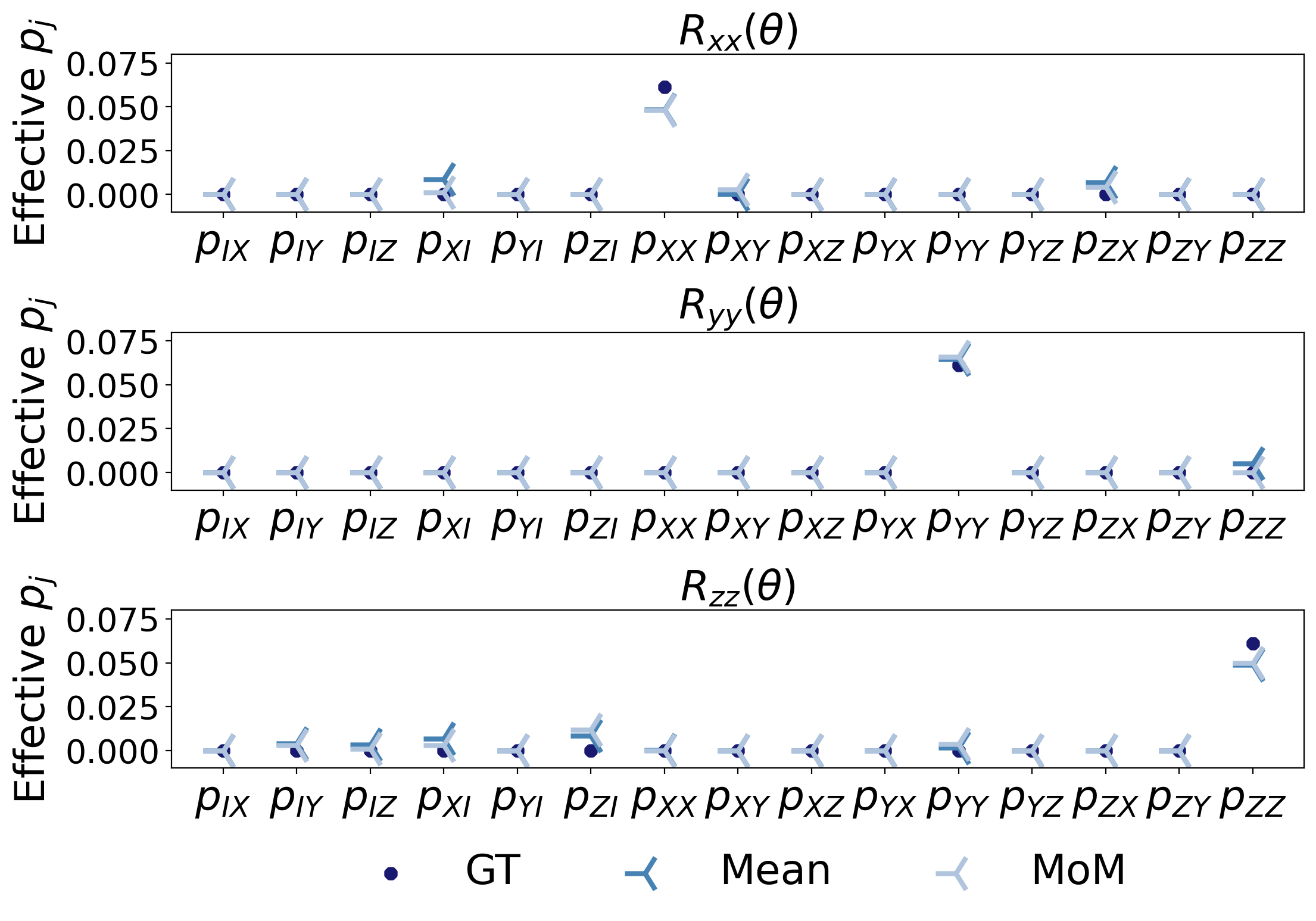}
\caption{Estimated Pauli error rates for the reconstructed unital marginals. The gate error model applied in the simulations corresponds to one of the following three two-qubit coherent gates: $R_{xx}(\theta)$ (top panel), $R_{yy}(\theta)$ (middle panel) and $R_{zz}(\theta)$ (bottom panel). For all three cases, the angle $\theta$ was fixed at $\theta=0.5$. The ground-truth (GT) values are derived using full knowledge on the simulated error model.} \label{fig_EPauli_rot}
\end{figure}

In Fig.~\ref{fig_EPauli_rot}, we show the effective Pauli error rates attained in postprocessing for the correlated coherent channels, while Tab.~\ref{tab:epsilon} shows the corresponding correlated RB coefficients $\varepsilon_i$~\cite{MckCroWoo2020}. These coefficients are defined as follows. As discussed in connection with Eq.(\ref{twirl_G}), the RB survival probabilities are governed by the effective noise channel resulting from averaging over the gate set (here, $\mathcal{C}_1\times\mathcal{C}_1$). In Ref.~\cite{MckCroWoo2020}, the coefficients $\varepsilon_i$ parametrize the average noise channel as a composition of fixed-weight depolarizing channels, whenever such a decomposition is possible. This parametrization makes it easier to identify whether the averaged noise over the gate set is dominated by local or higher weight error contributions~\cite{MckCroWoo2020}.

The explicit relation between the amplitudes $\{\varepsilon\}$ and the simultaneous RB decaying rates can be found in Refs.~\cite{MckCroWoo2020,SilGrep2025}. For the examples explored here, these amplitudes generally agree with the Pauli error rates obtained from the full marginal channel, in the sense that they correctly identify the dominant weight subspace contribution. However, contrary to the Pauli error rates obtained from the noise marginals, the amplitudes $\{\varepsilon\}$ do not provide information about the individual error components within each subspace.

\begin{figure}[t]
\centering
\includegraphics[width=1\linewidth]{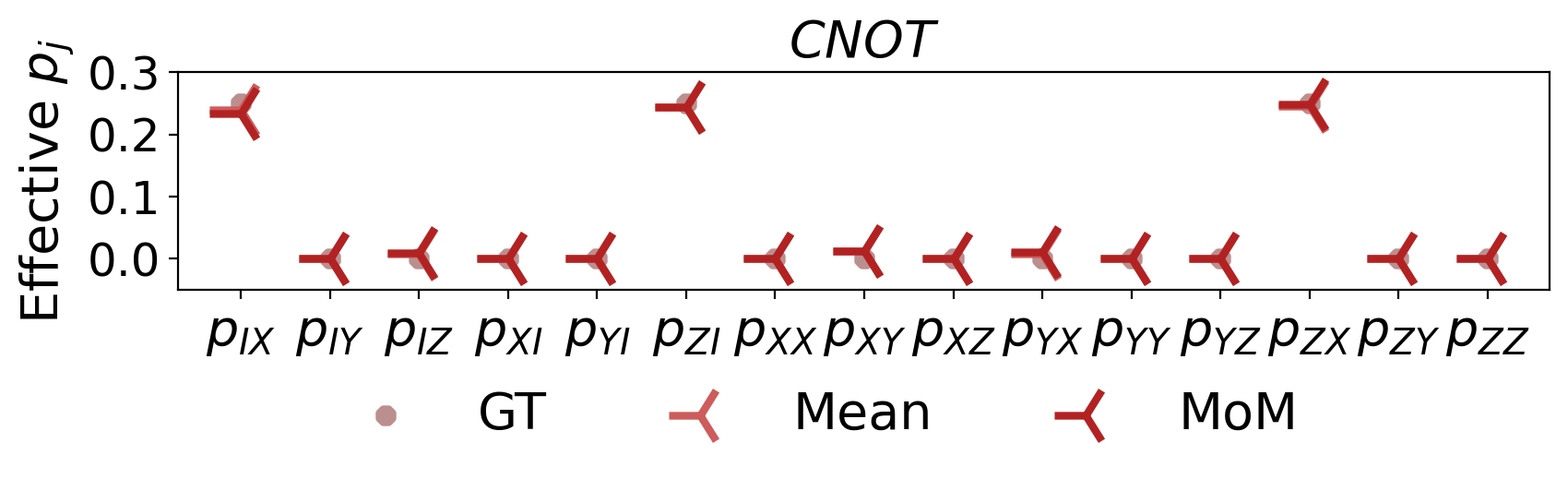}
\caption{Estimated Pauli error rates for the reconstructed unital marginals. The gate error channel applied in the simulations corresponds to a CNOT channel. The ground-truth (GT) values are derived using full knowledge on the simulated error model.} \label{fig_EPauli_cnot}
\end{figure}

A more striking limitation is provided by a CNOT-like error channel. This is a known example where correlated noise is not detected by the simultaneous RB correlation criterion~\cite{GamCorMer2013}, yielding instead $\delta \alpha=0$. As seen in Tab.~\ref{tab:epsilon}, the error amplitude in the joint subspace $\varepsilon_3$ similarly fails to capture the correlated character of this error. On the other hand, in the reconstructed Pauli representation, the CNOT-like error remains identifiable (see Fig.~\ref{fig_EPauli_cnot}).

\begin{table*}[t]\centering
\begin{tabular}{ccccccc}
\hline
\hline
 Gate Noise Model & $\hat{\varepsilon}_1$ & $\hat{\varepsilon}_2$ & $\hat{\varepsilon}_3$ &  $\{ \varepsilon_1, \varepsilon_2, \varepsilon_3\}$ \\
\hline
 $R_{xx}(\theta)$ & $0.002\;(+0.005/-0.002)$ & $0\;(+10^{-4})$ & $0.068\;(+0.002/-0.002)$ & $\{0,0,0.068\}$ \\
 $R_{yy}(\theta)$ & $0\;(+0.002)$ & $0\;(+0.003)$ & $0.07134\;(+0.002/-0.003)$ & $\{0,0,0.068\}$ \\
 $R_{zz}(\theta)$ & $0\;(+0.002)$ & $0.002\;(+0.003/-0.002)$ & $0.068\; (+0.003/-0.003)$ & $\{0,0,0.068\}$ \\
 CNOT & $0.684\;(+0.068/-0.069)$ & $0.662\;(+0.074/-0.075)$ & $0\;(+0.149/-0.281)$ & $\{2/3, 2/3, 0 \}$ \\
\hline
\end{tabular}
\caption{Estimated correlated RB amplitudes, $\{ \hat{\varepsilon} \}$. The estimates are obtained by simulating the gate-set shadow protocol using the same gate error models as in Figs.~\ref{fig_EPauli_rot}--\ref{fig_EPauli_cnot}. The reported values were obtained using the median-of-means (MoM) estimator. The last column shows the actual values for the $\{ \varepsilon \}$ coefficients for the simulated noise. The notation is as follows: $\varepsilon_1$ corresponds to a weight-1 coefficient, where all weight is in the qubit-1 subspace, $\varepsilon_2$ is a weight-1 coefficient, with all the weight in the qubit-2 subspace, and finally $\varepsilon_3$ is a weight-2 coefficient, supported jointly on both qubit subspaces. For more details on the evaluation of the amplitudes $\{ \hat{\varepsilon} \}$, see App.~\ref{unital_rec_app}.}
\label{tab:epsilon}
\end{table*}

Overall, these examples illustrate the complementary role provided by the effective Pauli error rates. The scalar correlated RB coefficients correctly identify the dominant error weight in the coherent rotation examples, but they do not resolve which Pauli components contribute within a given weight subspace. By contrast, the reconstructed marginals, once mapped to the effective Pauli error rates, retain this additional structure and can distinguish, for example, between $xx$, $yy$, and $zz$ coherent crosstalk. The CNOT example further illustrates how this Pauli resolved information can reveal correlated errors that are not directly visible from the $\{\varepsilon\}$ parametrization or the simultaneous RB correlation metric, $\delta \alpha$~\cite{GamCorMer2013}. We refer to Appendix~\ref{unital_rec_app} for additional details on the crosstalk metrics discussed in this section.

\section{Leakage RB}\label{leakage_RB_sec}
All the RB variants discussed so far operate under the assumption of no leakage outside the computational subspace. However, many promising quantum computing architectures, such as superconducting qubits~\cite{YanChuGuo2024} or qubits defined in quantum dot arrays~\cite{ZhaYanWan2018}, define their qubit states embedded in a multi-level environment~\cite{ClaRieWan2021, Fow2013}. In these settings, not properly accounting for leakage can lead to overestimating the average gate fidelity~\cite{CheBal2025}. Hence, quantifying leakage errors and their impact on the computational fidelity is crucial for the development of reliable quantum computation.

Several works have extended RB techniques to characterize leakage errors and their effects on gate fidelity. Early work on single-qubit RB showed that leakage modifies the standard single exponential decay to a multi-exponential model~\cite{EpsCrosMag2014}. The authors further showed that introducing an additional phase randomization step reduces the resulting signal to a bi-exponential decay model. The same strategy was later extended to multi-qubit systems, where the resulting RB signal generally remains multi-exponential~\cite{ChaWil2015}. Other protocols were developed to more directly estimate leakage rates. For example, Ref.~\cite{WalBarEme2016} provided a robust method for estimating the combined rate $L+S$, where $L$ denotes the the average probability of \emph{leaking out} of the computational subspace (leakage rate) and $S$ denotes the average probability of the reverse process (seepage rate). Reference~\cite{WooGam2018} further showed how the Clifford RB protocol can be modified to estimate leakage, seepage, and the computational fidelity as separate quantities.

Character RB provides another route to leakage characterization. In Ref.~\cite{ClaRieWan2021}, this approach was used to extend Ref.~\cite{WooGam2018} framework to more general gate sets with a group structure. A key advantage is that the protocol does not require independent control over the operations on the computational and leakage subspaces. Instead, the gate set may incorporate the action naturally induced by the computational gates on the leakage subspace~\cite{ClaRieWan2021}. Nevertheless, choosing an appropriate gate set generally requires some physically motivated assumptions about this induced action, which may depend on the particular qubit encoding and its implementation.

More recently, an interleaved leakage RB protocol based on Pauli randomization was proposed to estimate the leakage rate of a general multi-qubit gate, under physically motivated assumptions regarding the underlying leakage process~\cite{WuWanYu2024}. Reference~\cite{CheBal2025} revisited how leakage affects the RB fidelity estimates. It derived different survival probability models, valid in different error regimes and control assumptions, to improve estimates of the computational fidelity. In parallel, platform specific approaches have also been developed to characterize leakage errors~\cite{CheKelJul2016,AndJonRee2019,MunStoSta2025}.

In this work, we follow the character RB approach of Ref.~\cite{ClaRieWan2021} and recast it in the language of the gate-set shadow protocol. In addition to estimating the computational fidelity, we aim to individually characterize leakage ($L$) and seepage ($S$) rates.  We can define these with respect to the gate noise $\Lambda^{(\text{PTM})}$ as~\cite{ClaRieWan2021}
\begin{equation}
    \begin{split}
       & L = \frac{1}{d_1} \text{vec} \big( \mathbbm{1}_2 \big)^{\dagger} \; \Lambda^{(\text{PTM})} \; \text{vec} \big( \mathbbm{1}_1 \big) \;,  \\
       & S =  \frac{1}{d_2} \text{vec} \big( \mathbbm{1}_1 \big)^{\dagger} \; \Lambda^{(\text{PTM})} \; \text{vec} \big( \mathbbm{1}_2 \big)  \; ,
    \end{split}
    \label{LS_def}
\end{equation}
where $d_1$ and $d_2$ are, respectively, the dimensions of the computational and leakage subspace. The operators $\mathbbm{1}_1$ and $\mathbbm{1}_2$ denote the corresponding projectors.

For concreteness, we consider a singlet-triplet qubit encoded in the $S_z=0$ subspace of two electron spins~\cite{ClaRieWan2021}. Denoting the basis states by $|s,m \rangle$, where $s$ is the total spin and $m$ is its z-projection, the computational subspace is spanned by the states $|0,0 \rangle$ and $|1,0\rangle$, while the remaining states $|1, 1 \rangle$ and $|1,-1 \rangle$ form the leakage subspace. We will focus on the characterization of leakage and seepage rates in the case where the target computational gates are single-qubit Pauli gates. The more subtle problem of estimating the computational fidelity is left to App.~\ref{comp_fid_est_app}. With this aim in mind, we consider the gate-set shadow protocol implemented with the following gate set $G$~\cite{ClaRieWan2021}
\begin{equation}
    G = \langle X_C \oplus Z_L,  Z_C \oplus H_L \rangle \; .
    \label{leakage_group}
\end{equation}
Here, $X_C$ ($Z_C$) refers to the Pauli-X (Z) gate applied to the computational subspace, while $Z_L$ ($H_L$) refers to the Pauli-Z (Hadamard) gate applied to the leakage subspace. In the definition of $G$, the notation $\langle.\rangle$ indicates that all the elements in this gate set can be built from the elementary gates listed inside $\langle.\rangle$ (see App.~\ref{leakageRB_app_sec} for a description of all the elements in $G$). In total, $G$ contains 16 different elements. Note that the gate set $G$ cannot induce either leakage or seepage. Thus, we are aiming to characterize leakage and seepage caused by gate noise.

The gate set $G$ gives rise to multiplicity. Consequently, the resulting RB signals are generally governed by several decay rates. This remains true even when we employ filtering strategies akin to character RB. For estimating leakage ($L$) and seepage rates ($S$), however, character RB filtering allows for a single decay model, from which these quantities can be estimated from the fitted decay rate and amplitude coefficients. Consequently, estimates of these rates are not robust to SPAM errors, and rely on appropriate assumptions about the degree of SPAM noise. The multi-exponential structure associated with multiplicity nevertheless remains relevant when estimating the computational fidelity, as discussed in App.~\ref{comp_fid_est_app}.

In leakage RB, leakage and seepage are typically estimated from the computational subspace survival probability, $S_0(m)$~\cite{WooGam2018,ClaRieWan2021}:
\begin{equation}
    S_0(m) = \frac{S}{L+S} + \frac{L}{L+S} \; \lambda^{m} \; , 
\label{standard_leakageRB_formula}
\end{equation}
where $\lambda=1-L-S$. Within the gate-set shadow protocol, this survival probability can be viewed as one possible average sequence function, $k_A(m)$, constructed from the available experimental dataset by choosing a particular probe operator in postprocessing (see App.~\ref{LS_app_deriv} for further details).

To motivate the role of the probe operator in this setting, we start by introducing some notation. The vectorized computational and leakage projectors, $\text{vec}(\mathbbm{1}_i)$, can be expressed in the normalized Pauli basis as
\begin{equation}
\begin{split}
 & \text{vec}(\mathbbm{1}_1) = \frac{1}{2} \Big( \text{vec}(\mathbbm{1} \otimes \mathbbm{1}) + \text{vec}(Z \otimes \mathbbm{1}) \; \Big) , \\
 & \text{vec}(\mathbbm{1}_2) = \frac{1}{2} \Big( \text{vec}(\mathbbm{1} \otimes \mathbbm{1}) - \text{vec}(Z \otimes \mathbbm{1}) \Big) \; .
\end{split}  
\end{equation}
In this two-dimensional subspace, we consider two different choices of probe operator whose matrix form is equivalent to either a Pauli X ($A_x$) or a Pauli Z ($A_z$) matrix, namely:
\begin{equation}
\begin{split}
    &A_X = \frac{1}{4} \Big( \text{vec}(\mathbbm{1}\otimes \mathbbm{1}) \text{vec}(Z\otimes \mathbbm{1})^{\dagger} \\
    & +\text{vec}(Z\otimes \mathbbm{1})\text{vec}(\mathbbm{1}\otimes \mathbbm{1})^{\dagger} \Big) \; ,\\
  &  A_Z= \frac{1}{4} \Big( \text{vec}(\mathbbm{1}\otimes \mathbbm{1}) \text{vec}(\mathbbm{1} \otimes \mathbbm{1})^{\dagger}\\
  &-\text{vec}(Z\otimes \mathbbm{1})\text{vec}(Z\otimes \mathbbm{1})^{\dagger} \Big) \;.
\end{split}    
\end{equation}
Suppose we choose an initial state supported only on the computational subspace, $\rho=\frac{1}{2}\mathbbm{1}_1$, together with the POVM elements $E_x=\{E_0,E_1\}=\{\mathbbm{1}_1,\mathbbm{1}_2\}$. Fixing the probe operator to $A=A_X$ yields an average sequence function equivalent to Eq.(\ref{standard_leakageRB_formula}). We note that this equivalence is obtained in the ideal SPAM limit, which is also the regime in which Eq.(\ref{standard_leakageRB_formula}) is derived~\cite{ClaRieWan2021}. Keeping the same input state and POVM elements, but choosing $A=A_Z$, yields the alternative average sequence function:
\begin{equation}
k_A(m)=
    \begin{cases}
        S_0(m) \; , \; m\; \text{odd} \; , \\
        1-S_0(m) = L\; \frac{1-\lambda^m}{1-\lambda} \; , \; m\; \text{even} \; .
    \end{cases}
\label{k_A_Z_comp}    
\end{equation}
The even branch in Eq.(\ref{k_A_Z_comp}) is complementary to the standard leakage RB survival probability, $S_0(m)$, and therefore contains no independent information. More generally, for the fixed input state and POVM set considered here, different choices of probe operator $A$ yield average sequence functions of the form $\alpha(m)S_0(m)+\beta(m)$, where $\alpha(m)$ and $\beta(m)$ are fixed by the choice of $A$ (see App.~\ref{LS_app_deriv} for further details). Thus, different choices of $A$ can introduce sequence length dependent prefactors and offsets, but cannot arbitrarily modify the dependence of the signal on $L$ and $S$.

If we instead choose an input state that is fully mixed, $\rho \propto \mathbbm{1}_1 + \mathbbm{1}_2$, and set $A=A_X$, the average sequence function takes the form
\begin{equation}
    k_A(m) = \begin{cases}
       1  \;, \; m \; \text{odd} \\
       C\; \frac{1-\lambda^m}{1-\lambda} \;, \; m \; \text{even} \; ,
    \end{cases}
    \label{new_leakageRB_formula}
\end{equation}
where $\lambda$ is the same as in Eq.(\ref{standard_leakageRB_formula}) and $C=S-L$. Thus, by fitting even length sequences, we can estimate leakage and seepage as $L=\frac{(1-\lambda)-C}{2}$ and $S=\frac{(1-\lambda)+C}{2}$, respectively. Although preparing a maximally mixed state over the full Hilbert space may be unrealistic, this example illustrates how changing the input state allows us to directly estimate different, linearly independent combinations of $L$ and $S$. Such freedom cannot be achieved by varying the probe operator alone while keeping the original input state and measurement settings fixed.

\begin{figure}[t]
\centering
\includegraphics[width=1\linewidth]{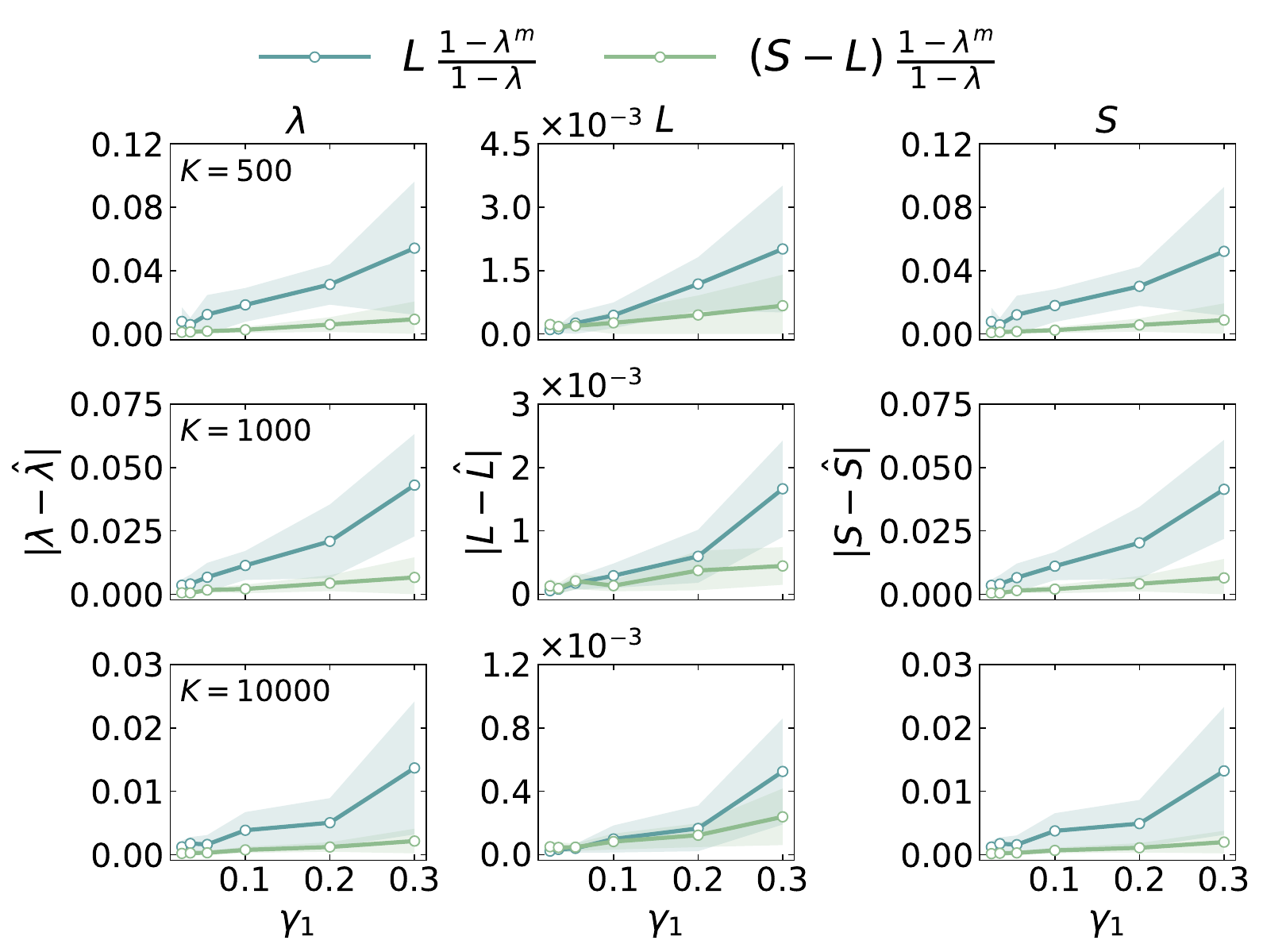}
\caption{Mean absolute error between the estimates of $\lambda$, $L$ and $S$ and their corresponding ground truth values, shown as a function of the amplitude damping parameter $\gamma_1$ and for different values of $K$. Here, $\theta=\frac{\pi}{4}$, corresponding to the regime where $S>L$. The columns display the mean absolute errors in the estimates of $\lambda$, $L$ and $S$, respectively, while the rows correspond to $K=500$, $K=1000$ and $K=10000$. For each value of $\gamma_1$ and $K$, the gate-set shadow experiment was repeated ten times. The markers indicate the mean absolute errors, and the shaded regions extend one standard deviation above and below the mean. The displayed errors correspond to estimates obtained by fitting the resulting average sequence functions (restricted to even sequence lengths) to Eq.(\ref{k_A_Z_comp}) and Eq.(\ref{new_leakageRB_formula}). The fitted parameters are then used to estimate the target figures of merit.} \label{Comparison_Fits_LS1}
\end{figure}

\begin{figure}[t]
\centering
\includegraphics[width=1\linewidth]{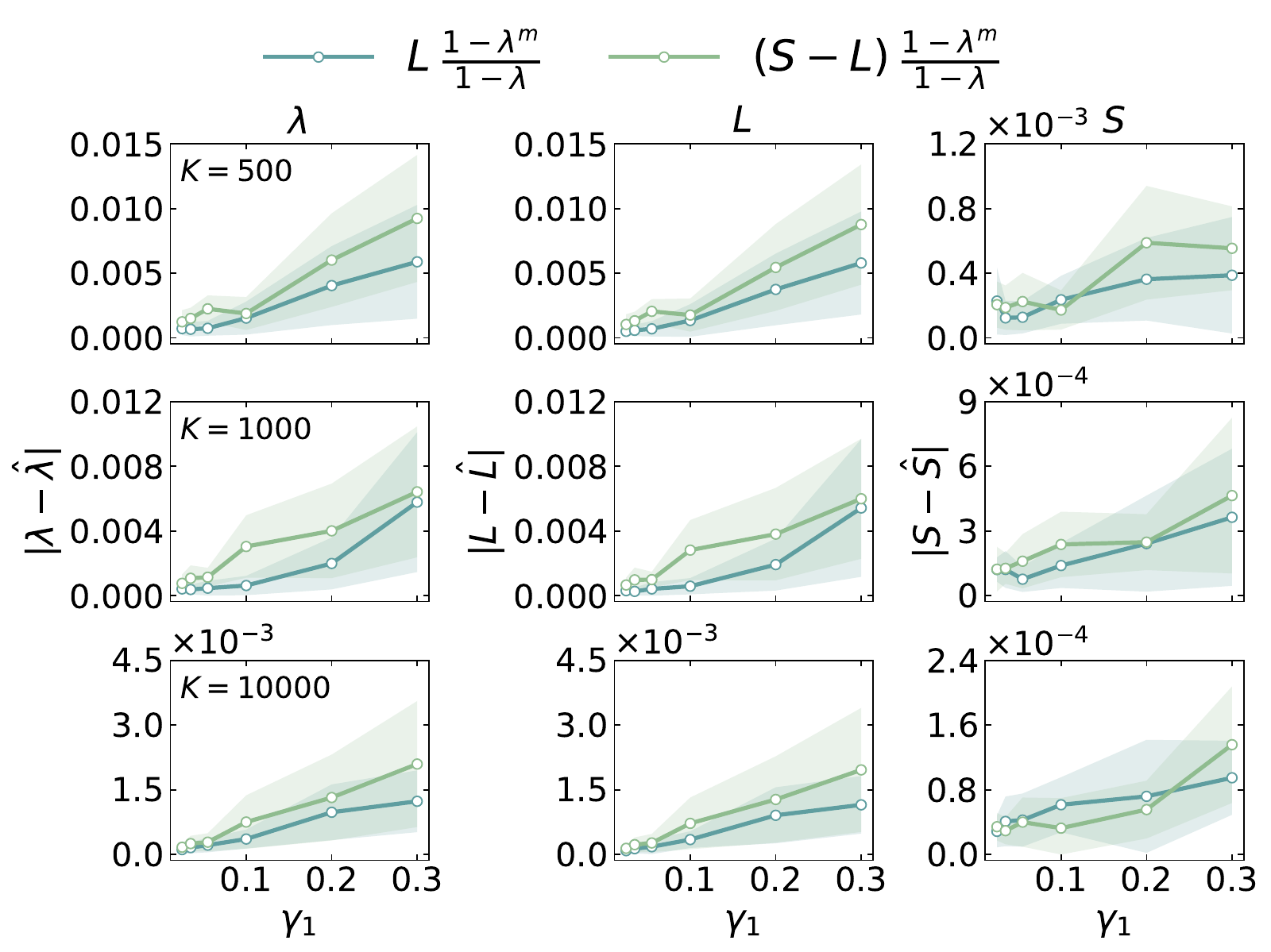}
\caption{Mean absolute error between the estimates of $\lambda$, $L$ and $S$ and their corresponding ground truth values, shown as a function of the amplitude damping parameter $\gamma_1$ and for different values of $K$. Here, $\theta=2.5$, corresponding to the regime where $L>S$. The columns display the mean absolute errors in the estimates of $\lambda$, $L$ and $S$, respectively, while the rows correspond to $K=500$, $K=1000$ and $K=10000$. For each value of $\gamma_1$ and $K$, the gate-set shadow experiment was repeated ten times. The markers indicate the mean absolute errors, and the shaded regions extend one standard deviation above and below the mean. The displayed errors correspond to estimates obtained by fitting the resulting average sequence functions (restricted to even sequence lengths) to Eq.(\ref{k_A_Z_comp}) and Eq.(\ref{new_leakageRB_formula}). The fitted parameters are then used to estimate the target figures of merit.} \label{Comparison_Fits_LS2}
\end{figure}

To examine how the different fitting models perform under finite sampling, we compare in Figs.~\ref{Comparison_Fits_LS1} and~\ref{Comparison_Fits_LS2} the estimation errors obtained from Eqs.(\ref{k_A_Z_comp}) and (\ref{new_leakageRB_formula}), for even sequence lengths. We use the rotated amplitude damping noise model defined in Eq.(\ref{noise_model_leakage_app}) in App.~\ref{def_A_LRB_app}, which depends on two amplitude damping parameters, $\gamma_1$ and $\gamma_2$, and a rotation angle $\theta$. In both figures, we fixed $\gamma_2=0.035$ and varied the amplitude damping strength $\gamma_1$. The two figures correspond to $\theta=\pi/4$ ($S>L$) and $\theta=2.5$ ($L>S$), respectively. For each value of $\gamma_1$ and for $K\in \{500,1000,10000\}$, we repeated the gate-set shadow experiment ten times and evaluated the mean absolute errors and their standard deviations for the estimates of $\lambda$, $L$ and $S$.

For even sequence lengths, both average sequence functions have essentially the same parametric form $k_A(m) = C \frac{1-\lambda^m}{1-\lambda}$, with $C=L$ in Eq.(\ref{k_A_Z_comp}) and $C=S-L$ in Eq.(\ref{new_leakageRB_formula}). Although both models contain in principle enough information to extract $L$ and $S$, their finite-sampling performance can differ depending on the noise regime. In Fig.~\ref{Comparison_Fits_LS1}, where $S>L$, the amplitude $L/(S+L)$ is small, making the signal modeled by Eq.(\ref{k_A_Z_comp}) weaker. Consequently, the corresponding estimates are more sensitive to finite sampling. By contrast, the model with $C=S-L$ yields more accurate estimates of $\lambda$, $L$ and $S$. This advantage persists throughout all values of $K$, although increasing the number of randomizations per sequence length naturally decreases the performance gap between the two fitting models.

On the other hand, in Fig.~\ref{Comparison_Fits_LS2}, both signal amplitudes have appreciable strength. Here, the two average sequence functions display a closer statistical performance, compared to the regime where $S>L$. Nonetheless, the fitting model provided by Eq.(\ref{k_A_Z_comp}) tends to yield more accurate estimates for $L$ and $\lambda$, whereas no consistent systematic advantage is observed for the estimation of $S$.

These examples show that, in the setting of leakage RB, different fitting models offer different finite sampling advantages under different noise regimes. However, the freedom to exploit these differences is constrained by the available preparation and measurement settings, since varying the probe operator alone does not generally provide the same flexibility in the presence of multiplicity.

\section{Conclusions} \label{conclusions_sec}
In this work, we have explored the gate-set shadow protocol as a unifying formalism for various RB protocols and formulated concrete examples illustrating how it can be used in practical experimental settings. We can view each RB variant as a different realization of the gate-set shadow protocol, with different choices for: input state, POVM elements, gate set and probability distribution defining the selection of random gates. For gate sets beyond the multi-qubit Clifford group, estimating a target fidelity can require access to several decay rates~\cite{GamCorMer2013,ClaRieWan2021}. Character RB can address this issue by introducing additional filtering mechanisms at the level of the experimental protocol. These mechanisms target specific decay factors and thereby facilitate the estimation of the desired fidelity~\cite{HelXueVan2019,ClaRieWan2021}. The examples presented in Sec.~\ref{inter_G4_main} illustrate how the gate-set shadow framework can instead implement analogous filtering at the postprocessing level, through the choice of probe operator. Thus, once a suitable dataset has been collected, different decay modes can be targeted, in a SPAM robust manner, without changing the data collection procedure. This same postprocessing flexibility also allowed us to extract effective Pauli error rates in the context of crosstalk characterization in Sec.~\ref{correlated_crosstalk_sec}. In this sense, our first contribution complements the main message of Ref.~\cite{HelIoaKit2023} by providing concrete examples of how the gate-set shadow postprocessing flexibility can be used, in a SPAM robust manner, to isolate and estimate multiple decay rates from a shared dataset.

Our second contribution is to explore the practical consequences of multiplicity in the gate-set shadow protocol. When multiplicity is present, the resulting average sequence function will generally contain multiple decay modes. This leads not only to a more complex fitting problem, but also to potential limitations in uniquely determining target quantities. In the subsystem fidelity example, the use of multiple input-output settings allowed us to use system identification techniques to reconstruct a representative of $\Phi$. Directly extracting target subsystem fidelities form the reconstruction of $\Phi$ is complicated by gauge ambiguity. Under the assumption the of uncorrelated noise, however, $\Phi$ allows us to identify the desired fidelity and characterize further properties of the noise acting on the idle quibit. In the leakage RB example, the restricted set of input states constrains the range of average sequence functions that can be constructed through the choice of probe operator. These two examples suggest that, for gate sets with multiplicity, varying the input state and measurement settings can reveal information that is not readily accessible by changing the probe operator alone.

The identifiability issues in the subsystem fidelity example naturally motivate examining which gauge invariant properties can still retain information about the underlying noise process. In this setting, the eigenvalues of the effective dynamics matrix $\Phi$ are the most immediate gauge invariant quantities. This suggests taking a similar perspective to techniques such as spectral quantum tomography~\cite{HelBatTer2019}, where spectral information is used to characterize the quality of an implemented gate. In our example, in the weakly correlated regime, the eigenvalues of the reconstructed effective dynamics can be related to the spectral features of the noise acting on the idle qubit. For small idle subsystems, this approach can thus serve as a complementary diagnostic to conventional RB.\\

\acknowledgments

This publication is part of the ’Quantum Inspire – the Dutch Quantum Computer in the Cloud’ project (with project number [NWA.1292.19.194]) of the NWA
research program ’Research on Routes by Consortia (ORC)’, which is funded by the Netherlands Organization for Scientific Research (NWO). This work is part of the project Engineered
Topological Quantum Networks (Project No.VI.Veni.212.278) of the research program NWO
Talent Programme Veni Science domain 2021 which is financed by the Dutch Research Council (NWO).
The authors acknowledge the use of computational resources of the DelftBlue supercomputer, provided by Delft High Performance Computing Centre~\cite{DHPC2024}. \\

\section*{Data and code availability} The code and data used in this work are openly available at~\cite{SilGrep2025_git,SilGrep_zenodo}.

\onecolumngrid      
\raggedbottom

\appendix

\section{Mathematical Preliminaries}
\subsection{Pauli transfer matrix formalism}\label{PTM_Rep_intro}
Throughout this work, we explicitly made use of the Pauli transfer matrix (PTM) representation of quantum operations (also known as the Liouville representation~\cite{DirHelWeh2019}). This representation exploits the fact that the Pauli matrices form an orthonormal basis, with respect to the Hilbert-Schmidt inner product, i.e.:
\begin{equation}
    \frac{1}{2} \text{Tr} \big( P^{\dagger}_i P_j  \big) =\delta_{i,j} \; , \; P_i,P_j \in \{\mathbbm{1},X,Y,Z\} \;.
    \label{app:ortho_paulis}
\end{equation}
This allows us to express any linear operator, such as a single qubit density matrix $\rho$, as a linear combination of the Pauli matrices. More generally, the same orthonormality property also holds true for the set of normalized $n$-qubit Pauli matrices, defined as: 
\begin{equation}
    \frac{1}{\sqrt{2^n}} \; \{ \mathbbm{1}, X, Y, Z \}^{\otimes n} \equiv \{ \sigma_0, \sigma_1, \sigma_2, \sigma_3 \}^{\otimes n} \;.
\label{n_qubit_paulis_app} 
\end{equation}
Hence, we can use the $n$-qubit normalized Pauli matrices as basis vectors, by employing the vectorization operation. This is a linear transformation that maps a matrix onto a vector. Since the $n-$qubit Pauli matrices are $d\times d$ matrices, their vectorization produces vectors with length $d^2$. Thus, the vectorization provides a linear mapping from the $d^2$-dimensional space of linear operators acting on the Hilbert space to a $d^2$-dimensional vector space, which we will often refer to as the Liouville space. It is common to denote the vectors in this space by employing the notation $|. \rangle \rangle$~\cite{DirHelWeh2019,ClaRieWan2021}. Although we did not use this convention in the main text, instead explicitly highlighting the vectorization transformation $\rho \to \text{vec}(\rho)$, we adhere to it here.

Using this notation, the orthogonality condition in Eq.(\ref{app:ortho_paulis}) becomes equivalent to the condition $\langle \langle \sigma_i| \sigma_j \rangle \rangle =\delta_{i,j}$, and linear operators, such as density matrices, become vectors, with the basis decomposition:
\begin{equation}
    |\rho \rangle \rangle = \sum^{d^2-1}_{i=0} \text{Tr} \big( \sigma^{\dagger}_i \rho \big) \; |\sigma_i \rangle \rangle = \sum^{d^2-1}_{i=0} \langle \langle\sigma_i | \rho \rangle \rangle \; |\sigma_i \rangle \rangle  \; ,
\end{equation}
where $d=2^n$.

Quantum operations, also known as quantum channels~\cite{NielChua2010}, are linear maps from density matrices to density matrices. We will use the two terms, quantum operations and quantum channels, interchangeably to refer to completely positive and trace-preserving (CPTP) maps. In the PTM representation, quantum channels are represented by matrices. In particular, a generic quantum channel $\mathcal{E}$ can be expressed in the PTM representation as:
\begin{equation}
\begin{split}
    \mathcal{E}^{(\text{PTM})}:&= \mathcal{R}(\mathcal{E}) = \sum^{d^2-1}_{i,j=0} \text{Tr} \big( \sigma^{\dagger}_i \mathcal{E}(\sigma_j) \big) \; |\sigma_i \rangle \rangle \langle \langle \sigma_j|\\ 
    &= \sum^{d^2-1}_{i,j=0} \big(\mathcal{R}({\mathcal{E}}) \big)_{i,j} \; |\sigma_i \rangle \rangle \langle \langle \sigma_j| \; ,
\end{split}    
\label{channel_PTM_app}    
\end{equation}
i.e., it is a matrix, with matrix elements given by $\big(\mathcal{R}({\mathcal{E}}) \big)_{i,j}$, or equivalently as $\big(\mathcal{R}({\mathcal{E}}) \big)_{i,j}=\text{Tr} \big( \sigma^{\dagger}_i \mathcal{E}(\sigma_j) \big)=\langle \langle \sigma_i|\mathcal{E}(\sigma_j)\rangle\rangle=\langle \langle \sigma_i|\mathcal{R}( \mathcal{E}) | \sigma_j \rangle \rangle $. In the main text, we simplified the notation by writing $\mathcal{R}(\mathcal{U})\equiv U$ for a unitary channel $\mathcal{U}$, so that $U$ in the main text denotes its PTM representation.

Any sequence of transformations induced by the action of several channels on a density matrix can now be fully expressed as a sequence of matrix transformations. Thus, the composition of any two channels becomes the product of their PTM matrices, and a tensor product channel is equivalent to the tensor product of the PTM matrices of the individual channels~\cite{DirHelWeh2019,SilGrep2025}:
\begin{equation}
|\rho' \rangle\rangle = |\mathcal{E}_b \big(\mathcal{E}_a(\rho) \big) \rangle \rangle = \mathcal{R}(\mathcal{E}_b) \mathcal{R}(\mathcal{E}_a) \; |\rho \rangle\rangle \;, 
\end{equation}
\begin{equation}
\begin{split}
    |\rho' \rangle\rangle &= | (\mathcal{E}_b \otimes \mathcal{E}_a) (\rho_1 \otimes \rho_2) \rangle \rangle 
   \\ &= \mathcal{R}(\mathcal{E}_b) \otimes \mathcal{R}(\mathcal{E}_a) \; |\rho_1 \otimes \rho_2 \rangle \rangle\\
      & = \big(\mathcal{R}(\mathcal{E}_b) \; |\rho_1 \rangle \rangle  \big) \otimes \big(\mathcal{R}(\mathcal{E}_a) \; |\rho_2 \rangle \rangle  \big) \;.
 \end{split}
\end{equation}

Throughout this work, we have exclusively focused on two-qubit systems ($d=4$). The explicit matrix representation of a generic two-qubit quantum channel is given as:
\begin{equation}
\mathcal{R}(\mathcal{E}) = 
\begin{blockarray}{ccc}
& |\tau_0 \rangle\rangle & | \tau_{i\neq 0} \rangle \rangle  \\
\begin{block}{c(cc)}
 \langle \langle \tau_0| & 1 & \vec{0}^\intercal  \\
\langle \langle \tau_{i\neq 0}| & \vec{w}_{\mathcal{E}} & R(\mathcal{E})  \\
\end{block}
\end{blockarray}
\;\; ,
\label{PTM_matrix_gen}
\end{equation}
where to make the notation more compact, we denote $|\tau_0\rangle\rangle:=|\sigma_0\otimes \sigma_0 \rangle \rangle$, $\vec{0}^\intercal$ is the zero row vector, $R(\mathcal{E})$ is a $15\times 15$ matrix, $\vec{w}_{\mathcal{E}}$ is a 15-dimensional column vector with possibly non-zero entries, and $| \tau_{i\neq 0} \rangle \rangle$ is shorthand notation for the set of vectors $|\tau_1 \rangle \rangle, |\tau_2\rangle\rangle,...,|\tau_{15}\rangle\rangle$, defined as:
\begin{equation}
\begin{split}
   & |\tau_1 \rangle \rangle := | \sigma_0 \otimes \sigma_1  \rangle\rangle \; , \\
   & |\tau_2 \rangle \rangle := | \sigma_0 \otimes \sigma_2  \rangle\rangle \; , \\
   & |\tau_3 \rangle \rangle := | \sigma_0 \otimes \sigma_3  \rangle\rangle \; , \\
   & |\tau_4 \rangle \rangle := | \sigma_1 \otimes \sigma_0  \rangle\rangle \; , \\
   & |\tau_5 \rangle \rangle := | \sigma_2 \otimes \sigma_0  \rangle\rangle \; , \\
   & |\tau_6 \rangle \rangle := | \sigma_3 \otimes \sigma_0  \rangle\rangle \; , \\
   & |\tau_7 \rangle \rangle := | \sigma_1 \otimes \sigma_1  \rangle\rangle \; , \\
   & |\tau_8 \rangle \rangle := | \sigma_1 \otimes \sigma_2  \rangle\rangle \; , \\
   & |\tau_9 \rangle \rangle := | \sigma_1 \otimes \sigma_3  \rangle\rangle \; , \\
   & |\tau_{10} \rangle \rangle := | \sigma_2 \otimes \sigma_1  \rangle\rangle \; , \\
   & |\tau_{11} \rangle \rangle := | \sigma_2 \otimes \sigma_2  \rangle\rangle \; , \\
   & |\tau_{12} \rangle \rangle := | \sigma_2 \otimes \sigma_3  \rangle\rangle \; , \\
   & |\tau_{13} \rangle \rangle := | \sigma_3 \otimes \sigma_1  \rangle\rangle \; , \\
   & |\tau_{14} \rangle \rangle := | \sigma_3 \otimes \sigma_2  \rangle\rangle \; , \\
   & |\tau_{15} \rangle \rangle := | \sigma_3 \otimes \sigma_3  \rangle\rangle \; .
\end{split}
\label{PTM_basis_choice1}
\end{equation}
Note that we are always free to re-order the vectors $| \tau_{i\neq 0} \rangle \rangle$ and redefine their labeling, without changing the general structure of the matrix in Eq.(\ref{PTM_matrix_gen}). The vector $\vec{w}_{\mathcal{E}}$ is formed by the matrix elements $\text{Tr} \big(\tau_{i\neq0} \; \mathcal{E} (\tau_0)  \big)=\frac{1}{2}\text{Tr} \big(\tau_{i\neq0} \; \mathcal{E} (\mathbbm{1}_{4\times4})  \big)$.  A unital channel is defined as a quantum operation that maps the identity to itself, $\mathcal{E}(\mathbbm{1})=\mathbbm{1}$. This requires its PTM representation to have $\vec{w}_{\mathcal{E}}=\vec{0}$. Any unitary channel is unital, while a noise process such as amplitude damping is non-unital.

\subsection{Simple introduction to group representations in the context of gate characterization}\label{RepGroups_intro}
This section provides a general introduction to the principles of finite groups and representation theory, which are used to derive analytical expressions for the fitting models in the main text. It is not intended to provide a comprehensive overview of the topic, but rather to clarify how some analytical results are obtained. Well-established results from representation theory of finite groups are introduced directly, without going over the proofs. For a more in-depth overview, see Refs.~\cite{Tinkham1964,FulHar2013}.

The PTM formalism has allowed us to think of gates as matrix operations acting on a vector state. Typically gates are generated by a composition of \emph{elementary} gate operations from a given gate set, where the latter contains a finite number of gates. In the PTM formalism, this composition is simply a series of matrix products. We can then consider all the different matrices that can be generated by taking any combination of products of \emph{elementary} matrices in the gate set, i.e.:

\begin{equation}
    \text{New gate} =  \mathcal{R}(g_{i_1}) \mathcal{R}(g_{i_2}) \cdots  \mathcal{R}(g_{i_k}) \; , \; g_{i_1}, g_{i_2} , \cdots, g_{i_k} \in \text{Gate set} \;.
    \label{gate_set_gen_app}
\end{equation}
If, through the procedure outlined in Eq.(\ref{gate_set_gen_app}), the number of distinct gates we are able to build satisfies the following requirements:
\begin{enumerate}
    \item The number of distinct elements (matrices) that can be generated is finite. After taking a certain number of products, we merely retrieve gates that have already been generated;
    \item The identity operation is contained within the set of elements we have created;
    \item For each element, its inverse operation is also contained in the set of distinct elements we have produced,
\end{enumerate}
then the collection of generated gates can be mathematically defined as a finite group~\cite{Tinkham1964}, and we can identify the gate set, $G$, as a matrix group.

More generally, group theory allows us to study the properties of a group at an abstract level, where we may infer properties and relations between its elements without ever needing to specify concrete algebraic objects to \emph{represent} them. In the present context, even if identifying the action of physical gates as matrices is already an abstract construction, we are still employing a concrete basis to express these matrices, namely the PTM representation. This means we are interested in exploring the group properties of the gate set in a concrete vector space. In very broad terms, representation theory provides the tools to study the properties of an abstract group by realizing its elements as invertible square matrices acting on a vector space, with matrix multiplication providing the corresponding group operation~\cite{Tinkham1964}. Concretely, let $g_a$, $g_b$ and $g_c$ be elements of a group, related as $g_c=g_a\times g_b$, and $M(g_a)$, $M(g_b)$ and $M(g_c)$ are square matrices. These matrices form an appropriate representation of the group if:
\begin{equation}
    M(g_a)M(g_b)=M(g_c) \;, \; \text{for all} \; g_a,g_b,g_c \in G \; .
\label{mat_rep_app}
\end{equation}

Thinking in terms of quantum gates, this means that, whenever the gate set $G$ forms a group, the PTMs of its gate operations form an actual matrix representation of the corresponding group. The PTM formalism implies a particular basis choice (the Pauli basis). However, Eq.(\ref{mat_rep_app}) does not imply there is a unique set of matrices forming an appropriate representation of a group. Indeed, any matrix similarity transformation acting over all matrices $M$ (such as a change of basis), $\tilde{M}(g)= SM(g)S^{-1}$, is an equally suitable representation, i.e. it likewise satisfies Eq.(\ref{mat_rep_app}). Since there are many possible matrix representations available, the study of how the group matrices relate to each other and how they transform vectors in their corresponding vector space cannot be dependent on our particular basis choice, but rather needs to be rooted in properties that remain invariant to it. Given the freedom in the choice of basis, we are always at liberty to find $S$ such that $SM(g)S^{-1}$, for all $M(g)$, renders the matrices to have a block diagonal form:
\begin{equation}
M(g)=
    \begin{pmatrix}
        M_1(g)  & \vec{0} & \dots  & \vec{0} \\
        \vec{0} & M_2(g)  & \dots  & \vec{0} \\
        \vdots  & \vdots  & \ddots & \vdots \\
        \vec{0} & \vec{0} & \dots  & M_k(g)
    \end{pmatrix}
    \; , \; \text{for all} \; g \in G \; .
\label{red_matrix_app}
\end{equation}
Each matrix $M_i(g)$ in Eq.(\ref{red_matrix_app}) is a lower-dimensional matrix. Its dimensionality can even be as low as a $1\times 1$ matrix, giving rise to a scalar. In representation theory, we search for an $S$ that can simultaneously transform all matrices $M(g)$ into block diagonal form, while at the same time guaranteeing that the resulting matrix blocks cannot be further collectively decomposed into smaller blocks by another similarity transformation. Note that for most purposes we never need to explicitly find the transformation $S$. Its existence is sufficient for us to directly think of the matrix representations of the group as matrices with the structure in Eq.(\ref{red_matrix_app}). Recalling that $M(g)$ is a matrix defined in some vector space $V$, what Eq.(\ref{red_matrix_app}) clarifies is that a group will naturally give rise to a decomposition of the underlying vector space, $V$, in terms of smaller subspaces $V_i$. For each vector contained in a subspace, $\vec{v}_i\in V_i$, the action of any group element on $\vec{v}_i$ is fully dictated by the action of the lower-dimensional matrix $M_i(g)$. Since there are no non-zero matrix elements outside the diagonal blocks, there is no way for the group operations to transform $\vec{v}_i$ into a vector belonging to a different subspace. Consequently, to understand how the group operations induce transformations on the basis vectors, we only need to study how each lower-dimensional matrix transforms the vectors in the corresponding subspaces.

The set of all matrices $M(g)$ form a reducible representation of $G$, and the lower dimensional matrices, $M_i(g)$, form its irreducible representations (irreps). The irreps can be equivalent or inequivalent. Two irreducible representations are said to be equivalent if there is a similarity transformation that can relate the two, $M_j(g)=\tilde{S} M_i(g)\tilde{S}^{-1}\;,\; \text{for all } g\in G$ (with $i\neq j$). In the context of finite groups, we distinguish between different irreps by their trace: equivalent irreps share the same trace for every element $g \in G$, while the same does not hold true for inequivalent irreps. The trace of a (matrix) representation is referred to as the character of the representation:
\begin{equation}
\begin{split}
    \text{Character of the \textit{ith} irrep}& := \chi_i(g) \\ 
    & = \text{Tr}\big( M_i(g) \big) \; ,
\end{split}    
\label{irrep_cha_app}
\end{equation}
\begin{equation}
\begin{split}
    \text{Character of a reducible rep.}& := \chi(g) \\
    &= \text{Tr}\big( M(g) \big) \; .
\label{rep_cha_app}
\end{split}
\end{equation}

There are gate sets that, upon arriving to Eq.(\ref{red_matrix_app}), result in a decomposition in terms of irreducible representations that are all different from one another (all irreps are inequivalent). These representations are called multiplicity-free representations. The $n-$qubit Clifford group ($\mathcal{C}_n$), the local Clifford group ($\mathcal{C}_1\times \mathcal{C}_1$), and the CNOT-dihedral group presented in the main text all have multiplicity-free representations in the Liouville space. On the other hand, some representations can lead to a decomposition that contains equivalent irreps, and are called non-multiplicity-free. The group $\mathcal{C}_1\times \mathcal{I}$, used in Section~\ref{op_crosstalk_sec}, gives rise to an example of a non-multiplicity-free representation. An important result in representation theory of finite groups is that the characters of irreducible representations satisfy the orthonormal condition~\cite{Tinkham1964,FulHar2013}:
\begin{equation}
    \frac{1}{|G|} \sum_{g \in G} \chi^*_i(g) \chi_j(g) = \delta_{i,j} \;,
\label{irrep_cha_ortho_app}
\end{equation}
where $|G|$ stands for the order of the group (the total number of distinct elements in $G$). The number of times the same irrep appears in Eq.(\ref{red_matrix_app}) is referred to as the multiplicity of the irrep. The fact that inequivalent irreps lead to different characters satisfying the orthogonality condition in Eq.(\ref{irrep_cha_ortho_app}) enables us to determine the multiplicity of an irrep with only knowledge on the character of the representation~\cite{FulHar2013}:
\begin{equation}
    \text{Multiplicity of the \textit{ith} irrep}:= a_i = \frac{1}{|G|} \sum_{g \in G} \chi^*_i(g) \chi(g) \; .
\label{multi_app}
\end{equation}
Whether or not the representation of a gate set is multiplicity-free is a relevant piece of information for any quantity resulting from averaging over group operations. This will soon become clearer when considering the averaged gate noise.

For any finite group, there is always one irreducible representation that assumes the form of an identity operation for all the elements in the group. In Eq.(\ref{red_matrix_app}), this irrep appears in the diagonal as a scalar equal to one, i.e. $M_i(g)=1$, for all $g \in G$. Since this irrep is the same for all elements in the group and will induce no transformations over the basis vectors in its corresponding vector subspace, it is often called the \emph{trivial} irrep. In contrast, all other irreps can induce different transformations on the basis of vectors spanning the corresponding subspace $V_i$, depending on which element $g$ we are considering. For this reason, they will sometimes be refered to as non-trivial irreps. The trivial irrep coincides with its character. This is true for any scalar irreducible representation. Note that since the irreps of the group give rise to a decomposition of the vector space into subspaces, the trivial irrep is associated with a one-dimensional subspace. To convey the relation between subspaces and irreps, we refer to the subspaces as simply irreducible subspaces, with the same labeling as the one ascribed to the irreps.

Weighted averages over group operations can be used to create projectors onto irreducible subspaces~\cite{FulHar2013,HelXueVan2019}. In particular, averaging over the group representation yields a projector onto the trivial subspace~\cite{FulHar2013}:
\begin{equation}
    \mathbb{P}_{\text{triv}} = \frac{1}{|G|} \sum_{g\in G} M(g) \; .
\label{trivial_proj1}
\end{equation}
It is important to note that if the trivial irrep appears with multiplicity greater than one, then the projector $\mathbb{P}_{\text{triv}}$ becomes higher-dimensional, instead of a simple one-dimensional projector. Projectors to other irreducible subspaces can also be obtain by a similar averaging procedure. The main difference compared to Eq.(\ref{trivial_proj1}) is that the representation matrices are weighted by the irrep character of the target subspace, i.e. $\chi_i(g)$. We will not make use of this result explicitly here, but note that it is the basis of why character RB is capable of extracting each decaying rate separately, whenever the corresponding representation is multiplicity-free~\cite{HelXueVan2019}.

Taking the Kronecker product between representations of a finite group $G$ provides another representation of $G$, acting on the tensor product of the original vector spaces. If $M_i$ and $M_j$ are two real and irreducible representations, then the result of averaging $M_i(g) \otimes M_j(g)$ over all group elements is zero, unless $M_i$ and $M_j$ are equivalent irreps. This result will be extensively used over the next sections. We can also consider what is the result of averaging an arbitrary constant matrix over all group operations. In that case, instead of a single projector, we get a weighted sum of all irreducible subspace projectors and, when multiplicities are present, additional operators mapping between equivalent copies of the same irrep~\cite{GamCorMer2013}:
\begin{equation}
\begin{split}
&\text{Multiplicity free group:\;}
    \frac{1}{|G|} \sum_{g\in G} M(g)\; B  \;M^{-1}(g) = \\
    &=\sum^{n}_{i=1} \frac{\text{Tr} \big( B \; \mathbb{P}_i \big)}{\text{Tr} \big(\mathbb{P}_i \big)} \; \mathbb{P}_i \\ 
    &= \sum^{n}_{i=1} \lambda_i(B) \; \mathbb{P}_i\;,
\label{RB_ave_nmg_app}
\end{split}
\end{equation}
\begin{equation}
\begin{split}
&\text{Non-multiplicity free group:\;}
    \frac{1}{|G|} \sum_{g\in G} M(g)\; B  \;M^{-1}(g) = \\ 
    &=\sum^{n}_{i=1} \sum^{a_i}_{k,k'=1} \frac{\text{Tr} \big( B \; \big(\mathbb{P}^{(k,k')}_i \big)^\intercal \big)}{\text{Tr} \Big( \big(\mathbb{P}^{(k,k')}_i \big)^\intercal \;  \mathbb{P}^{(k,k')}_i \Big)} \; \mathbb{P}^{(k,k')}_i \\
    &=\sum^{n}_{i=1} \sum^{a_i}_{k,k'=1} \lambda^{(k,k')}_i(B) \;\; \mathbb{P}^{(k,k')}_i \;.
\end{split}
\label{RB_ave_mg_app}
\end{equation}
In Eqs.(\ref{RB_ave_nmg_app}) and (\ref{RB_ave_mg_app}), $n$ denotes the number of distinct irreps, $a_i$ is the multiplicity of the \textit{ith}-irrep, $\mathbb{P}_i$ is the projector onto the \textit{ith} irreducible subspace, $\mathbb{P}^{(k,k')}_i$ is an operator mapping between equivalent copies of the $ith$-irrep: it maps the subspace associated with copy $k'$ onto the subspace associated with copy $k$. The symbol $\intercal$ denotes matrix transposition. In the context of the present work, we note that the RB procedure of generating random sequence of gates and then averaging over them is directly related with the group averages in Eq.(\ref{RB_ave_nmg_app}) (if the PTM representation of the gate set is multiplicity free) or Eq.(\ref{RB_ave_mg_app}) (if the PTM representation of the gate set is non-multiplicity free). This relationship exists because RB assumes that all gates experience the same noise, or at least that deviations from this limit are weak. Indeed, identifying the constant matrix $B$ as the gate noise $\mathcal{R}(\Lambda)$ makes it clear that the coefficients $\lambda_i$ are precisely the RB decay rates. For the more general case of a non-multiplicity-free representation, Eq.(\ref{RB_ave_mg_app}) must instead be treated as a genuinely PTM matrix of the averaged noise channel, with matrix elements given by the coefficients $\lambda^{(k,k')}_i$. Given that the RB sequence fidelity will depend on powers of this matrix, the resulting decay rates will correspond to the eigenvalues of this matrix. This also clarifies why certain gate sets can give rise to fitting models with different complexity.

\section{The mathematical construction of the fitting models}\label{fitting_models_app}
In this section, we derive the fitting models presented in the main text. We begin by revisiting the general construction of the gate-set shadow average sequence functions introduced in Ref.~\cite{HelIoaKit2023}, and then specialize this general formalism to the particular cases considered in the main text.

We start by rewriting the sequence correlation function, $f_A(x, \mathbf{g})$, using the PTM notation introduced in Section~\ref{PTM_Rep_intro} and the machinery of representation theory from Section~\ref{RepGroups_intro}:
\begin{equation}
    f_A(x,\mathbf{g}) = c^{(0)}_{f_A} \; \langle\langle E_x| \mathcal{R}_i(g_m)\prod^{m-1}_{j=1} A \mathcal{R}_i(g_j) |\rho \rangle\rangle \;.
    \label{seqcorrfun_def_app}
\end{equation}
In Eq.(\ref{seqcorrfun_def_app}), $g_j$ denotes a gate from the gate set $G$, and $\mathcal{R}(g_j)$ is its associated PTM matrix. Since the gate set is assumed to form a group, the collection $\{\mathcal{R}(g)\}_{g\in G}$ defines a representation on the full Liouville space (which we denote by $W$). In a basis adapted to the decomposition into invariant subspaces, each matrix $\mathcal R(g_j)$ takes a block-diagonal form, with blocks corresponding to the irreducible representations. Throughout this work, $\mathcal{R}_i(g)$ denotes the embedding into the full Liouville space $W$ of all blocks associated with the $ith$ irreducible representation (irrep). Namely, $\mathcal{R}_i(g)$ is a $d^2\times d^2$ block-diagonal matrix acting on the elements of $W$, whose only non-zero blocks correspond to the $a_i$ equivalent copies of the $ith$ irrep. To distinguish this full space representation from an individual irrep block, we denote by $\sigma_i(g)$ the $d_i \times d_i$ matrix representing the $ith$ irrep. Thus, $\mathcal{R}_i(g)$ is the embedding into $W$ of all $a_i$ equivalent copies of $\sigma_i(g)$.

In Ref.~\cite{HelIoaKit2023}, Eq.(\ref{seqcorrfun_def_app}) is defined using the irrep matrices $\sigma_i(g)$, whereas we use the full space embedding $\mathcal{R}_i(g)$. Our construction allows $|\rho\rangle \rangle$ and $|E_x \rangle \rangle$ to be identified directly with the vectorized ideal input state and POVM element, respectively, rather than requiring them to be restricted to the target irrep space, as in Ref.~\cite{HelIoaKit2023}.

Here, we assume that the gates are chosen uniformly at random and discuss interleaved gate sequences separately. The remaining terms in Eq.(\ref{seqcorrfun_def_app}) are the user-defined probe operator $A$ (expressed in the PTM formalism) and the normalization constant, $c^{(0)}_{f_A}$. Note that if $A$ is supported entirely on the $ith$ irrep subspace, namely $A=\mathbbm{P}_i A \mathbbm{P}_i$, then the interleaved product in Eq.(\ref{seqcorrfun_def_app}) remains confined to that subspace. Consequently, even if the PTM gates are written in their full reducible form, $\mathcal{R}(g)$, only the blocks associated with the $ith$ irrep contribute to sequence correlation function. This justifies using $\mathcal{R}_i(g)$ directly in Eq.(\ref{seqcorrfun_def_app}). Additionally, if $A$ commutes with every $\mathcal{R}_i(g)$ for all $g \in G$ and is idempotent, $A=A^2$, then $\mathcal{R}_i(g_m)A \mathcal{R}_i(g_{m-1})...A\mathcal{R}_i(g_2)A \mathcal{R}_i(g_1)=A\mathcal{R}_i(g_m)\mathcal{R}_i(g_{m-1})...\mathcal{R}_i(g_1)$. A particular example is when the probe operator corresponds to a projector onto an irreducible subspace, $A=\mathbbm{P}_i$. In that case, the role of the probe is simply to select which irrep enters in Eq.(\ref{seqcorrfun_def_app}).

The sequence function, $k_A(m)$, specifies the fitting model, and is defined as the expectation value of $f_A(x, \mathbf{g})$. Viewing $f_A(x, \mathbf{g})$ as a random variable, evaluating its expectation value requires assigning a probability distribution function, $p(x,\mathbf{g})$, to it:
\begin{equation}
    p(x,\mathbf{g}) = \langle\langle \tilde{E}_x| \prod^{m}_{j=1} \mathcal{R}(\Lambda) \mathcal{R}(g_j) | \tilde{\rho} \rangle \rangle \;.
    \label{probseq_def_app}
\end{equation}
The form of $p(x,\mathbf{g})$ merits a few observations. Firstly, we note that $p(x,\mathbf{g})$ is defined as the probability that, in a noisy experimental setup and given a sequence of gates $\mathbf{g}$, the measured outcome is $x$. As both gate noise and SPAM noise can affect the distribution of measurement outcomes, $p(x,\mathbf{g})$ incorporates the effects of both types of noise. This is accounted for by letting $\tilde{E}_x$ and $\tilde{\rho}$ be the noise implementation of the POVM element and input state, respectively. Secondly, $p(x,\mathbf{g})$ is define through the more general reducible (PTM) representation of the gates. Lastly, we adopt the usual RB convention for modeling the noisy gates, in which the noisy implementation of an ideal gate $\mathcal{R}(g)$ is given by $\mathcal{R}(\Lambda)\mathcal{R}(g)$. This is less general than taking, as in Ref.~\cite{HelIoaKit2023}, the noisy gates to be $\mathcal{R}(\Lambda_L)\mathcal{R}(g)\mathcal{R}(\Lambda_R)$. It, nevertheless, provides a more direct mapping to the RB fitting models. The sequence function is then given by the expectation value~\cite{HelIoaKit2023}:
\begin{equation}
\begin{split}
   & k_A(m) = \mathop{\mathlarger{\mathlarger{\mathlarger{\mathbb{E}}}}}_{\mathbf{g}\in G^{\times m}} \mathlarger{\sum}_{x\in \{0,1\}^n}  \; f_A(x,\mathbf{g}) \; p(x,\mathbf{g}) \\
   & = c^{(0)}_{f_A} \mathop{\mathlarger{\mathlarger{\mathlarger{\mathbb{E}}}}}_{\mathbf{g}\in G^{\times m}} \mathlarger{\sum}_{x\in \{0,1\}^n}  \; \langle\langle E_x| \mathcal{R}_i(g_m)\prod^{m-1}_{j=1} A \mathcal{R}_i(g_j) |\rho \rangle\rangle 
   \langle\langle \tilde{E}_x| \mathcal{R}(\Lambda) \mathcal{R}(g_m) \prod^{m-1}_{j=1} \mathcal{R}(\Lambda) \mathcal{R}(g_j) | \tilde{\rho} \rangle \rangle \\
   & = c^{(0)}_{f_A} \mathop{\mathlarger{\mathlarger{\mathlarger{\mathbb{E}}}}}_{\mathbf{g}\in G^{\times m}} \mathlarger{\sum}_{x\in \{0,1\}^n}  \; \text{Tr} \bigg[ | \rho \rangle \rangle \langle \langle E_x| \; \Big( \mathcal{R}_i(g_m) \prod^{m-1}_{j=1} A \mathcal{R}_i(g_j) \Big) \bigg] \text{Tr} \bigg[ | \tilde{\rho} \rangle \rangle \langle \langle \Lambda^*(\tilde{E}_x)| \; \Big( \mathcal{R}(g_m) \prod^{m-1}_{j=1} \mathcal{R} (\Lambda) \mathcal{R}(g_j) \Big) \bigg] \\
   & =c^{(0)}_{f_A} \mathop{\mathlarger{\mathlarger{\mathlarger{\mathbb{E}}}}}_{\mathbf{g}\in G^{\times m}} \mathlarger{\sum}_{x\in \{0,1\}^n}  \; \text{Tr} \bigg[ |\rho \otimes \tilde{\rho} \rangle \rangle \langle \langle E_x \otimes \Lambda^*(\tilde{E}_x)| \Big( \mathcal{R}_i(g_m) \otimes \mathcal{R}(g_m) \Big) \prod^{m-1}_{j=1} (A\otimes \mathcal{R} (\Lambda)) \Big(\mathcal{R}_i(g_j) \otimes \mathcal{R}(g_j) \Big) \bigg] \\
   &=c^{(0)}_{f_A} \mathlarger{\sum}_{x\in \{0,1\}^n} \; \text{Tr} \bigg[ \Omega(\rho,\tilde{\rho},E_x,\tilde{E}_x) \; \mathop{\mathlarger{\mathlarger{\mathlarger{\mathbb{E}}}}}_{g_m\in G} \Big(  \mathcal{R}_i(g_m) \otimes \mathcal{R}(g_m) \Big) \prod^{m-1}_{j=1} (A\otimes \mathcal{R} (\Lambda)) \mathop{\mathlarger{\mathlarger{\mathlarger{\mathbb{E}}}}}_{g_j\in G} \Big(\mathcal{R}_i(g_j) \otimes \mathcal{R}(g_j) \Big) \bigg] \;,
\end{split}   
\label{seqfun_app1}
\end{equation}
where we have defined $\Omega(\rho,\tilde{\rho},E_x,\tilde{E}_x) := |\rho \otimes \tilde{\rho} \rangle \rangle \langle \langle E_x \otimes \Lambda^*(\tilde{E}_x)|$. As discussed in Section~\ref{RepGroups_intro}, taking averages over representations gives rise to a projector. In particular, note that for representations~\cite{FulHar2013}:

\begin{equation}
    \mathcal{R}_i(g) \otimes \mathcal{R}(g) \cong \mathlarger{\mathlarger{ \mathop{\oplus}_{j}}} \big( \mathcal{R}_i(g) \otimes \mathcal{R}_j(g)  \big) \; .
\end{equation}
The averaged operator therefore decomposes into independent contributions from each term $\mathcal{R}_i(g) \otimes \mathcal{R}_j(g)$. As discussed in the previous section, uniformly averaging a representation gives rise to a projector onto the trivial invariant subspace. For unitary irreducible representations of a finite group, $\sigma_i(g) \otimes \sigma_j(g)$ contains the trivial irrep if and only if $\sigma_j \cong \sigma^*_i$~\cite{FulHar2013}, where $\sigma^*_i(g) = \big( \sigma_i(g^{-1}) \big)^{\intercal}$ denotes the dual representation and $\cong$ stands for equivalence of representations up to a similarity transformation. Since $\mathcal{R}_j(g)$ contains all copies of $\sigma_j(g)$, the same criterion determines which terms in the direct sum decomposition can survive the group average. For the gate sets considered here, the irreps admit real orthogonal realizations and are therefore self-dual (i.e., $\sigma_i\cong \sigma^*_i$). Thus, all terms with $j\neq i$ vanish upon averaging, and only $\mathcal{R}_i(g) \otimes \mathcal{R}_i(g)$ can contribute to $\mathop{\mathlarger{\mathlarger{\mathlarger{\mathbb{E}}}}}_{g\in G} \Big(\mathcal{R}_i(g) \otimes \mathcal{R}(g) \Big)$. We let $\mathlarger{\mathbb{P}}_{\text{triv}}$ denote the projector onto the trivial invariant subspace inside $\mathcal{R}_i(g) \otimes \mathcal{R}_i(g)$, 
\begin{equation}
\mathlarger{\mathbb{P}}_{\text{triv}} = \mathop{\mathlarger{\mathlarger{\mathlarger{\mathbb{E}}}}}_{g\in G} \Big(\mathcal{R}_i(g) \otimes \mathcal{R}(g) \Big)  \; .
 \label{trivial_proj_gss_def}
 \end{equation}
If the $ith$ irrep occurs in $W$ with multiplicity one, then $\mathcal{R}_i(g)$ is the full space embedding of a single copy of $\sigma_i(g)$. In this case, the trivial invariant subspace of $\mathcal{R}_i(g) \otimes \mathcal{R}_i(g)$ is one-dimensional and $\mathlarger{\mathbb{P}}_{\text{triv}}$ is a rank-1 projector. To see this, consider the invariant subspace $V_i\subset W$ on which $\sigma_i(g)$ acts irreducibly. Choosing an orthonormal basis for $V_i$ as the set $\{|v_l \rangle \rangle\}^{d_i}_{l=1}$, we define the normalized vector $|v\rangle \rangle$ as
\begin{equation}
    |v \rangle \rangle =  \frac{1}{\sqrt{d_i}} \sum^{d_i}_{l=1} |v_l \rangle \rangle \otimes |v_l \rangle \rangle \; ,
\label{gen_invvec_free}     
\end{equation}
with $d_i = \text{dim}(V_i)$. Since the restriction of $\mathcal{R}_i(g)$ to $V_i$ is the real and orthogonal matrix $\sigma_i(g)$, the vector $|v \rangle \rangle$ is left invariant under $\mathcal{R}_i(g) \otimes\mathcal{R}_i(g)$ for every $g \in G$. Therefore, when the irrep appears with multiplicity one,
\begin{equation}
    \mathlarger{\mathbb{P}}_{\text{triv}} = |v\rangle\rangle \langle \langle v| \; .
\end{equation}
On the other hand, suppose that $ith$ irrep occurs in $W$ with multiplicity $a_i>1$. In this case, $\mathcal R_i$ is the embedding of $a_i$ equivalent copies of $\sigma_i$. We choose an orthonormal basis $\{|e_{l,\mu} \rangle \rangle \}_{l,\mu}$, where $l=\{1, ..., d_i\}$ labels the basis states within each copy and $\mu= \{1,...,a_i\}$ labels the different equivalent copies. Thus, for fixed $\mu$, the vectors $\{|e_{l,\mu} \rangle \rangle \}^{d_i}_{l=1}$ span the subspace corresponding to $\mu$\textit{th} copy of the $ith$ irrep. Since the group acts identically on each copy and does not mix the copy labels, an orthonormal basis for the trivial invariant subspace is given by:
\begin{equation}
    |v_{\mu,\nu} \rangle \rangle =  \frac{1}{\sqrt{d_i}} \sum^{d_i}_{l=1} |e_{l,\mu} \rangle \rangle \otimes |e_{l,\nu} \rangle \rangle \; .
\label{gen_invvec_mul}     
\end{equation}
Hence, when the $ith$ irrep occurs with multiplicity $a_i>1$, the projector onto the trivial subspace becomes
\begin{equation}
    \mathlarger{\mathbb{P}}_{\text{triv}} = \sum^{a_i}_{\mu,\nu=1} |v_{\mu,\nu}\rangle\rangle \langle \langle v_{\mu,\nu}| \; ,
 \label{P_triv_general_app}   
\end{equation}
and the trivial invariant subspace of $\mathcal R_i(g)\otimes \mathcal R_i(g)$ has dimension $a^2_i$.

Providing a basis decomposition for the trivial projector inside the average sequence function requires then specifying the group and the choice of irrep. Keeping the discussion general for now, Eq.(\ref{seqfun_app1}) simplifies to:
\begin{equation}
   k_A(m)=c^{(0)}_{f_A} \mathlarger{\sum}_{x\in \{0,1\}^n} \; \text{Tr} \bigg[ \Omega(\rho,\tilde{\rho},E_x,\tilde{E}_x) \; \Big( \mathlarger{\mathbb{P}}_{\text{triv}} (A \otimes \mathcal{R} (\Lambda)) \mathlarger{\mathbb{P}}_{\text{triv}}  \Big)^{m-1} \bigg] \; .
   \label{seqfun_app2}
\end{equation}
Alternatively, by defining 
\begin{equation}
    \begin{split}
        & \Phi = \mathlarger{\mathbb{P}}_{\text{triv}} (A \otimes \mathcal{R} (\Lambda)) \mathlarger{\mathbb{P}}_{\text{triv}} \; ,\\
       & |W \rangle\rangle = |\rho \otimes\tilde{\rho} \rangle \rangle \;, \\
       & |M \rangle\rangle = \mathlarger{\sum}_{x\in \{0,1\}^n} |E_x \otimes \Lambda^*(\tilde{E}_x) \rangle \rangle \;, \\
    \end{split}
    \label{mapping_to_state_space_model_app1}
\end{equation}
we can express $k_A(m)$ more compactly as
\begin{equation}
     k_A(m)=c^{(0)}_{f_A} \; \langle\langle M| \Phi^{m-1} | W \rangle\rangle  \; .
     \label{mapping_to_state_space_model_app2}
\end{equation}

\subsection{Character RB as a special case of the gate-set shadow protocol}\label{char_and_gss_app}
In this section, we show how the character RB decay model can be recovered within the gate-set shadow framework. We first consider gate set groups with a multiplicity-free representation and later discuss the corresponding relation in the presence of multiplicity.

Before specializing to these two cases, let us first recall the mechanism by which character RB filters the RB signal and isolates specific decay contributions. Under the RB assumption of gate independent noise, averaging over many random sequences becomes equivalent to performing a group average as in Eq.(\ref{RB_ave_nmg_app}) (or Eq.(\ref{RB_ave_mg_app}) in the presence of multiplicities). This produces a decay model containing multiple decay factors, whose number depends on how the group decomposes into irreducible representations and whether multiplicities are present. As discussed in Section~\ref{rel_charRB}, the character RB survival probability results from two averaging procedures. One is an average over uniformly sampled RB sequences, whilst the other is a weighted average over the uniformly sampled initial gate, selected from a subgroup of the original gate set group. In character RB, the weights are chosen to be the characters associated with the irreducible representation whose decay contribution is to be isolated (see Eq.(\ref{irrep_cha_app}) for the definition of character). This weighted average provides an implementation of the projector onto the corresponding irreducible subspace~\cite{FulHar2013}:
\begin{equation}
    \Pi_i = \frac{d_i}{|G|}  \sum_{g \in G} \chi^{*}_i(g)\; g \; .
    \label{charRB_proj}
\end{equation}
thereby filtering out all contributions outside the selected subspace. A more detailed discussion of how the projector in Eq.(\ref{charRB_proj}) concretely emerges within the character RB procedure can be found in Refs.~\cite{HelXueVan2019, ClaRieWan2021}.

Having established the filtering mechanism underlying character RB, we now show how the corresponding decay model can be recovered within the gate-set shadow framework. The derivation presented here should be viewed as a specific instance of the general gate-set shadow protocol restricted to probe operators that project onto individual irreducible subspaces. Although this filtering mechanism is already implicit in the original formulation of Ref.~\cite{HelIoaKit2023}, making it explicit provides a direct connection to character RB and clarifies how the two protocols recover the same decay parameters despite their different operational implementations.

\subsubsection{Multiplicity-free representations}
For groups with multiplicity-free representations, the projector in Eq.(\ref{charRB_proj}) isolates a unique decay factor, reducing the character RB signal to a single exponential decay model:
\begin{equation}
    p_i(m) = C \; \lambda^{m}_i \; ,
    \label{surv_prob_char_mf}
\end{equation}
with $\lambda_i:=\frac{\text{Tr}(\mathbb{P}_i \mathcal{R}(\Lambda))}{\text{Tr}(\mathbb{P}_i)}$. The decay factor $\lambda_i$ is typically the quantity of interest, as it directly enters the estimation of the average fidelity of the gate set~\cite{HelXueVan2019}. The constant $C$ is a SPAM dependent prefactor. To recover a similar decay model within the gate-set shadow framework, we choose the probe operator $A$ to be the projector onto the \emph{ith} irreducible representation subspace, namely $A=\mathbb{P}_i$. Given this choice, the sequence correlation function defined in Eq.(\ref{seqcorrfun_def_app}) simplifies to
\begin{equation}
    \begin{split}
      f_A(x,\mathbf{g})& =   c^{(0)}_{f_A} \; \langle\langle E_x| \prod^{m}_{j=1} \mathcal{R}_i(g_j) |\rho \rangle\rangle \\
      & = c^{(0)}_{f_A} \; \langle\langle E_x|\mathcal{R}_i(g_m g_{m-1}...g_1) |\rho \rangle\rangle \\
      & = c^{(0)}_{f_A} \; \text{Tr} \Big( |\rho \rangle\rangle \langle\langle E_x| \;\; \mathcal{R}_i(g_m g_{m-1}...g_1)  \Big) \\
      & =  c^{(0)}_{f_A} \; \text{Tr} \Big( B_x \; \mathcal{R}_i(g_m g_{m-1}...g_1)  \Big) \;,
    \end{split}
\end{equation}
where we defined $B_x=|\rho \rangle\rangle \langle\langle E_x|$ as the rank-one operator constructed from the preparation state $\rho$ and POVM element $E_x$.

Let us introduce new group elements~\cite{ClaRieWan2021}
\begin{equation}
   \nu_1=g_1\;,  \; \nu_i = g_i g_{i-1} ... g_1 \implies \nu_i\nu^{-1}_{i-1}= g_i \; .
    \label{new_group_ele}
\end{equation}
Each $\nu_i$ represents the action of an RB sequence of length $i$. This provides an alternative representation of the group elements generated along the RB sequence without changing the underlying group average, allowing Eq.(\ref{seqfun_app1}) to be rewritten as follows
\begin{equation}
\begin{split}
     k_A(m) &= c^{(0)}_{f_A} \mathop{\mathlarger{\mathlarger{\mathlarger{\mathbb{E}}}}}_{\mathbf{g}\in G^{\times m}} \mathlarger{\sum}_{x\in \{0,1\}^n}  \; \langle\langle E_x| \mathcal{R}_i(g_m)\prod^{m-1}_{j=1} A \mathcal{R}_i(g_j) |\rho \rangle\rangle 
   \langle\langle \tilde{E}_x| \mathcal{R}(\Lambda) \mathcal{R}(g_m) \prod^{m-1}_{j=1} \mathcal{R}(\Lambda) \mathcal{R}(g_j) | \tilde{\rho} \rangle \rangle \\
   &= c^{(0)}_{f_A} \mathop{\mathlarger{\mathlarger{\mathlarger{\mathbb{E}}}}}_{\mathbf{g}\in G^{\times m}} \mathlarger{\sum}_{x\in \{0,1\}^n}  \; \text{Tr} \Big( B_x \; \mathcal{R}_i(g_m g_{m-1}...g_1) \Big) \; \langle\langle \tilde{E}_x| \mathcal{R}(\Lambda) \mathcal{R}(g_m) \prod^{m-1}_{j=1} \mathcal{R}(\Lambda) \mathcal{R}(g_j) | \tilde{\rho} \rangle \rangle \\
   &= \frac{c^{(0)}_{f_A}}{|G|^m}  \mathlarger{\sum}_{x\in \{0,1\}^n} \mathlarger{\sum}_{\nu_m,...,\nu_1} \langle\langle \tilde{E}_x| \text{Tr} \Big( B_x \; \mathcal{R}_i(\nu_m) \Big) \mathcal{R}(\Lambda) \mathcal{R}(\nu_m) \prod^{m-1}_{j=1} \mathcal{R}^{\dagger}(\nu_j) \mathcal{R}(\Lambda) \mathcal{R}(\nu_j)  | \tilde{\rho} \rangle \rangle \\
   &= \frac{c^{(0)}_{f_A}}{|G|} \mathlarger{\sum}_{x\in \{0,1\}^n} \mathlarger{\sum}_{\nu_m} \mathlarger{\sum}_j \; \lambda^{m-1}_j \; \langle\langle \tilde{E}_x|\mathcal{R}(\Lambda) \text{Tr} \Big( B_x \; \mathcal{R}_i(\nu_m) \Big) \mathcal{R}_j(\nu_m)  | \tilde{\rho} \rangle \rangle \; ,
\end{split}
\label{seq_fun_charrel_mf_1}
\end{equation}
where we made used of Eq.(\ref{RB_ave_nmg_app}). The sum $\sum_{\nu_m} \text{Tr} \Big( B_x \; \mathcal{R}_i(\nu_m) \Big) \mathcal{R}_j(\nu_m)$ can be simplified using the orthogonality relations between inequivalent irreducible representations. To see this explicitly, we examine its matrix elements:
\begin{equation*}
    \begin{split}
      \bigg \{ \sum_{\nu_m} \text{Tr} \Big( B_x \; \mathcal{R}_i(\nu_m) \Big) \mathcal{R}_j(\nu_m) \bigg \}_{k,l}& = \sum_{a,b} \big( B_x \big)_{a,b} \; \sum_{\nu_m} \big( \mathcal{R}_i(\nu_m) \big)_{b,a}    \big( \mathcal{R}_j(\nu_m) \big)_{k,l}  \\ 
      & = \sum_{a,b} \big( B_x \big)_{a,b} \; \frac{|G|}{d_i} \delta_{i,j} \; \delta_{b,k} \; \delta_{a,l} \\
      & = \frac{|G|}{d_i}\; \big( B_x \big)_{l,k} \; \delta_{i,j} \;.
    \end{split}
\end{equation*}
This result follows from the great orthogonality theorem~\cite{Tinkham1964}, here applied to real irreducible representations. Note that the indices $a,b$ label the matrix elements within the nonzero $ith$ irrep block of $\mathcal{R}_i(\nu_m)$, while $k,l$ label the matrix elements within the nonzero $jth$ block of $\mathcal{R}_j(\nu_m)$. Equivalently,
\begin{equation}
  \sum_{\nu_m} \text{Tr} \Big( B_x \; \mathcal{R}_i(\nu_m) \Big) \mathcal{R}_j(\nu_m) =  \frac{|G|}{d_i}\;  \mathbb{P}_i B^{\intercal}_x \mathbb{P}_i \;  \delta_{i,j} \;.
\end{equation}
This allows us to write the sequence function in Eq.(\ref{seq_fun_charrel_mf_1}) as
\begin{equation}
\begin{split}
    k_A(m) &= \frac{c^{(0)}_{f_A}}{d_i} \mathlarger{\sum}_{x\in \{0,1\}^n} \mathlarger{\sum}_j \; \lambda^{m-1}_j \; \langle\langle \tilde{E}_x|\mathcal{R}(\Lambda) \; \mathbb{P}_i B^{\intercal}_x \mathbb{P}_i \; \delta_{i,j}  | \tilde{\rho} \rangle \rangle \\
    &= \frac{c^{(0)}_{f_A}}{d_i}\;  \lambda^{m-1}_i \; \mathlarger{\sum}_{x\in \{0,1\}^n} \langle\langle \tilde{E}_x|\mathcal{R}(\Lambda) \; \mathbb{P}_i B^{\intercal}_x \mathbb{P}_i | \tilde{\rho} \rangle \rangle\\
    &= \tilde{C} \; \lambda^{m-1}_i \; ,
\end{split} 
\label{seq_fun_charrel_mf_2}
\end{equation}
where all SPAM-dependent terms have been absorbed into the prefactor $\tilde{C}$, defined as
\begin{equation*}
\tilde{C}=\frac{c^{(0)}_{f_A}}{d_i} \mathlarger{\sum}_{x\in \{0,1\}^n} 
\langle\langle \tilde{E}_x| \mathcal{R}(\Lambda) \mathbb{P}_i B_x^{\intercal}\mathbb{P}_i|\tilde{\rho}\rangle\rangle \;.
\end{equation*}
Comparing Eq.(\ref{seq_fun_charrel_mf_2}) with Eq.(\ref{surv_prob_char_mf}), we see that both models produce a single exponential decay model with the same decay factor $\lambda_i$. This agreement reflects the fact that both constructions combine a group average over the noise channel with representation theoretic orthogonality relations that selectively retain only the contributions associated with the target irreducible representation. In character RB, this filtering is implemented through character-weighted averages and originates from the orthogonality of characters. In the gate-set shadow framework, it is implemented through the choice of probe operator $A$ and follows from the orthogonality relations between irreducible representations. From this perspective, the examples discussed in Secs.~\ref{inter_G4_main} and~\ref{correlated_crosstalk_sec} can be viewed as instances where the gate-set shadow protocol effectively performs the same filtering operation as character RB, but without requiring additional subgroup sampling during the experimental stage.

\subsubsection{Non-multiplicity-free representations}
In the presence of multiplicities, character RB filters between inequivalent irreps but not between equivalent copies of the same irrep. In this setting, the resulting survival probability can contain up to $a_i$ decay factors, where $a_i$ is the multiplicity of the target irrep~\cite{ClaRieWan2021}:
\begin{equation}
    p_i(m) = \sum^{a_i}_{j=1} C_{i,j} \; \lambda^{m}_{i,j} \; .
\end{equation}

The decay factors $\lambda_{i,j}$ remain related to the noise channel and provide the means to estimate its trace. Knowing approximately the trace of the noise channel allows then to estimate the average fidelity of the gate set~\cite{ClaRieWan2021}. As previously, the constants $C_{i,j}$ are SPAM dependent prefactors. To recover a similar decay model within the gate-set shadow framework, we choose the probe operator $A$ to be the projector onto the \emph{ith} irreducible subspace, namely $A=\Pi_i= \sum^{a_i}_{k=1} \mathbb{P}^{(k,k)}_i$. Unlike the multiplicity-free case, this projector projects onto all $a_i$ equivalent copies of the $ith$ irrep. Here, $\mathbb{P}^{(k,k')}_i$ denotes the map from the subspace corresponding to copy $k'$ to the subspace associated with copy $k$. Given this choice of probe operator, and since $\Pi_i \mathcal{R}_i(g)=\mathcal{R}_i(g)$, the sequence correlation function defined in Eq.(\ref{seqcorrfun_def_app}) can be expressed as
\begin{equation}
    \begin{split}
      f_A(x,\mathbf{g})& = c^{(0)}_{f_A} \; \langle\langle E_x| \prod^{m}_{j=1} \mathcal{R}_i(g_j) |\rho \rangle\rangle \\
      & = c^{(0)}_{f_A} \; \langle\langle E_x|\mathcal{R}_i(g_m g_{m-1}...g_1) |\rho \rangle\rangle \\
      & = c^{(0)}_{f_A} \; \text{Tr} \Big( |\rho \rangle\rangle \langle\langle E_x| \;\; \mathcal{R}_i(g_m g_{m-1}...g_1)  \Big) \\
      & =  c^{(0)}_{f_A} \; \text{Tr} \Big( B_x \; \mathcal{R}_i(g_m g_{m-1}...g_1)  \Big) \\
      & = c^{(0)}_{f_A} \; \sum^{a_i}_{k=1} \; \text{Tr} \Big( B_x \; \mathcal{R}^{(k)}_i(g_m g_{m-1}...g_1) \Big) \;,
    \end{split}
\end{equation}
where $B_x=|\rho \rangle\rangle \langle\langle E_x|$ and $\mathcal{R}^{(k)}_i(g)$ denotes the full space embedding of the $kth$ copy of $\sigma_i(g)$.

More generally, the full Liouville space decomposes 
as~\cite{FulHar2013}
\begin{equation}
    W = \bigoplus^{N_\text{irrep}}_{i=1} W_i \; ,
\end{equation}
where $N_{\text{irrep}}$ denotes the number of inequivalent irreducible representations and $W_i$ contains all $a_i$ equivalent copies of the $ith$ irrep. Let $V_i$ denote the invariant subspace associated with a single copy of the $ith$ irrep. Since the group acts identically on all equivalent copies, $W_i$ can be written as
\begin{equation}
    W_i \cong \mathbb{C}^{a_i} \otimes V_i \;,
    \label{multiplicity_decomposition_app}
\end{equation}
where $\mathbb{C}^{a_i}$ denotes the multiplicity space. Informally, Eq.(\ref{multiplicity_decomposition_app}) states that the information identifying the copy can be separated from the irrep action itself.

Throughout this work, representations are embedded in the full Liouville space $W$, unless stated otherwise. Accordingly, each irrep contribution takes the tensor product form introduced in Eq.(\ref{multiplicity_decomposition_app}). Explicitly,
\begin{equation}
    \begin{split}
       & \mathcal{R}(g) = \bigoplus^{N_\text{irrep}}_{l=1} \big(\mathbbm{1}_{a_l} \otimes \sigma_l(g) \big) \; , \\
       & \mathcal{R}_i(g) = \Pi_i \mathcal{R}(g) \Pi_i = \bigoplus^{N_\text{irrep}}_{l=1} \delta_{l,i}  \big(\mathbbm{1}_{a_l} \otimes \sigma_l(g) \big) \;, \\
       &  \mathcal{R}^{(k)}_i(g) = \mathbf{0} \oplus ... \oplus \big( |k_i \rangle \langle k_i| \otimes \sigma_i(g) \big) \oplus \mathbf{0} \oplus ... \oplus \mathbf{0}  \;,
    \end{split}
    \label{reps_mult_char}
\end{equation}
where $\{|k_i \rangle\}^{a_i}_{k=1}$ is an orthonormal basis for the multiplicity space associated with the $ith$ irrep, and $\sigma_l(g)$ is the $d_l \times d_l$ matrix representing the $lth$ irrep.

Having introduced this notation, we turn our focus to the average sequence function. Following the same transformation to the new group elements introduced in Eq.(\ref{new_group_ele}), the average sequence function becomes
\begin{equation}
\begin{split}
    k_A(m) &= c^{(0)}_{f_A} \mathop{\mathlarger{\mathlarger{\mathlarger{\mathbb{E}}}}}_{\mathbf{g}\in G^{\times m}} \mathlarger{\sum}_{x\in \{0,1\}^n}  \; \langle\langle E_x| \mathcal{R}_i(g_m)\prod^{m-1}_{j=1} A \mathcal{R}_i(g_j) |\rho \rangle\rangle 
   \langle\langle \tilde{E}_x| \mathcal{R}(\Lambda) \mathcal{R}(g_m) \prod^{m-1}_{j=1} \mathcal{R}(\Lambda) \mathcal{R}(g_j) | \tilde{\rho} \rangle \rangle \\
   &= \frac{c^{(0)}_{f_A}}{|G|^m}  \mathlarger{\sum}_{x\in \{0,1\}^n} \mathlarger{\sum}_{\nu_m,...,\nu_1} \mathlarger{\sum}^{a_i}_{k=1} \;  \langle\langle \tilde{E}_x| \mathcal{R}(\Lambda) \text{Tr} \Big( B_x \;  \mathcal{R}^{(k)}_i(\nu_m) \Big) \mathcal{R}(\nu_m) \prod^{m-1}_{j=1} \mathcal{R}^{\dagger}(\nu_j) \mathcal{R}(\Lambda) \mathcal{R}(\nu_j)  | \tilde{\rho} \rangle \rangle \\
   &= \frac{c^{(0)}_{f_A}}{|G|^m}  \mathlarger{\sum}_{x\in \{0,1\}^n} \mathlarger{\sum}_{\nu_m,...,\nu_1} \mathlarger{\sum}^{a_i}_{k=1} \;  \langle\langle \tilde{E}_x| \mathcal{R}(\Lambda) \text{Tr} \Big(B^{(i)}_{x;kk} \;  \sigma_i(\nu_m) \Big) \mathcal{R}(\nu_m) \prod^{m-1}_{j=1} \mathcal{R}^{\dagger}(\nu_j) \mathcal{R}(\Lambda) \mathcal{R}(\nu_j)  | \tilde{\rho} \rangle \rangle \; .
\end{split} 
\label{seq_fun_charrel_nmf_1}
\end{equation}
The last equality follows from the support of $\mathcal{R}^{(k)}_i(\nu_m)$. Since $\mathcal{R}^{(k)}_i(\nu_m)$ acts only on copy $k$ of the $ith$ irreducible representation subspace, only the corresponding diagonal block of $B_x$ contributes to the trace. More explicitly, since $\mathbbm{1}_W=\sum_l \Pi_l$, any operator on W admits the following block matrix decomposition with respect to the irrep subspaces:
\begin{equation}
   B_x = \sum_{l,j} \Pi_l B_x \Pi_j \;.
\end{equation}
Hence 
\begin{equation}
     B_x \mathcal{R}^{(k)}_i(\nu_m) = \sum_l \Pi_l B_x \Pi_i  \; \mathcal{R}^{(k)}_i(\nu_m) \;,
\end{equation}
and, after taking the trace, only the diagonal contribution survives
\begin{equation}
   \text{Tr} \Big(B_x \mathcal{R}^{(k)}_i(\nu_m)\Big) = \text{Tr} \Big( \Pi_i B_x \Pi_i \; \mathcal{R}^{(k)}_i(\nu_m) \Big) = \text{Tr} \Big( B^{(i)}_x \; \mathcal{R}^{(k)}_i(\nu_m) \Big) \;,
\label{traceBx_multi_app}
\end{equation}
where we defined $ B^{(i)}_x =\Pi_i B_x \Pi_i$. Using the decomposition in Eq.(\ref{multiplicity_decomposition_app}), this block can be expanded in the multiplicity basis as
\begin{equation}
    B^{(i)}_x  =  \; \sum^{a_i}_{n,s=1} |n_i \rangle \langle s_i| \otimes B^{(i)}_{x;ns} \; ,
\end{equation}
where $B^{(i)}_{x;ns}: V_i \to V_i$ is the operator associated with the map from the $sth$ copy of $V_i$ to the $nth$ copy. Since, within the $ith$ irrep sector, $\mathcal{R}^{(k)}_i(\nu_m)=|k_i \rangle \langle k_i| \otimes \sigma_i( \nu_m)$, the product $B^{(i)}_x  \; \mathcal{R}^{(k)}_i(\nu_m)$ yields
\begin{equation}
\begin{split}
    B^{(i)}_x  \; \mathcal{R}^{(k)}_i(\nu_m) = \sum^{a_i}_{n=1} |n_i \rangle  \langle k_i| \; \otimes B^{(i)}_{x;nk} \sigma_i (\nu_m)  \; ,
\end{split}    
\end{equation}
and vanishes on all other remaining subspaces associated with irreps inequivalent to the $ith$ irrep. Therefore, Eq.(\ref{traceBx_multi_app}) can be written as
\begin{equation}
\text{Tr} \big(B_x \mathcal{R}^{(k)}_i(\nu_m) \big) = \text{Tr} \Big(B^{(i)}_x \; \mathcal{R}^{(k)}_i(\nu_m) \Big) = \text{Tr} \Big( B^{(i)}_{x;kk} \;  \sigma_i(\nu_m) \Big) \; .
\end{equation}
This form allows us to apply the great orthogonality theorem to the sum $\sum_{\nu \in G} \text{Tr} \Big(B^{(i)}_{x;kk} \;  \sigma_i(\nu) \Big) \mathcal{R}(\nu)$:
\begin{equation}
\begin{split}
 \mathlarger{\sum}_{\nu \in G} \text{Tr} \Big( B^{(i)}_{x;kk} \; \sigma_i(\nu) \Big) \mathcal{R}(\nu) &=
\mathlarger{\sum}_{\nu \in G} \text{Tr} \Big( B^{(i)}_{x;kk} \;  \sigma_i(\nu) \Big) \bigoplus_l \mathbbm{1}_{a_l} \otimes \sigma_l(\nu) \\
&= \bigoplus_l \mathbbm{1}_{a_l} \otimes  \mathlarger{\sum}_{\nu \in G} \text{Tr} \Big( B^{(i)}_{x;kk} \; \sigma_i(\nu) \Big) \sigma_l(\nu) \\\; .
\end{split}
\end{equation}
We evaluate the remaining group sum by examining its matrix elements:
\begin{equation}
\begin{split}
    \bigg \{\sum_{\nu \in G} \text{Tr} \Big( B^{(i)}_{x;kk} \; \sigma_i(\nu) \Big) \sigma_l(\nu) \bigg\}_{p,q} &=  \sum_{\alpha,\beta} \big( B^{(i)}_{x;kk} \big)_{\alpha,\beta} \sum_{\nu \in G} \big( \sigma_i(\nu)\big)_{\beta,\alpha} \big( \sigma_l(\nu)\big)_{p,q} \\
    &= \frac{|G|}{d_i} \; \sum_{\alpha,\beta} \big( B^{(i)}_{x;kk} \big)_{\alpha,\beta}  \delta_{\beta,p} \delta_{\alpha,q} \delta_{i,l} \\
    &= \frac{|G|}{d_i} \; \delta_{i,l} \; \big( B^{(i)}_{x;kk} \big)_{q,p} \;.
\end{split}    
\end{equation}
Thus,
\begin{equation}
   \mathlarger{\sum}_{\nu \in G} \text{Tr} \Big(B^{(i)}_{x;kk} \; \sigma_i(\nu) \Big) \mathcal{R}(\nu) = \frac{|G|}{d_i} \; \bigoplus^{N_{\text{irrep}}}_{l=1} \delta_{i,l} \; \Big( \mathbbm{1}_{a_l} \otimes \big( B^{(i)}_{x;kk} \big)^{\intercal} \Big) \; . 
   \label{got_mult_app}
\end{equation}
The result in Eq.(\ref{got_mult_app}) shows explicitly that the group sum removes all contributions from irreps inequivalent to the target irrep. Within the surviving irrep sector, it produces the operator $\big( B^{(i)}_{x;kk} \big)^{\intercal}$ on the irrep subspace while acting as the identity on the multiplicity space. Thus, the filtering isolates the target irrep but leaves its multiplicity structure intact.

The average sequence function is additionally influenced by the group average over the noise channel, namely
\begin{equation}
    \mathcal{T}(\Lambda) = \frac{1}{|G|} \sum_{\nu\in G} \mathcal{R}^{\dagger} (\nu) \mathcal{R}(\Lambda) \mathcal{R}(\nu) \; .
\end{equation}
By construction, $\mathcal{T}(\Lambda)$ commutes with the group action: $\mathcal{R}(h)\mathcal{T}(\Lambda)=\mathcal{T}(\Lambda)\mathcal{R}(h)$ for all $h \in G$. Such maps are called intertwining maps~\cite{Sim1996}.

As an operator on $W$, $\mathcal{T}(\Lambda)$ admits the block decomposition: $\mathcal{T}(\Lambda)=\sum_{l,j}\Pi_l \mathcal{T}(\Lambda) \Pi_j=\sum_{l,j} \mathcal{T}_{l,j}$. Each block component $\mathcal{T}_{l,j}$ defines a linear map $\mathbb{C}^{a_j} \otimes V_j \to \mathbb{C}^{a_l} \otimes V_l$ and is itself an intertwining map:
\begin{equation}
    \Big( \mathbbm{1}_{a_l} \otimes \sigma_l (\nu) \Big) \; \mathcal{T}_{l,j} = \mathcal{T}_{l,j} \; \Big( \mathbbm{1}_{a_j} \otimes \sigma_j (\nu) \Big) \; .
\end{equation}
This intertwining property, together with the irreducibility of $\sigma_l$ and $\sigma_j$, allows us to apply Schur's lemma~\cite{FulHar2013,Sim1996}. When $l\neq j$, the irreps $\sigma_l$ and $\sigma_j$ are inequivalent, and therefore $\mathcal{T}_{l,j}=0$~\cite{FulHar2013}. For $l=j$, applying Schur's lemma to each multiplicity block shows that the action on the irrep space must be proportional to the identity~\cite{FulHar2013}. The resulting proportionality constants can then be collected into an otherwise unrestricted linear map $Q_l$ acting on the multiplicity space. Consequently, 
\begin{equation}
    \mathcal{T}(\Lambda) = \bigoplus_l Q_l \otimes \mathbbm{1}_{d_l} \; ,
\end{equation}
where $Q_l$ is a $a_l \times a_l$ matrix acting on the multiplicity space.

If, as in Ref.~\cite{ClaRieWan2021}, we diagonalize $Q_l$ as $Q_l = \sum^{a_l}_{r=1} \lambda_{l,r} |e_{l,r} \rangle  \langle \bar{e}_{l,r}|$, then the average sequence function allows the simplification
\begin{equation}
    \begin{split}
        k_A(m) &= \frac{c^{(0)}_{f_A}}{d_i}  \mathlarger{\sum}_{x\in \{0,1\}^n} \mathlarger{\sum}^{a_i}_{k=1} \langle \langle \tilde{E}_x | \mathcal{R}(\Lambda) \Big[ \bigoplus_l \delta_{i,l} \; \Big( \mathbbm{1}_{a_i} \otimes \big( B^{(i)}_{x;kk} \big)^{\intercal} \Big) \Big] \Big[ \bigoplus_l  \mathlarger{\sum}^{a_l}_{r=1} \lambda^{m-1}_{l,r} |e_{l,r} \rangle \langle \bar{e}_{l,r}| \otimes \mathbbm{1}_{d_l} \Big] | \tilde{\rho} \rangle \rangle \\
        &= \frac{c^{(0)}_{f_A}}{d_i}  \mathlarger{\sum}_{x\in \{0,1\}^n} \mathlarger{\sum}^{a_i}_{k=1} \mathlarger{\sum}^{a_i}_{r=1} \lambda^{m-1}_{i,r} \langle \langle \tilde{E}_x | \mathcal{R}(\Lambda) \Big[ \bigoplus_l \delta_{i,l} \; \big( |e_{i,r} \rangle  \langle \bar{e}_{i,r}| \otimes \big( B^{(i)}_{x;kk} \big)^{\intercal}  \big) \Big] | \tilde{\rho} \rangle \rangle \\
        & = \mathlarger{\sum}^{a_i}_{r=1} \tilde{C}_{i,r} \lambda^{m-1}_{i,r} \; , 
    \end{split}
    \label{k_A(m)_charrel_mult_app_fin1}
\end{equation}
with the constant prefactor defined as
\begin{equation}
    \tilde{C}_{i,r} = \frac{c^{(0)}_{f_A}}{d_i}  \mathlarger{\sum}_{x\in \{0,1\}^n} \mathlarger{\sum}^{a_i}_{k=1} \langle \langle \tilde{E}_x | \mathcal{R}(\Lambda) \Big[ \bigoplus_l \delta_{i,l} \; \big( |e_{i,r} \rangle \langle \bar{e}_{i,r}| \otimes \big( B^{(i)}_{x;kk} \big)^{\intercal}  \big) \Big] | \tilde{\rho} \rangle \rangle \; .
    \label{k_A(m)_charrel_mult_app_fin2}
\end{equation}
We note that in Eqs.(\ref{got_mult_app}), (\ref{k_A(m)_charrel_mult_app_fin1}) and (\ref{k_A(m)_charrel_mult_app_fin2}), the direct sum with $\delta_{i,l}$ denotes the embedding of the displayed operator into the $ith$ irrep sector of $W$: the $l=i$ block is nonzero, while all blocks corresponding to $l\neq i$ are understood to be the zero matrices with the appropriate dimensions.

The decay factors $\lambda_{i,r}$ coincide with those obtained in character RB, since they originate from the same mechanism: they are the eigenvalues of the multiplicity space matrix arising from the group average of the noise channel. The prefactors may nevertheless differ between protocols.

\subsection{Standard RB}\label{fitting_models_staRB_app}
\subsubsection{Multi-qubit Clifford group}\label{C2groupRB}
In Liouville space, the representation of the multi-qubit Clifford group is multiplicity-free and decomposes into two inequivalent irreducible representations~\cite{HelIoaKit2023,GroAudEis2007}. This statement applies to any $n-$qubit system, but to simplify the discussion, we focus on a $2-$qubit case.

The emergence of these two irreps can be made more evident by recalling several properties of the PTM formalism. First, we work in a vector space spanned by the vectorized set of all $2$-qubit Pauli operators. Second, any CPTP map in this space has the matrix structure in Eq.(\ref{PTM_matrix_gen}). Furthermore, up to a sign, the Clifford group acts by permuting the normalized Pauli vectors
\begin{equation}
    \mathcal{R}(g)|\tau_i\rangle\rangle=\pm |\tau_l\rangle\rangle\;, \; g \in \mathcal{C}_n \;.
\end{equation}
The vectorized identity is a fixed point of this action, while the traceless Pauli vectors are permuted among themselves. It follows that $\mathcal{R}_{0,0}(g)=1$, $\mathcal{R}_{0,i}(g)=\mathcal{R}_{i,0}(g)=0 \; (i\neq0)$, and that the subspace spanned by the traceless Pauli vectors is invariant under $\mathcal{R}(g)$ for every $g\in \mathcal{C}_n$. Hence, in the PTM basis, the decomposition into the two irreducible representations takes the explicit form
\begin{equation}
\mathcal{R}(g) = 
\begin{blockarray}{ccc}
& |\tau_0 \rangle\rangle & | \tau_{i\neq 0} \rangle \rangle  \\
\begin{block}{c(cc)}
 \langle \langle \tau_0| & 1 & \vec{0}^\intercal  \\
\langle \langle \tau_{i\neq 0}| & \vec{0} & R(g)  \\
\end{block}
\end{blockarray}
\; .
\label{PTM_matrix_cliffordn}
\end{equation}
From the discussion in Section~\ref{RepGroups_intro}, we recognize $\mathcal{R}(g)$ in Eq.(\ref{PTM_matrix_cliffordn}) as a reducible representation of the group. Its two irreps are immediately transparent: the scalar irrep (the trivial irrep), associated with the one-dimensional vector space spanned by the single vector $|\tau_0\rangle\rangle$, and a non-trivial irrep, operating on the subspace spanned by the set of vectors $\{|\tau_i \rangle \rangle\}_{i \neq0}$. In this example, $R(g)$ is the matrix realization of $\sigma_i(g)$ introduced in App.~\ref{fitting_models_app}, with $i$ labeling the nontrivial irreducible subspace.

To recover the the decay parameter associated with standard RB, and hence the corresponding average gate set fidelity, the following settings should be applied: the gate set is fixed to be the multi-qubit Clifford group; the random sequences are constructed by uniformly sampling the group; $f_A(x,\mathbf{g})$ is evaluated using only the non-trivial irreducible representation; and $A$ is fixed as the projector onto the corresponding non-trivial subspace. Using the basis vectors specified in Eq.(\ref{PTM_basis_choice1}), the projector onto the non-trivial subspace is given as:
\begin{equation}
    \mathbb{P}_1=\sum^{15}_{i=1} |\tau_i \rangle \rangle \langle \langle\tau_i| \; .
\end{equation}

The projector onto the trivial representation contained in the tensor product representation appearing in Eq.(\ref{trivial_proj_gss_def}) is now evaluated using the nontrivial irrep $R(g)$, or equivalently its full-space embedding $\mathcal{R}_1(g)$
\begin{equation}
   \mathcal{R}_1(g)= 
  \begin{pmatrix}
      0 & \vec{0}^\intercal \\
      \vec{0} & R(g)
  \end{pmatrix}
  \;.
\end{equation}
Since $R(g)$ operates in the subspace spanned by the vectors $\{|\tau_i \rangle \rangle\}_{i \neq0}$, the matrix representation $R^{\otimes2}(g)$ has a natural decomposition in terms of the basis vectors $\{|\tau_i  \rangle \rangle\  \otimes |\tau_j \rangle \rangle \}^{15}_{i,j=1 }$. Considering that $R(g)$ permutes, with phases $\pm$, the basis vectors  $\{|\tau_i \rangle \rangle\}_{i \neq0}$, an invariant vector under the action of $R^{\otimes2}(g)$, for any Clifford element $g$, is given by a uniform superposition of the vectors $\{|\tau_i  \rangle \rangle\  \otimes |\tau_i \rangle \rangle \}^{15}_{i\neq 0}$, i.e. $|v \rangle \rangle$ in Eq.(\ref{gen_invvec_free}) takes on the specific form:
\begin{equation}
    |v \rangle \rangle = \frac{1}{|\mathbb{P}_1|^{1/2}}  \sum^{15}_{i=1} |\tau_i \rangle\rangle \otimes |\tau_i \rangle \rangle = \frac{1}{|\mathbb{P}_1|^{1/2}} \sum^{15}_{i=1} \text{vec} \big(|\tau_i \rangle \rangle \langle \langle \tau_i| \big) =  \frac{1}{|\mathbb{P}_1|^{1/2}}\; |\mathbb{P}_1 \rangle\rangle \; .
\end{equation}
Note that we have used the relation between the Kronecker and outer products: $|i \rangle \rangle \otimes | j \rangle \rangle =\text{vec} \Big(|j \rangle \rangle\langle\langle i| \Big)$, where $\text{vec}(\cdot)$ denotes the usual column-stacking vectorization convention~\cite{Mele2024}. In the equation above, the constant $|\mathbb{P}_1|=\text{Tr}(\mathbb{P}_1)$ is introduced as a normalization factor. This allows us to express the trivial projector in Eq.(\ref{trivial_proj_gss_def}) as:

\begin{equation}
\begin{split}
    \mathlarger{\mathbb{P}}_{\text{triv}} &=   | v \rangle\rangle \langle \langle v| = \frac{1}{|\mathbb{P}_1|} \;  |\mathbb{P}_1 \rangle\rangle  \langle\langle \mathbb{P}_1 | \;.
\label{trivial_proj_c2_app}
\end{split}
\end{equation}

The fitting model in Eq.(\ref{seqfun_app2}) then reduces to:
\begin{equation}
\begin{split}
 k_A(m) &=   c^{(0)}_{f_A} \mathlarger{\sum}_{x\in \{0,1\}^2} \; \text{Tr} \bigg[ \Omega(\rho,\tilde{\rho},E_x,\tilde{E}_x) \; \Big( \frac{1}{|\mathbb{P}_1|^2} \;  \langle \langle \mathbb{P}_1|(A \otimes \mathcal{R} (\Lambda))| \mathbb{P}_1 \rangle \rangle \; | \mathbb{P}_1 \rangle \rangle \langle \langle \mathbb{P}_1|  \Big)^{m-1} \bigg] \\
 &=  c^{(0)}_{f_A} \mathlarger{\sum}_{x\in \{0,1\}^2} \; \text{Tr} \bigg[ \Omega(\rho,\tilde{\rho},E_x,\tilde{E}_x) \; \Big( \frac{1}{|\mathbb{P}_1|} \; \langle\langle \mathbb{P}_1 | \mathcal{R} (\Lambda) \mathbb{P}_1 A^\intercal \rangle\rangle \; \mathbb{P}_{\text{triv}} \Big)^{m-1} \bigg] \\
 & = c^{(0)}_{f_A} \; \lambda^{m-1}(A^\intercal, \Lambda) \;  \mathlarger{\sum}_{x\in \{0,1\}^2}  \frac{1}{|\mathbb{P}_1|} \;  \langle \langle \mathbb{P}_1| \Omega(\rho,\tilde{\rho},E_x,\tilde{E}_x ) | \mathbb{P}_1 \rangle \rangle \; ,
 \end{split}
\label{seqfun_app_srb} 
\end{equation}
with $\lambda(A^\intercal, \Lambda) =\frac{1}{|\mathbb{P}_1|} \;  \text{Tr}\big( \mathbb{P}_1 A^\intercal \mathbb{P}_1 \mathcal{R}(\Lambda) \big)$. Note that in Eq.(\ref{seqfun_app_srb}), we used the matrix vectorization identity~\cite{WoBiCo2015}:
\begin{equation}
   |BCA^{\intercal} \rangle\rangle = \big( A \otimes B \big) | C \rangle \rangle \; ,
\label{ABC_rule}
\end{equation} where \(|X\rangle\rangle=\text{vec}(X)\) is defined using the same column-stacking convention as before. Letting $A=\mathbb{P}_1$, the resulting decay rate can be directly related to the average gate set fidelity (see Eq.(\ref{average_fid})):
\begin{equation}
    \lambda(\mathbb{P}_1, \Lambda) = \frac{1}{|\mathbb{P}_1|} \; \text{Tr}\big(\mathbb{P}_1 \mathcal{R}(\Lambda)  \big) = \frac{d \bar{F}(\Lambda, \mathcal{I})-1}{d-1} \; ,
\end{equation}
with $d$ the dimension of the Hilbert space. The decay rate $ \lambda(\mathbb{P}_1, \Lambda)$ is the typical figure of merit measured in standard RB.

We note that, in this example, a convenient choice of POVM operators and input state is to set them as operators defined in the $z$-computational basis, i.e.:
\begin{equation}
    \begin{split}
       & |\rho \rangle \rangle = |00 \rangle \rangle= \frac{1}{2} \Big( |\tau_0 \rangle\rangle + |\tau_3 \rangle\rangle + |\tau_6 \rangle\rangle + |\tau_{15} \rangle\rangle \Big) = | E_{00} \rangle \rangle \; ,\\
       & | E_{01} \rangle \rangle = \frac{1}{2} \Big( |\tau_0 \rangle\rangle - |\tau_3 \rangle\rangle + |\tau_6 \rangle\rangle - |\tau_{15} \rangle\rangle \Big) \\
       & | E_{10} \rangle \rangle =  \frac{1}{2} \Big( |\tau_0 \rangle\rangle + |\tau_3 \rangle\rangle - |\tau_6 \rangle\rangle - |\tau_{15} \rangle\rangle \Big) \\
        & | E_{11} \rangle \rangle =  \frac{1}{2} \Big( |\tau_0 \rangle\rangle - |\tau_3 \rangle\rangle - |\tau_6 \rangle\rangle + |\tau_{15} \rangle\rangle \Big) \; .
    \end{split}
\label{z_compbasis_def}
\end{equation}

It only remains to determine the normalization constant $c^{(0)}_{f_A}$. This constant is defined such that, in the absence of SPAM and gate noise, $k_A(1)=1$. In this regime, Eq.(\ref{seqfun_app_srb}) reads:
\begin{equation}
    \begin{split}
        k_A(1)&= c^{(0)}_{f_A} \; \mathlarger{\sum}_{x\in \{0,1\}^2}\frac{1}{|\mathbb{P}_1|} \;  \langle \langle \mathbb{P}_1| \Omega(\rho,\tilde{\rho},E_x,\tilde{E}_x ) | \mathbb{P}_1 \rangle \rangle \\
        &= c^{(0)}_{f_A} \; \mathlarger{\sum}_{x\in \{0,1\}^2}\frac{1}{|\mathbb{P}_1|} \; \sum_{i \neq 0} \langle \langle \tau_i| \rho \rangle \rangle^2 \; \sum_{j \neq 0} \langle \langle E_x| \tau_j \rangle \rangle^2 \\
        &= \frac{3}{4}c^{(0)}_{f_A} \; \mathlarger{\sum}_{x\in \{0,1\}^2}\frac{1}{|\mathbb{P}_1|} \; \sum_{j \neq 0} \langle \langle E_x| \tau_j \rangle \rangle^2 = \frac{3^2}{4}c^{(0)}_{f_A} \; \frac{1}{|\mathbb{P}_1|} \implies c^{(0)}_{f_A}=\frac{20}{3} \; .
    \end{split}
\end{equation}
The value of the normalization constant $c^{(0)}_{f_A}$ is in agreement with numerical simulations of the protocol, guaranteeing $k_A(1)=1$ in the noiseless regime.

\subsubsection{CNOT-Dihedral group $G(4)$}\label{G4groupRB}
In Liouville space, the representation of the CNOT-Dihedral group $G(4)$ is multiplicity-free. The group is generated by the elements in the set $\mathcal{S}$, up to a global phase factor $w$ (here $\mathbb{C}$ denotes the space of complex numbers). This means that every element of the group can be obtained as a finite product of elements in $\mathcal{S}$, while two elements are considered identical if they differ only by a scalar phase factor.

\begin{equation}
    G(4) = \langle \mathcal{S} \rangle/ \langle w \mathbbm{1}: w \in \mathbb{C} \rangle \; ,\; \mathcal{S} = \big \{ X \otimes \mathbbm{1}, \mathbbm{1} \otimes X,  S \otimes \mathbbm{1}, \mathbbm{1} \otimes S, \text{CNOT}_{0,1}, \text{CNOT}_{1,0} \big \} \;.
    \label{all_ele_G4}
\end{equation}
Here, $S$ corresponds to the phase gate. The group $G(4)$ has 768~\cite{CrosMagBis2016} distinct elements, and is a subgroup of the Clifford group. To construct all its elements, we make use of the class decomposition proposed in Ref.~\cite{GarCros2020}. Note that in Eq.(\ref{all_ele_G4}) the group elements are represented as unitary operations acting on a $2-$qubit Hilbert space. We move to the PTM description below.

It is clear that all elements of $G(4)$ map the identity to itself, $g (\mathbbm{1}\otimes \mathbbm{1})g^{\dagger}=(\mathbbm{1}\otimes \mathbbm{1})$ for all $g \in G(4)$. The subspace 
\begin{equation}
  W_0=\text{span}\{|\tau_0 \rangle \rangle \}  
\end{equation}
is therefore an invariant subspace associated with the trivial irrep.

A further invariant subspace is 
\begin{equation}
    W_1=\text{span} \{ |\tau_3 \rangle \rangle, |\tau_6 \rangle \rangle, |\tau_{15} \rangle \rangle \}= \text{span} \{| \mathbbm{1}\otimes Z \rangle \rangle, |Z \otimes\mathbbm{1} \rangle \rangle, |Z \otimes Z \rangle \rangle \} \; .
\end{equation}
To see that $W_1$ is invariant, we examine the action of the elements in $\mathcal{S}$ on the corresponding Pauli operators. Defining $Z_1=Z\otimes \mathbbm{1}$ and $Z_2=\mathbbm{1}\otimes Z$, with $(i_1,i_2) \in \{0,1\}^2 \setminus \{(0,0)\}$, the phase gate generators leave each $(Z_1)^{i_1}(Z_2)^{i_2}$ unchanged, while the $X$ generators map it to $\pm (Z_1)^{i_1}(Z_2)^{i_2}$. The CNOT generators permute the three Pauli operators within $W_1$, namely $\text{CNOT}_{0,1}: \{\mathbbm{1}\otimes Z, Z \otimes\mathbbm{1}, Z \otimes Z\} \mapsto \{Z \otimes Z, Z \otimes\mathbbm{1}, \mathbbm{1}\otimes Z\}$, while $\text{CNOT}_{1,0}: \{\mathbbm{1}\otimes Z, Z \otimes\mathbbm{1}, Z \otimes Z\} \mapsto \{\mathbbm{1}\otimes Z, Z \otimes Z, Z \otimes\mathbbm{1}\}$. Hence, every generator maps vectors in $W_1$ back to $W_1$. Since all elements of $G(4)$ are formed by finite products of these generators, $W_1$ is invariant under the action of $G(4)$.

The remaining subspace can be written as $W_2=\text{span} \big\{|\tau_i\rangle \rangle: i\in \{1,2,...,15\} \setminus \{3,6,15\} \big\}$. This is the orthogonal complement to $W_0 \oplus W_1$ and is invariant under the group action. This follows from the fact that $G(4)$ is a subgroup of the Clifford group, which is the normalizer of the Pauli group~\cite{HelWalWeh2018}. Hence, conjugation by any $g \in G(4)$ permutes Pauli operators among themselves, up to possible signs. Since the Pauli operators spanning $W_0 \oplus W_1$ are mapped within $W_0 \oplus W_1$, the remaining twelve Pauli operators must likewise be permuted among themselves, and hence remain in $W_2$.

The irreducibility of these invariant subspaces can be verified from the character of the corresponding PTM representation (recall the definition of character in Eq.(\ref{rep_cha_app})). To see this, we write the character of the reducible representation as $\chi(g)=\sum_i a_i \chi_i(g)$, where $\chi_i(g)$ is the character of the $ith$ irrep and $a_i$ is its multiplicity. The orthogonality relation between irreducible characters (see Eq.(\ref{irrep_cha_ortho_app})), together with the fact that every $\mathcal{R}(g)$ is real, give rise to $\frac{1}{|G(4)|} \sum_{g \in G(4)} \chi^*(g) \chi(g)=\frac{1}{|G(4)|} \sum_{g \in G(4)} \chi^2(g)=\sum_i a^2_i$. For $G(4)$, one can verify numerically that $\frac{1}{|G(4)|} \sum_{g \in G(4)} \chi^2(g)=3$. Since the multiplicities $a_i$ are positive integers, $\sum_i a^2_i=3$ is possible only if three inequivalent irreps occur, each with multiplicity one. Consequently, the three invariant subspaces $W_0$, $W_1$ and $W_2$ are irreducible, in agreement with Ref.~\cite{CrosMagBis2016}.

Identifying the irreps is important for benchmarking the fidelity of the gate set. The standard RB figure of merit, $\bar{F}(\Lambda, \mathcal{I})$, can be decomposed in terms of the projectors onto the two non-trivial irreducible subspaces as follows:
\begin{equation}
\begin{split}
   \bar{F}(\Lambda, \mathcal{I}) &= \frac{1}{d} + \frac{1}{d(d+1)} \bigg( \text{Tr} \big( \mathbb{P}_1 \mathcal{R}(\Lambda) \big) + \text{Tr} \big( \mathbb{P}_2 \mathcal{R}(\Lambda) \big) \bigg) \\ 
   &= \frac{1}{d} + \frac{1}{d(d+1)} \bigg(|\mathbb{P}_1| \; \lambda(\mathbb{P}_1,\Lambda) \; + \; |\mathbb{P}_2| \; \lambda(\mathbb{P}_2,\Lambda) \bigg) \;,
\label{Fid_cnot_group}
\end{split}
\end{equation}
with
\begin{equation}
    \mathbb{P}_1 = \sum_{i=\{3,6,15\}} |\tau_i \rangle \rangle \langle \langle \tau_i| \; , \;\;\; \mathbb{P}_2 = \sum_{i=\{1,2,...,15\} \setminus \{3,6,15\}} |\tau_i \rangle \rangle \langle \langle \tau_i| \;.
\label{invariant_projs_cnot_group}
\end{equation}
The gate set fidelity in Eq.(\ref{Fid_cnot_group}) shows that gauging its value relies in acquiring an estimate for both $\lambda(\mathbb{P}_1,\Lambda)$ and $\lambda(\mathbb{P}_2,\Lambda)$. The fitting model for each of these quantities follows essentially the same steps as those discussed in Section~\ref{C2groupRB}. Indeed, Eqs.(\ref{trivial_proj_c2_app}) to (\ref{seqfun_app_srb}) carry through, with the exception that the projectors are now either $\mathbb{P}_1$ or $\mathbb{P}_2$, as defined in Eq.(\ref{invariant_projs_cnot_group}). The choice between the two depends on whether we are constructing the fitting model for $\lambda(\mathbb{P}_1,\Lambda)$ or $\lambda(\mathbb{P}_2,\Lambda)$. We note, however, that the convenient choice of input state and POVM elements is no longer the $z-$computational basis, as for the Clifford group. We can understand this as a consequence of the irrep decomposition of the CNOT-Dihedral group $G(4)$. When expressed in the PTM basis, states in the $z$-computational basis lie entirely within $W_0 \oplus W_1$, and therefore have no overlap with the second irreducible subspace $W_2$. Consequently, this choice of measurement basis cannot probe the decay associated with $W_2$. This highlights that, in order to estimate all non-trivial decay factors, the choice of input state and POVM elements must provide nonzero overlap with each of the corresponding invariant subspaces. For this reason, all simulations employing the CNOT-Dihedral group $G(4)$ were run with the input and POVM elements set to a mixed basis, where the first qubit is measured in the $z-$computational basis and the second qubit in the $x-$computational basis, i.e.:
\begin{equation}
    \begin{split}
     &|\rho \rangle \rangle  = \frac{1}{2} \Big( |\tau_0 \rangle\rangle + |\tau_1 \rangle\rangle + |\tau_6 \rangle\rangle + |\tau_{13} \rangle\rangle \Big) = |E_{0+} \rangle\rangle \; , \\
     & | E_{0-} \rangle\rangle = \frac{1}{2} \Big( |\tau_0 \rangle\rangle - |\tau_1 \rangle\rangle + |\tau_6 \rangle\rangle - |\tau_{13} \rangle\rangle \Big) \; , \\
     & | E_{1+} \rangle\rangle = \frac{1}{2} \Big( |\tau_0 \rangle\rangle + |\tau_1 \rangle\rangle - |\tau_6 \rangle\rangle - |\tau_{13} \rangle\rangle \Big) \; , \\
     & | E_{1-} \rangle \rangle = \frac{1}{2} \Big( |\tau_0 \rangle\rangle - |\tau_1 \rangle\rangle - |\tau_6 \rangle\rangle + |\tau_{13} \rangle\rangle \Big) \; .
    \end{split}
\label{POVM_and_input_G4}    
\end{equation}

\subsection{Interleaved RB}\label{fitting_models_intRB_app}
The goal of interleaved RB is to extract information on the average error rate of individual gates~\cite{MagGamJoh2012}. This is achieved through constructing interleaved sequences, where the target benchmarking gate is intertwined with randomly selected gates. We adopt the same noise modeling convention as in interleaved RB, whereby the noisy random gates and the noisy target gate, $\tilde{\mathcal{U}}$, are modeled respectively as~\cite{MagGamJoh2012}:

\begin{equation}
    \mathcal{R}(\tilde{g}) = \mathcal{R}(\Lambda)\mathcal{R}(g) \;  , \; \mathcal{R}(\tilde{\mathcal{U}})= \mathcal{R}(\mathcal{U}) \mathcal{R}(\Lambda_\mathcal{U}) \; .
\end{equation}
Given this noise model and interleaved sequences, the probability distribution, $p(x,\mathbf{g})$, introduced in Eq.(\ref{probseq_def_app}) is modified to:

\begin{equation}
\begin{split}
    p(x,\mathbf{g}) & =  \langle\langle \tilde{E}_x| \mathcal{R}(\Lambda)\mathcal{R}(g_m) \mathcal{R}(\mathcal{U}) \mathcal{R}(\Lambda_\mathcal{U})  \mathcal{R}(\Lambda)\mathcal{R}(g_{m-1}) \mathcal{R}(\mathcal{U}) \mathcal{R}(\Lambda_\mathcal{U})\mathcal{R}(\Lambda)...   \mathcal{R}(g_{1})| \tilde{\rho} \rangle \rangle   \\
    &=\langle\langle \tilde{E}_x| \mathcal{R}(\Lambda) \mathcal{R}(g_m)  \prod^{m-1}_{j=1} \mathcal{R}(\tilde{\Lambda}) \mathcal{R}(g_j) | \tilde{\rho} \rangle \rangle \;,
\end{split}    
    \label{probseq_int_def_app}
\end{equation}

where we defined:

\begin{equation}
    \mathcal{R}(\tilde{\Lambda}) :=   \mathcal{R}(\mathcal{U}) \mathcal{R}(\Lambda_\mathcal{U})  \mathcal{R}(\Lambda) \; .
\end{equation}
The sequence function model is then given by the expectation value:
\begin{equation}
\begin{split}
   & k_A(m) = \mathop{\mathlarger{\mathlarger{\mathlarger{\mathbb{E}}}}}_{\mathbf{g}\in G^{\times m}} \mathlarger{\sum}_{x\in \{0,1\}^n}  \; f_A(x,\mathbf{g}) \; p(x,\mathbf{g}) \\ 
   & = c^{(0)}_{f_A}\mathop{\mathlarger{\mathlarger{\mathlarger{\mathbb{E}}}}}_{\mathbf{g}\in G^{\times m}} \mathlarger{\sum}_{x\in \{0,1\}^n} \langle\langle E_x| \mathcal{R}_i(g_m)\prod^{m-1}_{j=1} A \mathcal{R}_i(g_j) |\rho \rangle\rangle \langle\langle \tilde{E}_x| \mathcal{R}(\Lambda) \mathcal{R}(g_m)  \prod^{m-1}_{j=1} \mathcal{R}(\tilde{\Lambda}) \mathcal{R}(g_j) | \tilde{\rho} \rangle \rangle \;,
\end{split}   
\end{equation}
which has a very similar structure to Eq.(\ref{seqfun_app1}). This allows us to directly write:
\begin{equation}
   k_A(m)=c^{(0)}_{f_A} \mathlarger{\sum}_{x\in \{0,1\}^n} \; \text{Tr} \bigg[ \Omega(\rho,\tilde{\rho},E_x,\tilde{E}_x) \; \Big( \mathlarger{\mathbb{P}}_{\text{triv}} (A \otimes \mathcal{R} (\tilde{\Lambda})) \mathlarger{\mathbb{P}}_{\text{triv}}  \Big)^{m-1} \bigg] \; .
   \label{seqfun_int_app2}
\end{equation}
To further simplify the fitting model, we need to specify both the choice of gate set and the probe operator. In accordance with the main text, we examine the fitting model produced by the multi-qubit Clifford group, and the CNOT-dihedral group $G(4)$.

\subsubsection{Multi-qubit Clifford group}\label{C2groupRB_int}
The similarity between Eq.(\ref{seqfun_int_app2}) and Eq.(\ref{seqfun_app2}) allows us to immediately write the sequence function for this gate set as:
\begin{equation}
k_A(m)=c^{(0)}_{f_A} \; \bigg(\frac{1}{|\mathbb{P}_1|} \;  \text{Tr}\big( \mathbb{P}_1 A^{\intercal}  \mathbb{P}_1 \mathcal{R}(\tilde{\Lambda})  \big) \bigg)^{m-1} \;  \mathlarger{\sum}_{x\in \{0,1\}^2}  \frac{1}{|\mathbb{P}_1|} \;  \langle \langle \mathbb{P}_1| \Omega(\rho,\tilde{\rho},E_x,\tilde{E}_x ) | \mathbb{P}_1 \rangle \rangle \; ,
\label{seqfun_int_C2_1}
\end{equation}

Suppose that $\mathcal{U}$ is one of the Clifford gates in the gate set. Then choosing the probe operator to be $A=\mathcal{R}(\mathcal{U})$, renders the interleaved RB decay factor as:

\begin{equation}
\begin{split}
\lambda(\mathcal{R}^{ \dagger}(\mathcal{U}),\tilde{\Lambda})&= \frac{1}{|\mathbb{P}_1|}\text{Tr}\big( \mathbb{P}_1 \mathcal{R}^{ \dagger}(\mathcal{U})  \mathbb{P}_1 \mathcal{R}(\tilde{\Lambda})  \big) \\ 
&= \frac{1}{|\mathbb{P}_1|}  \text{Tr}\big( \mathbb{P}_1 \mathcal{R}^{ \dagger}(\mathcal{U})  \mathbb{P}_1 \mathcal{R}(\mathcal{U}) \;  \mathcal{R}(\Lambda_\mathcal{U})  \mathcal{R}(\Lambda)  \big) \\ 
&= \frac{1}{|\mathbb{P}_1|} \text{Tr}\big(\mathbb{P}_1\mathcal{R}(\Lambda_\mathcal{U})  \mathcal{R}(\Lambda)  \big) \; .
\end{split}
\end{equation}
Although the probe operator entering in the sequence correlation function is $A=\mathcal{R}(\mathcal{U})$, we label the corresponding decay factor by $A^{\intercal}=\mathcal{R}^{ \dagger}(\mathcal{U})$, since the decay rate depends explicitly on $A^{\intercal}$, and $\mathcal{R}^{\intercal}(\mathcal{U})=\mathcal{R}^{\dagger}(\mathcal{U})$ in the PTM formalism.

Using Eq.(\ref{average_fid}), the decay factor $\lambda(\mathcal{R}^{ \dagger}(\mathcal{U}),\tilde{\Lambda})$ can be directly related to average fidelity $\bar{F}(\Lambda_\mathcal{U} \circ \Lambda,\mathcal{I})$. As stated in the main text, having both knowledge on $\bar{F}(\Lambda,\mathcal{I})$ and $\bar{F}(\Lambda_\mathcal{U} \circ \Lambda,\mathcal{I})$ is sufficient to bound the value of $\bar{F}(\tilde{\mathcal{U}},\mathcal{U})$. This result arises from the fact that (see Ref.~\cite{CarWalEm2019} for proof) for any two quantum channels, $\mathcal{A}$ and $\mathcal{B}$, the following inequality holds true:
\begin{equation}
    |\chi^{(\mathcal{A}\mathcal{B})}_{00}- \chi^{(\mathcal{A})}_{00}\chi^{(\mathcal{B})}_{00}-(1-\chi^{(\mathcal{A})}_{00})(1-\chi^{(\mathcal{B})}_{00})| \leq 2 \sqrt{\chi^{(\mathcal{A})}_{00}\chi^{(\mathcal{B})}_{00}(1-\chi^{(\mathcal{A})}_{00})(1-\chi^{(\mathcal{B})}_{00})} \; ,
 \label{bound_interleaved_gate}   
\end{equation}
where $\chi^{(\mathcal{A})}_{00}=\frac{(d+1)\bar{F}(\mathcal{A},\mathcal{I})-1}{d}$, and likewise for $\chi^{(\mathcal{B})}_{00}$ and $\chi^{(\mathcal{A}\mathcal{B})}_{00}$. By identifying $\mathcal{A}=\Lambda_{\mathcal{U}}$, $\mathcal{B}=\Lambda$, and thus $\mathcal{AB}=\Lambda_{\mathcal{U}} \circ \Lambda$, the inequality can be inverted to obtain a bound for $\bar{F}(\Lambda_{\mathcal{U}}, \mathcal{I})=\bar{F}(\tilde{\mathcal{U}},\mathcal{U})$. Note that this approach does not require any assumptions on the specific type of noise affecting the gates, i.e. whether it is a depolarizing channel, a unitary channel, or any other type.

\subsubsection{CNOT-Dihedral group $G(4)$}\label{G4groupRB_int}
Interleaved RB choosing the CNOT-Dihedral group $G(4)$ as gate set follows similarly to the multi-qubit Clifford group, with the exception that two decay factors contribute to the average fidelity $\bar{F}(\Lambda_{\mathcal{U}} \circ \Lambda,\mathcal{I})$ (see Eq.(\ref{Fid_cnot_group})).

The choice of probe operator remains the same, $A=\mathcal{R}(\mathcal{U})$, provided that $\mathcal{R}(\mathcal{U})$ preserves the invariant subspaces of the gate set $G(4)$. If $\mathcal{U}$ belongs to the gate set, this condition is automatically satisfied, although membership is not strictly necessary. In the main text, the target gate was a CNOT, which belongs to both $G(4)$ and the Clifford group.

To extract the two decay factors separately, two sets of sequence correlation functions must be constructed, using respectively the two non-trivial irreps  $\mathcal{R}_1(g)$ and $\mathcal{R}_2(g)$, embedded in the full PTM space.

The remaining difference with respect to the multi-qubit Clifford group as gate set is the choice of POVMs and input state, which are now given by Eq.(\ref{POVM_and_input_G4}). The desired fidelity $\bar{F}(\tilde{\mathcal{U}},\mathcal{U})$ can then be bounded with Eq.(\ref{bound_interleaved_gate}), from estimates of both $\bar{F}(\Lambda_\mathcal{U} \circ \Lambda,\mathcal{I})$ (obtained from interleaved sequences) and $\bar{F}(\Lambda,\mathcal{I})$ (obtained from fully random sequences).

\subsection{Simultaneous RB}\label{fitting_models_simRB_app}
\subsubsection{The local Clifford group $\mathcal{C}_1\times \mathcal{C}_1$}
The local Clifford group on two qubits, $\mathcal{C}_1\times \mathcal{C}_1$, provides another example of a gate set whose PTM representation is multiplicity-free. Recall that the PTM representation of the single-qubit Clifford group decomposes into the trivial and non-trivial irreps, which we denote by $\mathcal{R}_0$ and $\mathcal{R}_1$, respectively. Since the irreducible representations of a direct product group are obtained as tensor products of the irreducible representations of the individual groups~\cite{Tinkham1964}, the PTM representation of $\mathcal{C}_1\times \mathcal{C}_1$ decomposes as $\mathcal{R}_0 \otimes \mathcal{R}_0, \mathcal{R}_0 \otimes \mathcal{R}_1, \mathcal{R}_1 \otimes \mathcal{R}_0, \mathcal{R}_1 \otimes \mathcal{R}_1$. These act on the four invariant subspaces listed in Tab.~\ref{tab_c1xc1_proj}.
\begin{table}[htbp]
\centering
\begin{tabular}{ccc}
\hline
\hline
 Irreducible Representation & Invariant Subspace & Projector  \\
\hline
Trivial irrep & $W_0=\text{Span} \big\{ |\tau_0 \rangle \rangle \big \}$ & $\mathbb{P}_0 = |\tau_0 \rangle \rangle \langle \langle \tau_0|$\\
1st non-trivial irrep & $W_1=\text{Span} \big\{|\tau_4 \rangle \rangle, | \tau_5 \rangle\rangle , | \tau_6 \rangle \rangle \big\}$ & $\mathbb{P}_1 = \sum_{i\in \{4,5,6\}} |\tau_i \rangle\rangle \langle \langle \tau_i |$\\
2nd non-trivial irrep & $W_2=\text{Span} \big\{|\tau_1 \rangle \rangle, | \tau_2 \rangle\rangle , | \tau_3 \rangle \rangle \big\}$ &  $\mathbb{P}_2= \sum_{i\in \{1,2,3\}} |\tau_i \rangle\rangle \langle \langle \tau_i |$ \\
3rd non-trivial irrep & $W_3=\text{Span} \big\{|\tau_7 \rangle \rangle, ... , | \tau_{15} \rangle \rangle \big\}$ &  $\mathbb{P}_3= \sum^{15}_{i=7} |\tau_i \rangle\rangle \langle \langle \tau_i |$ \\
\hline
\end{tabular}
\caption{\label{tab_c1xc1_proj} List of irreps of the $\mathcal{C}_1\times \mathcal{C}_1$ group. The table also includes the corresponding irreducible subspaces and the projectors onto those subspaces. Note that we adhere to the vector labeling convention introduced in Eq.(\ref{PTM_basis_choice1}).}
\end{table}

In simultaneous RB, the sequence fidelity contains three decay parameters~\cite{GamCorMer2013}. The gate-set shadow protocol allows us to extract each of these parameters separately. To derive the fitting model for each decay rate, the sequence correlation function $f_A(x,\mathbf{g})$ is constructed using the corresponding embedded irrep. The resulting fitting model is given by Eq.(\ref{seqfun_app2}), with the trivial projector, $\mathbb{P}_{\text{triv}}$, now related to the invariant subspace of the chosen irrep, i.e. $\mathbb{P}_{\text{triv}}=\frac{1}{|\mathbb{P}_i|} |\mathbb{P}_i\rangle\rangle \langle\langle\mathbb{P}_i|$. This leads to a decaying model similar to Eq.(\ref{seqfun_app_srb}). By fixing the probe operator $A$ as the projector onto the invariant subspace of the chosen irrep, we can derive the corresponding fitting models:
\begin{equation}
    k_A(m) = c_i \; \Bigg(\frac{\text{Tr}(\mathbb{P}_i\Lambda)}{|\mathbb{P}_i|} \Bigg)^{m-1} \;.
\end{equation}
The constant $c_i$ is a SPAM dependent parameter. From Eq.(\ref{seqfun_app_srb}), we can immediately conclude that $c_i$ is given by:
\begin{equation}
  c_i = c^{(0)}_{f_{A_i}} \; \sum_{x \in \{0,1 \}^2} \frac{1}{|\mathbb{P}_i|} \langle \langle \mathbb{P}_i| \Omega(\rho,\tilde{\rho}, E_x, \tilde{E}_x) |\mathbb{P}_i \rangle \rangle \; ,
\end{equation}
with $\Omega(\rho,\tilde{\rho}, E_x, \tilde{E}_x) $ defined as in section~\ref{fitting_models_app}. The set of normalization constants $\{c^{(0)}_{f_{A_i}}\}$ is determined by the limit of no SPAM and gate errors. A priori, each irrep can give rise to a different normalization constant. In this case, and using the input state and POVM elements defined in Eq.(\ref{z_compbasis_def}), we have:
\begin{equation}
\begin{split}
   A= \mathbb{P}_1: \; k_A(1) &= c^{(0)}_{f_{\mathbb{P}_1}} \; \sum_{x \in \{0,1 \}^2} \frac{1}{|\mathbb{P}_1|} \langle \langle \mathbb{P}_1| \Omega(\rho,\tilde{\rho}, E_x, \tilde{E}_x) |\mathbb{P}_1 \rangle \rangle \\
   &= \frac{c^{(0)}_{f_{\mathbb{P}_1}}}{3} \;  \sum_{i \in \{4,5,6\}} \langle \langle \tau_i| \rho \rangle \rangle^2 \; \sum_{x \in \{0,1 \}^2} \sum_{j \in \{4,5,6\}}  \langle \langle E_x| \tau_j \rangle \rangle^2\\
   &= \frac{c^{(0)}_{f_{\mathbb{P}_1}}}{12} \implies c^{(0)}_{f_{\mathbb{P}_1}}=12 \;.  \;
\end{split}
\end{equation}

\begin{equation}
\begin{split}
   A= \mathbb{P}_3: \; k_A(1) &= c^{(0)}_{f_{\mathbb{P}_3}} \; \sum_{x \in \{0,1 \}^2} \frac{1}{|\mathbb{P}_3|} \langle \langle \mathbb{P}_3| \Omega(\rho,\tilde{\rho}, E_x, \tilde{E}_x) |\mathbb{P}_3 \rangle \rangle \\
   &= \frac{c^{(0)}_{f_{\mathbb{P}_3}}}{9} \;  \sum^{15}_{i=7} \langle \langle \tau_i| \rho \rangle \rangle^2 \; \sum_{x \in \{0,1 \}^2} \sum^{15}_{j=7}  \langle \langle E_x| \tau_j \rangle \rangle^2\\
   &= \frac{c^{(0)}_{f_{\mathbb{P}_3}}}{36} \implies c^{(0)}_{f_{\mathbb{P}_3}}=36 \;.  \;
\end{split}
\end{equation}
Following similar arguments, $c^{(0)}_{f_{\mathbb{P}_2}}=12$.

\subsubsection{The group $\mathcal{C}_1\times \mathcal{I}$}\label{group_c1xI_sec}
The PTM representation of the group $\mathcal{C}_1\times \mathcal{I}$, which is a subgroup of the local Clifford group $\mathcal{C}_1 \times \mathcal{C}_1$, is not multiplicity-free. In particular, each of the two inequivalent irreps appearing in the single-qubit Clifford PTM representation occurs with multiplicity four. To motivate this statement, we consider the operations $\mathcal{R}(g) \otimes \mathbbm{1}$, with $g\in \mathcal{C}_1$, and their action on the invariant subspaces of the local Clifford group listed in Tab.~\ref{tab_c1xc1_proj}.

We begin with the trivial subspace, spanned by $|\sigma_0 \otimes \sigma_0 \rangle\rangle$, which is mapped to itself under all $\mathcal{R}(g) \otimes \mathbbm{1}$. The second non-trivial invariant subspace is instead decomposed into three one-dimensional invariant subspaces, since each of the vectors $|\sigma_0 \otimes \sigma_1 \rangle\rangle$, $|\sigma_0 \otimes \sigma_2 \rangle\rangle$, and $|\sigma_0 \otimes \sigma_3 \rangle\rangle$ is individually invariant under the action of all $\mathcal{R}(g) \otimes \mathbbm{1}$. Together with $|\sigma_0 \otimes \sigma_0 \rangle\rangle$, these provide four copies of the trivial irrep.

The invariant subspace associated with the first non-trivial irrep of the local Clifford group remains irreducible under the action of $\mathcal{C}_1\times \mathcal{I}$. The third non-trivial invariant subspace, however, becomes reducible, decomposing into three subspaces
\begin{equation}
    \text{Span}\big \{ |\sigma_i \otimes \sigma_1 \rangle \rangle \big \}_{i \in \{1,2,3\}} \; , \; 
    \text{Span}\big \{ |\sigma_i \otimes \sigma_2 \rangle \rangle \big \}_{i \in \{1,2,3\}} \;\; \text{and} \;\;
    \text{Span}\big \{ |\sigma_i \otimes \sigma_3 \rangle \rangle \big \}_{i \in \{1,2,3\}} \; .
\end{equation}
These three subspaces, together with the first non-trivial invariant subspace, carry equivalent copies of the same non-trivial irrep: in each case, the group action permutes, up to a sign, the non-identity Pauli basis operators on the first qubit. Thus, the non-trivial irrep also occurs with multiplicity four.

We are interested in the average sequence functions $k_A(m)$ associated with the non-trivial irrep. To simplify notation, let us re-label the basis vectors in Eq.(\ref{PTM_basis_choice1}) as follows:
\begin{equation}
    \begin{split}
    &\text{Non-trivial irrep, copy 1:}\\
    & | \sigma_1 \otimes \sigma_0 \rangle\rangle \equiv | \tau^{(1)}_1 \rangle\rangle \; , \; | \sigma_2 \otimes \sigma_0 \rangle\rangle \equiv | \tau^{(1)}_2 \rangle\rangle \; , \; | \sigma_3 \otimes \sigma_0 \rangle\rangle \equiv | \tau^{(1)}_3 \rangle\rangle \; , \\
    & \text{Non-trivial irrep, copy 2:}\\
    & | \sigma_1 \otimes \sigma_1 \rangle\rangle \equiv | \tau^{(2)}_1 \rangle\rangle \; , \; | \sigma_2 \otimes \sigma_1 \rangle\rangle \equiv | \tau^{(2)}_2 \rangle\rangle \; , \; | \sigma_3 \otimes \sigma_1 \rangle\rangle \equiv | \tau^{(2)}_3 \rangle\rangle \; , \\
    & \text{Non-trivial irrep, copy 3:}\\
    & | \sigma_1 \otimes \sigma_2 \rangle\rangle \equiv | \tau^{(3)}_1 \rangle\rangle \; , \; | \sigma_2 \otimes \sigma_2 \rangle\rangle \equiv | \tau^{(3)}_2 \rangle\rangle \; , \; | \sigma_3 \otimes \sigma_2 \rangle\rangle \equiv | \tau^{(3)}_3 \rangle\rangle \; . \\
     & \text{Non-trivial irrep, copy 4:}\\
    & | \sigma_1 \otimes \sigma_3 \rangle\rangle \equiv | \tau^{(4)}_1 \rangle\rangle \; , \; | \sigma_2 \otimes \sigma_3 \rangle\rangle \equiv | \tau^{(4)}_2 \rangle\rangle \; , \; | \sigma_3 \otimes \sigma_3 \rangle\rangle \equiv | \tau^{(4)}_3 \rangle\rangle \; . \\
    \end{split}
    \label{new_basis_c1xI}
\end{equation}
In Eq.(\ref{new_basis_c1xI}), the basis vectors spanning the trivial irrep are not included. This is justified, since the trivial projector $ \mathlarger{\mathbb{P}}_{\text{triv}}$ will not have any contribution from this irrep. For this reason, we neglect the trivial irrep and its copies for the remainder of this section. Given our choice of gate set, irrep, and basis vectors, the invariant vectors in Eq.(\ref{gen_invvec_mul}) now acquire the explicit form
\begin{equation}
    |v_{\mu,\nu} \rangle \rangle = \frac{1}{\sqrt{|\mathbbm{P}_1|}} \sum^{|\mathbbm{P}_1|}_{l=1} |\tau^{(\mu)}_l \rangle \rangle \otimes | \tau^{(\nu)}_l \rangle \rangle = \frac{1}{\sqrt{|\mathbbm{P}_1|}} \sum^{|\mathbbm{P}_1|}_{l=1} \text{vec} \big(| \tau^{(\nu)}_l \rangle \rangle \langle \langle \tau^{(\mu)}_l|  \big) = \frac{1}{\sqrt{|\mathbbm{P}_1|}} \sum^{|\mathbbm{P}_1|}_{l=1} | \mathbbm{P}^{(\nu,\mu)}_l  \rangle \rangle =  |\mathbbm{P}^{(\nu,\mu)} \big \rangle \rangle \; ,
\end{equation}
where we have defined $\mathbbm{P}^{(\nu,\mu)}_l  := | \tau^{(\nu)}_l \rangle \rangle \langle \langle \tau^{(\mu)}_l|$, $\mathbbm{P}^{(\nu,\mu)} := \frac{1}{\sqrt{|\mathbbm{P}_1|}} \sum^{|\mathbbm{P}_1|}_{l=1} \mathbbm{P}^{(\nu,\mu)}_l $, and have employed the Kronecker and outer product relations using the usual column-stacking vector convention. The resulting trivial projector is then given by 
\begin{equation}
    \mathbb{\mathbb{P}}_{\text{triv}} = \sum^4_{\mu,\nu=1} | \mathbbm{P}^{(\nu,\mu)} \rangle \rangle \langle \langle \mathbbm{P}^{(\nu,\mu)}|  \; .
    \label{Ptriv_gen_C1xI_app}
\end{equation}
As a consequence
\begin{equation}
    \begin{split}
        \mathbb{\mathbb{P}}_{\text{triv}} \Big( A \otimes \mathcal{R}(\Lambda) \Big) \mathbb{\mathbb{P}}_{\text{triv}} &= \sum_{\mu,\nu, \mu',\nu'}  \langle \langle \mathbbm{P}^{(\nu,\mu)}|  A \otimes \mathcal{R}(\Lambda) | \mathbbm{P}^{(\nu',\mu')}  \rangle \rangle \; | \mathbbm{P}^{(\nu,\mu)} \big \rangle \rangle \langle \langle \mathbbm{P}^{(\nu',\mu')}| \\
        &= \sum_{\mu,\nu, \mu',\nu'} \text{Tr} \big(\mathbbm{P}^{(\mu,\nu)} \mathcal{R}(\Lambda) \mathbbm{P}^{(\nu',\mu')} A^{\intercal} \big) \; | \mathbbm{P}^{(\nu,\mu)}  \rangle \rangle \langle \langle \mathbbm{P}^{(\nu',\mu')}| \\
        &= \frac{1}{|\mathbbm{P}_1|} \sum_{\mu,\nu, \mu',\nu'}  \sum^{|\mathbbm{P}_1|}_{p,n=1} \langle \langle \tau^{(\mu')}_p | A^{\intercal} | \tau^{(\mu)}_n \rangle \rangle \langle\langle \tau^{(\nu)}_n | \mathcal{R}(\Lambda) | \tau^{(\nu')}_p \rangle \rangle \; | \mathbbm{P}^{(\nu,\mu)} \rangle \rangle \langle \langle \mathbbm{P}^{(\nu',\mu')}| \; .
    \end{split}
    \label{Phi_C1xI_1}
\end{equation}
Now suppose we set 
\begin{equation}
A= \mathbbm{P}_1=\sum^3_{l=1} |\tau^{(1)}_l \rangle\rangle\langle\langle \tau^{(1)}_l| =   \sum^3_{l=1}|\sigma_l \otimes \sigma_0 \rangle \rangle \langle \langle \sigma_l \otimes \sigma_0| \; .
\label{P1_proj_C1xI}
\end{equation}
Then Eq.(\ref{Phi_C1xI_1}) simplifies to
\begin{equation}
    \begin{split}
     \mathbb{\mathbb{P}}_{\text{triv}} \Big( \mathbbm{P}_1 \otimes \mathcal{R}(\Lambda) \Big) \mathbb{\mathbb{P}}_{\text{triv}} &=   \frac{1}{|\mathbbm{P}_1|} \sum_{\mu,\nu, \mu',\nu'}  \sum^{|\mathbbm{P}_1|}_{l,p,n=1} \delta_{p,l} \delta_{l,n} \delta_{\mu',1} \delta_{\mu,1} \langle\langle \tau^{(\nu)}_n | \mathcal{R}(\Lambda) | \tau^{(\nu')}_p \rangle \rangle \; | \mathbbm{P}^{(\nu,\mu)} \rangle \rangle \langle \langle \mathbbm{P}^{(\nu',\mu')}| \\
     & =\frac{1}{|\mathbbm{P}_1|} \sum_{\nu, \nu'} \sum_l \langle\langle \tau^{(\nu)}_l | \mathcal{R}(\Lambda) | \tau^{(\nu')}_l \rangle \rangle \;  | \mathbbm{P}^{(\nu,1)} \rangle \rangle \langle \langle \mathbbm{P}^{(\nu',1)}| \\
     & = \sum_{\nu, \nu'} \Phi_{\nu,\nu'} \; | \mathbbm{P}^{(\nu,1)} \rangle \rangle \langle \langle \mathbbm{P}^{(\nu',1)}| \; ,
    \end{split}
    \label{Phi_C1xI_2}
\end{equation}
with 
\begin{equation}
   \Phi_{\nu,\nu'}:= \frac{1}{|\mathbbm{P}_1|} \sum^3_{l=1}  \langle\langle \tau^{(\nu)}_l | \mathcal{R}(\Lambda) | \tau^{(\nu')}_l \rangle \rangle \; .
   \label{Phi_C1xI_matrixele_app}
\end{equation}

In Eq.(\ref{lambda1_decay}) in the main text, we defined the decay rate $\lambda_1$. Comparing that expression with the matrix elements $\Phi_{\nu,\nu'}$, we can conclude that $\lambda_1=\Phi_{1,1}$. The average sequence function $k_{\mathbbm{P}_1}(m)$, however, contains in general multiple decay rates, corresponding to the eigenvalues of $\Phi$. Even if these decay rates are successfully extracted through a multi-exponential fit, they do not, in general, uniquely determine the matrix element $\Phi_{1,1}$.

We can, nonetheless, proceed with determining the resulting average sequence function model for the choice $A= \mathbbm{P}_1$. Using Eq.(\ref{seqfun_app2}) and Eq.(\ref{Phi_C1xI_2}), we arrive at the model: 
\begin{equation}
    \begin{split}
        k_{\mathbbm{P}_1}(m) = \frac{c^{(0)}_{f_{\mathbbm{P}_1}}}{|\mathbbm{P}_1|} \; \sum_{\nu,\nu'} \big[\Phi^{m-1} \big]_{\nu,\nu'} \; \sum^3_{l,n=1} \sum_x \langle\langle\tau^{(1)}_l| \rho \rangle \rangle \langle\langle\tau^{(\nu')}_l| \tilde{\rho} \rangle \rangle \langle\langle E_x| \tau^{(1)}_n \rangle \rangle \langle \langle \tilde{E}_x| \mathcal{R}(\Lambda) |\tau^{(\nu)}_n \rangle \rangle = \sum_{\nu,\nu'}C_{\nu,\nu'} \; \big[\Phi^{m-1} \big]_{\nu,\nu'}  \; .
    \end{split}
    \label{Phi_C1xI_3}
\end{equation}
The coefficients $C_{\nu,\nu'}$ can be further simplified after specifying the input state and set of POVM elements. Suppose we choose
\begin{equation}
\begin{split}
   & |\rho \rangle \rangle = \frac{1}{2} \Big( | \sigma_0 \otimes \sigma_0 \rangle \rangle + |\tau^{(1)}_3 \rangle \rangle \Big) \; , \\
   & |E_0 \rangle \rangle=2|\rho \rangle \rangle \;, \\
   & |E_1 \rangle \rangle=  |\sigma_0 \otimes \sigma_0 \rangle \rangle - |\tau^{(1)}_3 \rangle \rangle \; ,
\end{split}
\label{POVM_input_C1xI_app}
\end{equation}
then $C_{\nu,\nu'}$ reduces to $C_{\nu,\nu'}=\frac{c^{(0)}_{f_{\mathbbm{P}_1}}}{2|\mathbbm{P}_1|} \; \langle\langle\tau^{(\nu')}_3| \tilde{\rho} \rangle \rangle \langle \langle (\tilde{E}_0-\tilde{E}_1)| \mathcal{R}(\Lambda) |\tau^{(\nu)}_3 \rangle \rangle$. Further simplification can be achieved in the absence of SPAM errors. In that case, Eq.(\ref{Phi_C1xI_3}) admits the form
\begin{equation}
    k_{\mathbbm{P}_1}(m) = \frac{c^{(0)}_{f_{\mathbbm{P}_1}}}{2|\mathbbm{P}_1|} \; \sum_{\nu} \big[ \Phi^{m-1} \big]_{\nu,1} \langle \langle \tau^{(1)}_3| \mathcal{R}(\Lambda) |\tau^{(\nu)}_3 \rangle \rangle \; .
    \label{Phi_C1xI_4}
\end{equation}
While Eq.(\ref{Phi_C1xI_4}) can in principle contain up to four decay modes, certain noise processes may nevertheless produce an effective single decay signal. Under the assumption of ideal state preparation and measurements (used to obtain Eq.(\ref{Phi_C1xI_4})), any uncorrelated gate noise channel reduces the sequence function to a single-decay model $k_{\mathbbm{P}_1}(m) \propto \lambda^{m-1}_1$.

If we instead choose $A=\mathbbm{1}$ in Eq.(\ref{Phi_C1xI_1}) and follow the approach outlined in Sec.~\ref{op_crosstalk_sec}, where we make use of multiple preparation states and measurement operators, then in postprocessing each setting allows us to estimate the average sequence function through the model
\begin{equation}
    \begin{split}
        k^{(\mu,\nu)}(m)&=c^{(0)}_{f_{\mathbbm{1};\mu,\nu}} \sum_{i,j,k} \big[\Phi^{m-1}\big]_{j,k} \sum_x \text{Tr} \Big[  \Omega\big(\rho^{(\nu)}, \tilde{\rho}^{(\nu)}, E^{(\mu)}_x, \tilde{E}^{(\mu)}_x \big) \; | \mathbbm{P}^{(j,i)} \rangle \rangle \langle \langle \mathbbm{P}^{(k,i)}| \Big] \\
        &=c^{(0)}_{f_{\mathbbm{1};\mu,\nu}} \sum_{i,j,k} \big[\Phi^{m-1}\big]_{j,k} \sum_x \text{Tr} \Big[ | \rho^{(\nu)} \otimes \tilde{\rho}^{(\nu)} \rangle \rangle \langle \langle E^{(\mu)}_x \otimes \Lambda^*(\tilde{E}^{(\mu)}_x)| \mathbbm{P}^{(j,i)} \rangle \rangle \langle \langle \mathbbm{P}^{(k,i)}| \Big] \\
        &=\frac{c^{(0)}_{f_{\mathbbm{1};\mu,\nu}}}{|\mathbbm{P}_1|} \sum_{i,j,k} \big[\Phi^{m-1}\big]_{j,k} \sum_x \sum_{l,l'} \langle \langle \tau^{(i)}_l | \rho^{(\nu)} \rangle \rangle \langle \langle  \tau^{(k)}_l|\tilde{\rho}^{(\nu)} \rangle \rangle \langle \langle E^{(\mu)}_x| \tau^{(i)}_{l'} \rangle \rangle \langle \langle \tilde{E}^{(\mu)}_x | \mathcal{R}(\Lambda) | \tau^{(j)}_{l'} \rangle \rangle \;,
    \end{split}
    \label{kmunu_c1xI_1}
\end{equation}
where we have used Eq.(\ref{seqfun_app2}) and Eq.(\ref{Ptriv_gen_C1xI_app}). In Eq.(\ref{kmunu_c1xI_1}), $i,j$ and $k$ label the different equivalent copies of the non-trivial irrep, while the indices $l,l'$ label the basis elements within each copy of the irrep. In contrast, $\mu$ and $\nu$ label the different measurement and state preparation settings, respectively.

Using the set of input states and POVM elements defined in Eq.(\ref{States_and_POVMs_C1xI}), we can determine the overlaps with the idealized preparation states and measurement operators
\begin{equation}
    \begin{split}
     \langle \langle \tau^{(i)}_l | \rho^{(\nu)} \rangle \rangle &= \frac{1}{\sqrt{2}} \text{Tr} \bigg( \Big(\sigma_l \otimes \sigma_{i-1}\Big) \Big( \big(\sigma_0+\sigma_3 \big) \otimes \big(\frac{1}{2}(\mathbbm{1}+\mathbf{r}^{\intercal}_{\nu}\mathbf{P})\big) \Big) \bigg)\\
     &=\frac{1}{\sqrt{2}} \; \delta_{l,3} \;\text{Tr} \bigg( \sigma_{i-1} \big(\frac{1}{2}(\mathbbm{1}+\mathbf{r}^{\intercal}_{\nu}\mathbf{P})\big)\bigg) \\
     &= \frac{1}{\sqrt{2}} \; \delta_{l,3} \; u_{i,\nu} \; ,
    \end{split}
\end{equation}
and
\begin{equation}
    \langle \langle E^{(\mu)}_x|\tau^{(i)}_{l'} \rangle \rangle = (-1)^x \delta_{3,l'} \delta_{\mu,i} \;,
\end{equation}
allowing Eq.(\ref{kmunu_c1xI_1}) to simplify to
\begin{equation}
    \begin{split}
        k^{(\mu,\nu)}(m)&=\frac{c^{(0)}_{f_{\mathbbm{1}};\mu,\nu}}{\sqrt{2}|\mathbbm{P}_1|} \sum^4_{j,k=1} \big[\Phi^{m-1}\big]_{j,k} \; u_{\mu,\nu} \sum_x (-1)^x \; \langle \langle  \tau^{(k)}_3|\tilde{\rho}^{(\nu)} \rangle \rangle \langle \langle \tilde{E}^{(\mu)}_x | \mathcal{R}(\Lambda) | \tau^{(j)}_{3} \rangle \rangle\\
        &= \frac{c^{(0)}_{f_{\mathbbm{1}};\mu,\nu}}{\sqrt{2}|\mathbbm{P}_1|} \; u_{\mu,\nu} \; \sum_{j,k} \big[\Phi^{m-1}\big]_{j,k} \;\langle \langle  \tau^{(k)}_3|\tilde{\rho}^{(\nu)} \rangle \rangle \; \langle \langle (\tilde{E}^{(\mu)}_0-\tilde{E}^{(\mu)}_1 )| \mathcal{R}(\Lambda) | \tau^{(j)}_{3} \rangle \rangle \\
        &= \frac{c^{(0)}_{f_{\mathbbm{1}};\mu,\nu}}{\sqrt{2}|\mathbbm{P}_1|} \; u_{\mu,\nu} \; \sum_{j,k} M^{(\mu)}_j \; \big[\Phi^{m-1}\big]_{j,k} \; W^{(\nu)}_k = \frac{c^{(0)}_{f_{\mathbbm{1}};\mu,\nu}}{\sqrt{2}|\mathbbm{P}_1|} \; u_{\mu,\nu} \; Y_{\mu,\nu}(m)  \; ,
    \end{split}
    \label{kmunu_c1xI_2}
\end{equation}
where we defined $M^{(\mu)}_j:=\langle \langle (\tilde{E}^{(\mu)}_0-\tilde{E}^{(\mu)}_1 )| \mathcal{R}(\Lambda) | \tau^{(j)}_{3} \rangle \rangle$, $W^{(\nu)}_k:=\langle \langle  \tau^{(k)}_3|\tilde{\rho}^{(\nu)} \rangle \rangle$ and $Y_{\mu,\nu}(m):=\sum_{j,k} M^{(\mu)}_j \; \big[\Phi^{m-1}\big]_{j,k} \; W^{(\nu)}_k$. Arranging the sequence functions $Y_{\mu,\nu}(m)$, obtained from all preparation and measurement settings, into a matrix $Y(m)$ allows the problem to be expressed in the form 
\begin{equation}
    Y(m) = M \; \Phi^{m-1} \; W\; ,
\end{equation}
where $M$, $\Phi$ and $W$ are the $4\times4$ matrices
\begin{equation}
M=
    \begin{pmatrix}
        \big(\vec{M}^{(1)} \big)^{\intercal} \\
        \big(\vec{M}^{(2)} \big)^{\intercal} \\
        \big(\vec{M}^{(3)} \big)^{\intercal} \\
        \big(\vec{M}^{(4)} \big)^{\intercal}
    \end{pmatrix}
    \;,
\end{equation}
\begin{equation}
    \Phi^{m-1} = 
    \begin{pmatrix}
       \big(\Phi^{m-1} \big)_{1,1}  &  \big(\Phi^{m-1} \big)_{1,2}  & \cdots & \big(\Phi^{m-1} \big)_{1,4} \\
       \big( \Phi^{m-1} \big)_{2,1} &  \big( \Phi^{m-1} \big)_{2,2} & \cdots &  \big( \Phi^{m-1} \big)_{2,4} \\
        \vdots           &    \vdots        & \ddots &    \vdots        \\
        \big(\Phi^{m-1} \big)_{4,1} & \big(\Phi^{m-1} \big)_{4,2}   &  \cdots & \big(\Phi^{m-1} \big)_{4,4} 
    \end{pmatrix}
\end{equation}
\;,
\begin{equation}
W=
    \begin{pmatrix}
       \vec{W}^{(1)} &   \vec{W}^{(2)} &  \vec{W}^{(3)} &  \vec{W}^{(4)}
    \end{pmatrix}
    \;,
\end{equation}
with each $\big(\vec{M}^{(\mu)} \big)^{\intercal}$ a row vector with components given by $M^{(\mu)}_j$ and $\vec{W}^{(\nu)}$ a column vector with elements $W^{(\nu)}_k$. This matrix formulation allows the effective dynamics to be reconstructed using standard system identification techniques. In this work, we employ the Eigensystem Realization Algorithm (ERA)~\cite{GueSinMaj2021,JuaPap1985}. Its implementation can be summarized as follows: 
\begin{enumerate}
    \item Estimate all $k^{(\mu,\nu)}(m)$ average sequence functions and assemble the $4\times 4$ matrices $Y(m)$.
    \item Define the Hankel matrices $H_0 \in \mathbb{R}^{4L \times 4L}$ and $H_1 \in \mathbb{R}^{4L \times 4L}$ as
    \begin{equation}
        H_0 =
        \begin{pmatrix}
            Y(1)   & Y(2) & \cdots & Y(L) \\
            Y(2)   & Y(3) & \cdots  & Y(L+1) \\
            \vdots & \vdots & \ddots & \vdots \\
            Y(L)   &  Y(L+1) & \cdots & Y(2L-1)
        \end{pmatrix} 
        \; , \;  H_1= 
        \begin{pmatrix}
            Y(2)   & Y(3) & \cdots & Y(L+1) \\
            Y(3)   & Y(4) & \cdots  & Y(L+2) \\
            \vdots & \vdots & \ddots & \vdots \\
            Y(L+1)   &  Y(L+2) & \cdots & Y(2L)
        \end{pmatrix}
        \;.
    \end{equation}
The parameter $L$ controls the size of the Hankel matrices and therefore the number of sequence lengths included in the reconstruction. The effective dynamics considered in this work admits a minimal realization of dimension $n=4$, corresponding to the multiplicity of the non-trivial irrep. The preparation states and measurement operators are chosen such that the full multiplicity space is accessible, allowing the $4 \times 4$ matrix $\Phi$ to be reconstructed (up to similarity transformations). In this work, we chose $L=5$ and used a total of $2L=10$ equally spaced sequence lengths, namely $m\in\{1,2,...,10\}$, for the ERA reconstruction. Note that here the Hankel matrices are constructed from consecutive sequence lengths, $m_{i+1}=m_i+1$, following the standard ERA formulation.
\item Apply the Singular Value Decomposition (SVD) to $H_0$
    \begin{equation}
        H_0=U \Sigma V^{\intercal} = \big( U \Sigma^{1/2} \big) \big(\Sigma^{1/2} V^{\intercal} \big) \approx \big( U_n \Sigma^{1/2}_n \big) \big(\Sigma^{1/2}_n V^{\intercal}_n \big) = O_n C_n \;,
    \end{equation}
In theory, the rank of the matrix $H_0$ equals the minimal realization dimension (here $n=4$). However, finite-sampling can introduce additional small singular values that are not representative of the underlying physical dynamics. For this reason, ERA performs a truncated SVD decomposition, retaining only the $n$ largest singular values. This then allows us to construct the truncated matrices $O_n:=U_n \Sigma^{1/2}_n$ and $C_n:=\Sigma^{1/2}_n V^{\intercal}_n$.
   \item The shifted Hankel matrix $H_1$ satisfies the relation:
   \begin{equation}
       H_1= O\; \Phi \; C \approx O_n \; \Phi\; C_n \; .
   \end{equation}
To account for the most general case in which $C_n$ and $O_n$ are rectangular matrices, a common practice is to use the Moore-Penrose pseudoinverse to obtain the least-squares estimate
\begin{equation}
    \Phi \approx O^{+}_nH_1C^{+}_n
\end{equation}
This step yields an estimate of a realization of $\Phi$. For a perfect reconstruction, the ERA realization coincides with $\Phi$, up to a possible similarity transformation.
\end{enumerate}

If the underlying noise channel is uncorrelated, then $\mathcal{R}(\Lambda)=\mathcal{R}(\Lambda_1) \otimes \mathcal{R}(\Lambda_2)$, with $\mathcal{R}(\Lambda_1)$ and $\mathcal{R}(\Lambda_2)$ acting independently on the operator spaces of the first and second qubit, respectively. In this limit, $\Phi$ acquires a simplified form, namely $\Phi=\lambda_1 \; \mathcal{R}(\Lambda_2)$. This follows automatically from Eq.(\ref{Phi_C1xI_matrixele_app}): substituting the tensor-product form of the noise channel causes the trace in Eq.(\ref{Phi_C1xI_matrixele_app}) to factorize into two independent contributions. The contribution related to the noise channel acting on the operator space associated to the first qubit reduces to the scalar $\lambda_1$, while the remaining matrix elements are precisely the components of $\mathcal{R}(\Lambda_2)$. This simplification suggests a direct route for estimating $\lambda_1$. Since a trace preserving channel has at least one eigenvalue equal to one, $\lambda_1$ is necessarily one of the eigenvalues of $\Phi=\lambda_1 \; \mathcal{R}(\Lambda_2)$. Given that the eigenvalues remain invariant under similarity transformations, $\lambda_1$ remains an eigenvalue for any ERA realization of $\Phi$, in the uncorrelated regime. Additionally, because the eigenvalues of a CPTP channel are restricted to lie in the unit disk~\cite{Wolf2012}, $|\lambda_1|$ can be estimated from the ERA reconstruction as the maximum absolute value of its eigenvalues. When operations on the first qubit have sufficiently high fidelity such that $\lambda_1>0$, this approach can serve as a direct estimator of $\lambda_1$. Indeed, this observation has motivated our choice of spectral estimator for $\lambda_1$: within a prescribed numerical tolerance, we identify it as the largest real eigenvalue of the ERA realization of $\Phi$, and otherwise use the spectral radius when no such real eigenvalue can be selected.

In the uncorrelated limit, the factorized form of $\Phi$ is preserved under gauge transformations. However, the matrix $R=G^{-1} \mathcal{R}(\Lambda_2) G$ is not guaranteed to represent a CPTP channel. This motivates focusing on what information can be extracted solely from the eigenspectrum of $\Phi$. In particular, we would like to assess whether the spectrum allows us to infer the presence of correlated noise. In the uncorrelated limit, the spectrum of $\Phi/\lambda_1$ must be compatible with that of single-qubit CPTP channel. Spectral incompatibility can therefore indicate the presence of correlated noise, whereas compatibility alone cannot establish its absence. The latter is illustrated by the choice of correlated noise in the main text, where the entangling gate $R_{XX}(\theta_{\text{corr}})$ gives rise to:
\begin{equation}
\Phi = \text{diag} \bigg(\frac{1+2\cos(\theta_{\text{corr}})}{3}, \frac{1+2\cos(\theta_{\text{corr}})}{3} , \frac{2+\cos(\theta_{\text{corr}})}{3} , \frac{2+\cos(\theta_{\text{corr}})}{3} \bigg) \; .
\end{equation}
Identifying $\lambda_1$ as the spectral radius, it follows that $\text{Spec}\Big( \frac{\Phi}{\lambda_1} \Big) = \Big\{1,1, \frac{1+2\cos(\theta_{\text{corr}})}{2+\cos(\theta_{\text{corr}})}, \frac{1+2\cos(\theta_{\text{corr}})}{2+\cos(\theta_{\text{corr}})} \Big\}$. This, nevertheless, can also correspond to the eigenspectrum of a single-qubit bit-flip channel, with error probability $p_{\text{bit-flip}}= \frac{1-\cos(\theta_{\text{corr}})}{2(2+\cos(\theta_{\text{corr}}))}$. Thus, even though  $R_{XX}(\theta_{\text{corr}})$ induces correlations, the resulting re-scaled spectrum of its $\Phi$ matrix can be made compatible with a valid single-qubit CPTP channel.

However, this does not need to be the case for all possible noise models. To see this, consider the following two-qubit noise channels (expressed in the Kraus representation):
\begin{equation}
\begin{split}
   & \Lambda^{(1)}_{\text{corr}} (\rho) = (1-q_1) \big( \mathbbm{1} \otimes \mathbbm{1} \big) \rho \big( \mathbbm{1} \otimes \mathbbm{1} \big) + \frac{q_1}{3} \sum_{P \in \{X,Y,Z\}} \big( X \otimes P \big) \rho \big( X \otimes P \big)  \; ,  \\
   & \Lambda^{(2)}_{\text{corr}} (\rho) = (1-q_2) \big( \mathbbm{1} \otimes R_z(\theta) \big) \rho \big( \mathbbm{1} \otimes R^{\dagger}_z(\theta) \big) + q_2 \; \big( X \otimes R_z(-\theta) \big) \rho \big( X \otimes R^{\dagger}_z(-\theta) \big) \;.
\end{split}    
\end{equation}
Neither channel can be expressed as a tensor product of single-qubit channels, and hence both describe correlated noise. Imposing the same two-qubit average fidelity, $\bar{F}$, for both channels, we can re-express the parameters $q_1$ and $\theta$ as 
\begin{equation}
 q_1 = \frac{5}{4} \Big(1-\bar{F} \Big) \; \text{and} \; \theta = 2 \arccos \bigg( \sqrt{\frac{5\bar{F}-1}{4(1-q_2)}} \bigg) \;.
\end{equation}
The corresponding simultaneous RB correlation coefficients are then $\delta \alpha^{(1)} = \frac{16}{9}q_1(1-q_1)$ and $\delta \alpha^{(2)} = 0$, respectively. Hence, while the correlations induced by $\Lambda^{(1)}_{\text{corr}}$ are detected by simultaneous RB~\cite{GamCorMer2013}, those induced by $\Lambda^{(2)}_{\text{corr}}$ are not.

We can also compute the resulting effective dynamical matrix $\Phi$ for both noise channels:
\begin{equation}
\begin{split}
    &\Phi^{(1)} = \text{diag} \bigg( 1 - \frac{4 q_1}{3},1 - \frac{8 q_1}{9}, 1 - \frac{8 q_1}{9}, 1 - \frac{8 q_1}{9}  \bigg) \; , \\ 
    &\Phi^{(2)} = 
    \begin{pmatrix}
    \frac{1}{3} (3-4q_2) & 0 & 0 & 0 \\
    0 & \frac{1}{3} (3-4q_2) \cos(\theta) & -\frac{1}{3} (3-2q_2) \sin(\theta) & 0 \\
    0 &\frac{1}{3} (3-2q_2) \sin(\theta)  & \frac{1}{3} (3-4q_2) \cos(\theta)  & 0 \\
    0 & 0 & 0 & \frac{1}{3} (3-4q_2) 
    \end{pmatrix}
    \; .
\end{split}    
\end{equation}
Fixing $q_2=q_1/2$ and choosing $\bar{F}\approx0.974339$, if we again attempt to factorize each $\Phi$ as $\Phi = \lambda_1 R$, with $\lambda_1$ the largest real eigenvalue, the resulting $R$ matrices have the eigenspectrum:  
\begin{equation}
    \text{Spec} \big(R^{(1)} \big) = \Big\{1,1,1, \frac{9-12q_1}{9-8q_1} \Big\} \; , \; \text{Spec} \big(R^{(2)} \big) = \Big\{1,1, \cos(\theta) + i \; \frac{3-2q_2}{3-4q_2} \sin(\theta), \cos(\theta) - i \; \frac{3-2q_2}{3-4q_2} \sin(\theta) \Big\} \; .
\end{equation}
Denoting the eigenvalues of $R^{(i)}$ by $\{\phi^{(i)}_0, \phi^{(i)}_1, \phi^{(i)}_2, \phi^{(i)}_3\}$, with $\phi^{(i)}_0=1$, their compatibility with the spectrum of a single-qubit CPTP channel is determined by the simultaneous fulfillment of the following conditions~\cite{WolGar2010}:
\begin{equation}
\begin{split}
   & 1+ s_1 + s_2 + s_3 \geq 0 \\
   & 1+ s_1 - s_2 - s_3 \geq 0  \\
   & 1- s_1 + s_2 - s_3 \geq 0  \\
   & 1- s_1 - s_2 + s_3 \geq 0  \; ,
\end{split}    
\end{equation}
with $s_k = \phi^{(i)}_k$ if $\phi^{(i)}_k$ is real, and $s_k = |\phi^{(i)}_k|$ if $\phi^{(i)}_k$ is complex. Since the three non-trivial eigenvalues should correspond to the eigenvalues of a real $3\times 3$ matrix, they must either all be real or consist of one real eigenvalue and a complex-conjugate pair. In the latter case, the eigenvalue compatibility conditions listed above simplify to: $1+\phi \pm 2\rho \geq 0$ and $1-\phi \geq 0$, with $\phi$ denoting the real non-trivial eigenvalue and $\rho$ the common modulus of the complex conjugate pair. Applying these conditions to the eigenspectrum of $R^{(1)}$ and $R^{(2)}$, and denoting by $D$ the left-hand side of the corresponding violated condition, we get $D = \phi_1 -1 = (9-12q_1)/(9-8q_1)-1 \approx -0.0146743$ for $R^{(1)}$ and $D = 1+\phi-2\rho = 2(1-\rho) \approx -0.0014086$ for $R^{(2)}$. Thus, both noise models violate the eigenspectrum compatibility conditions.

To test how these predictions survive finite sampling, we simulated the gate-set shadow protocol employing the gate-set $\mathcal{C}_1\times \mathcal{C}_1$ to obtain an estimate for $\delta \alpha$ (along with its 95 \% bootstrap confidence interval) for both noise channels, and separately employed the gate set $\mathcal{C}_1\times \mathcal{I}$ to reconstruct $\Phi$. The results are listed in Table~\ref{tab_eig_incomp}.
\begin{table}[htbp]
\centering
\begin{tabular}{ccc}
\hline
\hline
 Noise Model & $\delta \hat{\alpha}$ (95 \% \; C.I.) & $\hat{D} \pm \sigma_D$  \\
\hline
$\Lambda^{(1)}_{\text{corr}}$ & 0.054649 \; $\big([0.05232, 0.05691]\big)$  &  -0.012969 $\pm$ 0.004376 \\
$\Lambda^{(2)}_{\text{corr}}$ & 0.000378 \; $\big([-0.00198, 0.00263]\big)$ &  -0.000728 $\pm$ 0.004515 \\
\hline
\end{tabular}
\caption{\label{tab_eig_incomp} Every simulation employed $K=15000$ and no SPAM noise. The 95 \% confidence interval for the estimate of $\delta \alpha$ ($\delta \hat{\alpha}$) was evaluated via the non-parametric bootstrap percentile method~\cite{Eftib1993}. The estimates for the incompatible eigenspectrum condition, $\hat{D}$, were obtained by repeating the reconstruction of $\Phi$ over five independent realizations of the protocol. The reported values denote the mean over these five repetitions, while the uncertainties reflect the resulting standard deviation ($\sigma_D$). The results are consistent with the theoretical expectations: for $\Lambda^{(1)}_{\text{corr}}$, $\delta \hat{\alpha}$ is positive, thus excluding an uncorrelated channel, while $\hat{D}$ is consistent with the violation of the eigenspectrum conditions; for $\Lambda^{(2)}_{\text{corr}}$, the uncertainty over $\delta \hat{\alpha}$ does not exclude the possibility of an uncorrelated channel (as the 95 \% C.I. still includes zero), while the average value of $\hat{D}$ is compatible with a violation of the eigenspectrum conditions, yet its uncertainty prevent us from confidently establishing the presence of correlations.}
\end{table}
While these results are consistent with our theoretical analysis, they also highlight the limitations imposed by finite sampling. In particular, although the eigenspectrum of $R^{(2)}$ is theoretically incompatible with that of a single-qubit CPTP channel, the corresponding violation is too small to be reliably resolved with the sampling parameters considered here.

We conclude this subsection by discussing the normalization constants appearing in Eq.(\ref{kmunu_c1xI_2}) and Eq.(\ref{Phi_C1xI_4}). These constants are fixed by requiring that the corresponding average sequence functions reproduce their ideal noiseless value. In this limit, a sequence of length $m=1$ should satisfy $k(1)=1$. Examining Eq.(\ref{kmunu_c1xI_2}) in the noiseless limit and letting $m=1$, it follows that
\begin{equation}
\begin{split}
    k^{(\mu,\nu)}(1) &= \frac{c^{(0)}_{f_{\mathbbm{1}};\mu,\nu}}{\sqrt{2}|\mathbbm{P}_1|} u_{\mu,\nu} \sum_{j,k} \; \delta_{j,k} \;\langle \langle \tau^{(k)}_3| \rho^{(\nu)} \rangle \rangle \langle \langle E^{(\mu)}_0-E^{(\mu)}_1| \tau^{(j)}_3 \rangle\rangle \\ 
    &= \frac{c^{(0)}_{f_{\mathbbm{1}};\mu,\nu}}{|\mathbbm{P}_1|} u_{\mu,\nu} \; \sum_k u_{k,\nu} \; \delta_{k,\mu} \\
    &= \frac{c^{(0)}_{f_{\mathbbm{1}};\mu,\nu}}{|\mathbbm{P}_1|} u^2_{\mu,\nu} \implies c^{(0)}_{f_{\mathbbm{1}};\mu,\nu} = \frac{|\mathbbm{P}_1|}{u^2_{\mu,\nu}} \; .
\end{split}    
\end{equation}
Using a similar argument, it is possible to show that $c^{(0)}_{f_{\mathbbm{P}_1}}=2|\mathbbm{P}_1|$.

\subsection{Leakage RB}\label{leakageRB_app_sec}
The aim of leakage RB is two-fold: to characterize the leakage and seepage rates and to estimate the average gate fidelity restricted to the computational subspace. In this work, we analyze the feasibility of both goals.

The leakage and seepage rates are defined as~\cite{ClaRieWan2021}:
\begin{equation}
    \begin{split}
       & L  = \frac{1}{d_1} \langle \langle \mathbbm{1}_2 | \mathcal{R}(\Lambda) | \mathbbm{1}_1 \rangle \rangle \;,  \\
       & S =\frac{1}{d_2} \langle \langle \mathbbm{1}_1 | \mathcal{R}(\Lambda) | \mathbbm{1}_2 \rangle \rangle \; ,
    \end{split}
    \label{LS_def_app}
\end{equation}
 where $d_1$ and $d_2$ denote the dimensions of the computational and leakage subspace, and $| \mathbbm{1}_1 \rangle \rangle, | \mathbbm{1}_2 \rangle \rangle$ are the vectorized identity operators with support restricted to the computational and leakage subspaces, respectively. Recall that the full Hilbert space decomposes as $\mathcal{H}=\mathcal{H}_C \oplus \mathcal{H}_L$, where $\mathcal{H}_C$ is the computational subspace used to encode the qubit and $\mathcal{H}_L$ is the leakage subspace. For the particular qubit encoding used as an example in the main text~\cite{ClaRieWan2021}, $\mathcal{H}_C$ is spanned by the total-spin states $|0,0 \rangle$ and $|1, 0 \rangle$ of a two-spin system, while $\mathcal{H}_L$ is spanned by the remaining two basis states. Hence, both subspaces are two-dimensional and therefore $d_1=d_2=2$.

In this case, every operator $B$ on the Hilbert space admits the block decomposition:
\begin{equation}
    B =  \begin{pmatrix}
        B_{CC} & B_{CL} \\
        B_{LC} & B_{LL}
    \end{pmatrix}
    \;,
\label{bloch_decomp_leakage}    
\end{equation}
where $B_{CC}$ acts only within the computational subspace, $B_{LL}$ acts only within the leakage subspace, and 
$B_{CL}$ and $B_{LC}$ map between the two subspaces. At the same time, the gate set introduced in the main text forms a group, whose generators are given in Eq.(\ref{leakage_group})~\cite{ClaRieWan2021}. The group contains 16 distinct elements. A possible choice of representatives for the group elements is given by
\begin{equation}
\begin{split}
    G = \{ & \mathbbm{1} \oplus \mathbbm{1}, \mathbbm{1} \oplus (-\mathbbm{1}), X \oplus Z, X \oplus (-Z), Z \oplus H, Z \oplus (-H), \\
    & X \oplus X,  X \oplus (-X), \mathbbm{1} \oplus iY, \mathbbm{1} \oplus (-iY), iY \oplus HZ, \\
    & iY \oplus (-HZ), iY \oplus HX, iY \oplus (-HX), Z \oplus iHY , Z \oplus (-iHY)\} \; ,
\end{split}
\label{leakage_group_elements}
\end{equation} 
where elements differing only by a global phase are identified as the same group element. It then follows that the action of any group element $g \in G$ on any operator $B$ preserves the block decomposition in Eq.(\ref{bloch_decomp_leakage}). Thus, the block decomposition defines four invariant operator subspaces that cannot mix under the group action. As will soon become clear, the representation of $G$ on the vectorized operator space is not multiplicity free. To see this, let us introduce the operators:
\begin{equation}
\begin{split}
    &\mathbbm{P}_0 = \begin{pmatrix}
        \mathbbm{1}_{2 \times 2} & 0_{2 \times 2} \\
        0_{2 \times 2} & 0_{2 \times 2} 
    \end{pmatrix} = \frac{1}{2} \Big(\mathbbm{1} \otimes \mathbbm{1} + Z \otimes \mathbbm{1}  \Big) \;\; , \; \; \mathbbm{P}_1 = \begin{pmatrix}
        0_{2 \times 2} & 0_{2 \times 2} \\
        0_{2 \times 2} & \mathbbm{1}_{2 \times 2}  
    \end{pmatrix}
    = \frac{1}{2} \Big(\mathbbm{1} \otimes \mathbbm{1} - Z \otimes \mathbbm{1} \Big) \; , \\
& S_{+} = \begin{pmatrix}
        0_{2 \times 2} & \mathbbm{1}_{2 \times 2} \\
        0_{2 \times 2} & 0_{2 \times 2}  
    \end{pmatrix}
    = \frac{1}{2} \Big( X \otimes \mathbbm{1} + i Y \otimes \mathbbm{1}  \Big) \;\;, \;\; S_{-} = \begin{pmatrix}
        0_{2 \times 2} & 0_{2 \times 2} \\
        \mathbbm{1}_{2 \times 2} & 0_{2 \times 2}  
    \end{pmatrix}
=\frac{1}{2} \Big( X \otimes \mathbbm{1} - i Y \otimes \mathbbm{1}  \Big) \; .
\end{split}
\label{operator_subspaces_leakage_encoding}
\end{equation}
Note that $\mathbbm{P}_0$ and $\mathbbm{P}_1$ are the projectors onto the computational and leakage subspace, respectively, while $S_{+}$ maps states from the leakage subspace to the computational subspace, and $S_{-}$ performs the reverse. All these operators can be expressed in the normalized tensor product Pauli basis $\{\sigma_i \otimes \sigma_j\}$. The group action over $\{\sigma_i \otimes \sigma_j\}$ allows to identify seven inequivalent irreducible representations, with the corresponding invariant subspaces listed in Tab.~\ref{irreps_of_leakage_group_table}. 
\begin{table}[htbp]
\centering
\begin{tabular}{ccc}
\hline
\hline
 Irreducible Representation & Invariant Subspace (copy 1) & Invariant Subspace (copy 2)   \\
\hline
Trivial irrep & $\text{Span} \big\{ |\sigma_0 \otimes \sigma_0  \rangle \rangle \big \}$ & $\text{Span} \big\{ |\sigma_3 \otimes \sigma_0  \rangle \rangle \big \}$ \\
1st non-trivial irrep & $\text{Span} \big\{ |\sigma_0 \otimes \sigma_2  \rangle \rangle \big \}$ & $\text{Span} \big\{ |\sigma_3 \otimes \sigma_2  \rangle \rangle \big \}$\\
2nd non-trivial irrep & $\text{Span} \big\{ |\sigma_1 \otimes \sigma_0  \rangle \rangle, |\sigma_2 \otimes \sigma_2  \rangle \rangle \big \}$ & $\text{Span} \big\{ |\sigma_2 \otimes \sigma_0  \rangle \rangle , |\sigma_1 \otimes \sigma_2  \rangle \rangle \big \}$ \\
3rd non-trivial irrep & $\text{Span} \big\{ \frac{1}{\sqrt{2}} \big( |\sigma_0 \otimes \sigma_1  \rangle \rangle -|\sigma_3 \otimes \sigma_1  \rangle \rangle \big), \frac{1}{\sqrt{2}} \big( |\sigma_0 \otimes \sigma_3  \rangle \rangle - |\sigma_3 \otimes \sigma_3  \rangle \rangle \big) \big \}$ & -  \\
4th non-trivial irrep & $\text{Span} \big\{ \frac{1}{\sqrt{2}} \big(| \sigma_0 \otimes \sigma_1 \rangle \rangle + | \sigma_3 \otimes \sigma_1 \rangle \rangle \big) \big\} $  & - \\
5th non-trivial irrep & $\text{Span} \big\{ \frac{1}{\sqrt{2}} \big(| \sigma_0 \otimes \sigma_3 \rangle \rangle + | \sigma_3 \otimes \sigma_3 \rangle \rangle \big) \big \} $  & - \\
6th non-trivial irrep & $\text{Span} \big\{ | \sigma_1 \otimes \sigma_1 \rangle \rangle , | \sigma_1 \otimes \sigma_3 \rangle \rangle \big\} $  & $\text{Span} \big\{ | \sigma_2 \otimes \sigma_3 \rangle \rangle , | \sigma_2 \otimes \sigma_1 \rangle \rangle \big\} $ \\
\hline
\end{tabular}
\caption{ Irreducible representations (irreps) of the PTM representation of the gate set group $G$ defined in Eq.(\ref{leakage_group_elements}) and their corresponding invariant vector subspaces. The trivial irrep and the non-trivial irreps 1, 2 and 6 appear with multiplicity two, while the remaining irreps occur only once. The total number of inequivalent irreps can be verified independently by computing the conjugacy classes of the group elements listed in Eq.(\ref{leakage_group_elements}), since the number of conjugacy classes is equal to the number of inequivalent irreps~\cite{Tinkham1964}.}
\label{irreps_of_leakage_group_table}
\end{table}
Table~\ref{irreps_of_leakage_group_table} clarifies that, when expressed in the vectorized operator space, the operators introduced in Eq.(\ref{operator_subspaces_leakage_encoding}) contain components belonging to different copies of equivalent irreps. In particular, $\mathbbm{P}_0$ and $\mathbbm{P}_1$ have components in the two copies of the trivial irrep, while $S_{+}$ and $S_{-}$ have components in the two copies of the second non-trivial irrep. As a result, even when the average sequence functions are restricted to a particular subspace, they generally receive contributions from multiple copies of the same irrep. The corresponding average sequence functions are therefore expected, in general, to be described by multi-exponential decay models. In the following subsections, we examine the consequences of multiplicity for the two objectives of leakage RB: estimating the leakage and seepage rates and estimating the computational fidelity.

\subsubsection{Leakage and seepage rates} \label{LS_app_deriv}

The quantities required to estimate leakage ($L$) and seepage ($S$) rates are entirely contained within the invariant subspaces associated with the trivial irrep. To make this explicit, it is convenient to express the vectorized identity operators $|\mathbbm{1}_1\rangle\rangle, |\mathbbm{1}_2\rangle\rangle$ and the resulting leakage ($L$) and seepage ($S$) rates in the PTM basis:
\begin{gather}
 |\mathbbm{1}_1\rangle \rangle = |\sigma_0 \otimes \sigma_0 \rangle \rangle + |\sigma_3 \otimes \sigma_0 \rangle \rangle = |t_1 \rangle \rangle + |t_2 \rangle \rangle \; , \; |\mathbbm{1}_2\rangle \rangle = |\sigma_0 \otimes \sigma_0 \rangle \rangle - |\sigma_3 \otimes \sigma_0 \rangle \rangle = |t_1 \rangle \rangle - |t_2 \rangle \rangle  \label{leakage_mixed_states_app} \\
   L = \frac{1}{2} \Big( 1 - \langle \langle t_2 | \mathcal{R}\big(\Lambda \big) | t_1 \rangle \rangle - \langle\langle t_2| \mathcal{R}\big(\Lambda \big) | t_2 \rangle \rangle \Big)  \label{leakage_PTM_app} \\ 
 S= \frac{1}{2} \Big( 1 + \langle \langle t_2 | \mathcal{R}\big(\Lambda \big) | t_1 \rangle \rangle - \langle\langle t_2 | \mathcal{R}\big(\Lambda \big)  | t_2 \rangle \rangle \Big) \; , \label{seepage_PTM_app} 
\end{gather}
where $\mathcal{R}\big(\Lambda \big)$ denotes the PTM representation of the CPTP noise channel $\Lambda$, and the trace preserving property was used to simplify the expressions for $L$ and $S$. Note that we have defined $|t_1 \rangle\rangle:=| \sigma_0 \otimes \sigma_0 \rangle \rangle$ and $|t_2 \rangle \rangle:= | \sigma_3 \otimes \sigma_0 \rangle \rangle$. From Eqs.(\ref{leakage_PTM_app}) and (\ref{seepage_PTM_app}), it follows that estimating $L$ and $S$ requires determining the matrix elements $\langle \langle t_2 | \mathcal{R}\big(\Lambda \big)| t_1 \rangle \rangle$ and $\langle \langle t_2 | \mathcal{R}\big(\Lambda \big)| t_2 \rangle\rangle$. These matrix elements require only the restriction of $\mathcal{R}\big(\Lambda \big)$ to the two-dimensional subspace spanned by $|t_1 \rangle \rangle$ and $|t_2 \rangle \rangle$, with each vector spanning one of the two copies of the trivial irrep (see Tab.~\ref{irreps_of_leakage_group_table}).

Specializing Eq.(\ref{P_triv_general_app}) to the trivial irrep yields
\begin{equation}
  \mathlarger{\mathbb{P}}_{\text{triv}} = \sum^2_{i,j=1} |t_i \otimes t_j \rangle\rangle \langle \langle t_i \otimes t_j| = \sum^2_{i,j=1} | \mathbbm{T}^{(j,i)} \rangle \rangle \langle \langle\mathbbm{T}^{(j,i)}| \; ,
\end{equation}
where we defined $\mathbbm{T}^{(j,i)}:= |t_j \rangle \rangle \langle \langle t_i|$. The decaying modes of the average sequence function are determined by the eigenvalues of the matrix $\Phi$, which depends on both the noise and the choice of probe operator, through $\Phi = \mathlarger{\mathbb{P}}_{\text{triv}} (A \otimes \mathcal{R} (\Lambda)) \mathlarger{\mathbb{P}}_{\text{triv}}$. In this case, the matrix elements of $\Phi$ are given by
\begin{equation}
    \Phi_{(j,i),(k,l)} = \langle\langle t_j| \mathcal{R}\big(\Lambda \big) | t_k \rangle \rangle \langle \langle t_l | A^{\intercal}| t_i \rangle\rangle \; , 
\label{phi_leakageRB_gen_app}    
\end{equation}
The average sequence function then reads
\begin{equation}
    k_A(m) = c^{(0)}_{f_A} \;  \sum_x \sum^2_{i,j,k,l=1} \big[\Phi^{m-1} \big]_{(j,i),(k,l)} \;  \langle \langle t_l| \rho \rangle\rangle \langle\langle E_x| t_i \rangle \rangle \langle \langle t_k| \tilde{\rho} \rangle \rangle \langle\langle \tilde{E}_x| \mathcal{R}\big(\Lambda \big) | t_j \rangle \rangle \; .
\label{ave_seq_leakageRB_gen_app}    
\end{equation}
In standard leakage RB, the extraction of leakage and seepage rates assumes that SPAM errors are sufficiently small such that the fitting model remains informative about the underlying leakage dynamics~\cite{WooGam2018,ClaRieWan2021}. We apply the same assumption here and proceed to identify the choice of input state, POVM elements, and probe operator $A$ for which the gate set shadow average sequence function directly encodes the information required to estimate both $L$ and $S$ in the low SPAM regime.

Suppose we choose
\begin{equation}
    \begin{split}
        & |\rho \rangle \rangle = \frac{1}{2} | t_1 \rangle \rangle \; , \;
        | E_0 \rangle \rangle = | t_1 \rangle \rangle + | t_2 \rangle \rangle = | \mathbbm{1}_1 \rangle \rangle \; , \;
        | E_1 \rangle \rangle = | t_1 \rangle \rangle - | t_2 \rangle \rangle = | \mathbbm{1}_2 \rangle \rangle \; ,
    \end{split}
    \label{LS_input_POVM_app}
\end{equation}
and take as probe operator the $16 \times 16$ matrix in the PTM representation (denoted as $A_X$ in the main text)
\begin{equation}
A=
\begin{pmatrix}
  \begin{matrix}
  0 & 1 \\
  1 & 0
  \end{matrix}
  & \rvline & \bigzero_{2 \times 14} \\
\hline
\rule{0pt}{1.2em}
  \bigzero_{14 \times 2} & \rvline &
  \bigzero_{14 \times 14}
\end{pmatrix}
\implies A^n = \begin{pmatrix}
  \begin{matrix}
  1 & 0 \\
  0 & 1
  \end{matrix}
  & \rvline & \bigzero_{2 \times 14} \\
\hline
\rule{0pt}{1.2em}
  \bigzero_{14 \times 2} & \rvline &
  \bigzero_{14 \times 14}
\end{pmatrix} \; \text{(for $n$ even and $n \geq2$)} \;, \;
A^n = A \; \text{(for $n$ odd)} \; ,
\label{def_A_LRB_app}    
\end{equation}
then in the absence of SPAM noise, the average sequence function $k_A(m)$ simplifies to
\begin{equation}
    k_A(m) = \begin{cases}
       1  \;, \; m \; \text{odd} \\
       C\; \frac{1-\lambda^m}{1-\lambda} \;, \; m \; \text{even} \; ,
    \end{cases}
    \label{new_leakageRB_formula_app}
\end{equation}
with $\lambda=\langle \langle t_2| \mathcal{R}\big(\Lambda \big) | t_2 \rangle \rangle=1-S-L$, $C=S-L$ and where we used the fact that in this case $c^{(0)}_{f_A}=2$. Note that the only non-zero block in Eq.(\ref{def_A_LRB_app}) acts on the subspace spanned by the basis vectors $\{|t_1 \rangle \rangle, |t_2 \rangle \rangle\}$. If this $2 \times 2$ block is instead expressed in the basis $\{|\mathbbm{1}_1 \rangle \rangle, |\mathbbm{1}_2 \rangle \rangle\}$ (through Eq.(\ref{leakage_mixed_states_app})), the non-zero block defining the probe $A$ becomes diagonal with eigenvalues $+1$ and $-1$, respectively, and therefore assigns a different sign to the computational and leakage subspace projectors.

The parameters $C$ and $\lambda$ directly determine the leakage and seepage rates
\begin{equation}
  L=\frac{1-\lambda-C}{2} \; , \;   S=\frac{1-\lambda+C}{2} \;.
  \label{LS_formulas_app}
\end{equation}
For the choice of input state in Eq.(\ref{LS_input_POVM_app}), they can be estimated by applying the gate-set shadow protocol using only even sequence lengths and fitting the corresponding average sequence function to extract $C$ and $\lambda$. However, this procedure relies on the ability to prepare the fully mixed state in Eq.(\ref{LS_input_POVM_app}), which may be experimentally challenging.

Recall that the standard leakage RB fitting model is given by~\cite{ClaRieWan2021}
\begin{equation}
    k_A(m)= \frac{S}{S+L} + \frac{L}{S+L} \; \lambda^m \equiv S_0(m) \; .
    \label{standard_leakageRB_formula_app}
\end{equation}
This fitting model can be recovered using the gate-set shadow protocol by choosing $A$ as in Eq.(\ref{def_A_LRB_app}), $\rho=\frac{1}{2}\mathbbm{1}_1$, and by constructing the average sequence function ($f_A(x,\mathbf{g})$) using the same POVM elements as in Eq.(\ref{LS_input_POVM_app}), with $c^{(0)}_{f_A}=1$. Note that with these settings, $S_0(m)$ is recovered, independently of the parity of $m$. Additionally, Eq.(\ref{k_A_Z_comp}) in the main text is retrieved if we instead choose the probe operator to be
\begin{equation}
A=
\begin{pmatrix}
  \begin{matrix}
  1 & 0 \\
  0 & -1
  \end{matrix}
  & \rvline & \bigzero_{2 \times 14} \\
\hline
\rule{0pt}{1.2em}
  \bigzero_{14 \times 2} & \rvline &
  \bigzero_{14 \times 14}
\end{pmatrix}
\;.
\end{equation}

Let us now consider how Eq.(\ref{ave_seq_leakageRB_gen_app}) simplifies when using the computational state as input, together with the measurement settings in Eq.(\ref{LS_input_POVM_app}), under negligible SPAM noise, but allowing for a generic probe operator $A$ instead. Letting $\mathcal{B}:=A^{m-1}$, $\mathcal{S}_{m-2}=\sum^{m-2}_{n=0} \lambda^n$, Eq.(\ref{ave_seq_leakageRB_gen_app}) reads:
\begin{equation}
    \begin{split}
      \big(c^{(0)}_{f_A} \Big)^{-1}  k_A(m) &= \frac{1}{2} \sum_k \Big( \Phi^{m-1}_{(1,1),(k,1)} + \Phi^{m-1}_{(1,1),(k,2)} \Big) + \frac{1}{2} \sum_{j,k} \Big(\Phi^{m-1}_{(j,2),(k,1)} + \Phi^{m-1}_{(j,2),(k,2)} \Big) \langle \langle t_2 | \mathcal{R}(\Lambda) | t_j \rangle \rangle \\
        &= \frac{1}{2} \Big(\mathcal{B}_{1,1} + \mathcal{B}_{1,2} \Big) + \frac{\mathcal{R}(\Lambda)_{2,1}}{2} \Big(\mathcal{B}_{2,1} + \mathcal{B}_{2,2}  \Big) + \frac{\mathcal{R}(\Lambda)_{2,2}}{2} \mathcal{B}_{2,1} \Big( C\mathcal{S}_{m-2} + \big[\mathcal{R}(\Lambda)^{m-1} \big]_{2,2} \Big) \\
        &+\frac{\mathcal{R}(\Lambda)_{2,2}}{2} \mathcal{B}_{2,2} \Big( C \mathcal{S}_{m-2} + \big[\mathcal{R}(\Lambda)^{m-1} \big]_{2,2} \Big) \\
        &=\big(\mathcal{B}_{2,1} +\mathcal{B}_{2,2} \big) S_0(m) + \frac{1}{2} \Big(\mathcal{B}_{1,1} + \mathcal{B}_{1,2} - \mathcal{B}_{2,1} -\mathcal{B}_{2,2} \Big) \;,
    \end{split}
\end{equation}
where we have used the identity $\frac{\mathcal{R}(\Lambda)_{2,2}}{2} \Big( C\mathcal{S}_{m-2} + \big[\mathcal{R}(\Lambda)^{m-1} \big]_{2,2} \Big)=S_0(m)-\frac{1}{2}(1+\mathcal{R}(\Lambda)_{2,1})$. Since $\mathcal{B}=A^{m-1}$, its matrix elements are generally dependent on the sequence length $m$, but not on the leakage and seepage parameters. Thus, the general form of $k_A(m)$ clarifies how different choices of the probe operator can produce different $m$ dependent functions of the survival probability $S_0(m)$. Nevertheless, with the input state fixed to the computational subspace, no choice of $A$ that is independent of the unknown noise parameters can reproduce Eq.(\ref{new_leakageRB_formula_app}). If we instead fix the input state to the fully mixed state and proceed similarly, we get $k_A(m) = \frac{c^{(0)}_{f_A}}{2} \Big( \mathcal{B}_{1,1}+\mathcal{B}_{2,1}\; C\Big(\frac{1-\lambda^m}{1-\lambda}\Big) \Big)$. These observations highlight the importance of the choice of input state and POVMs in the gate-set shadow protocol, as well as the limitations of what can be achieved by only varying the probe operator when multiplicity is present. Although the probe provides some freedom to build different average sequence functions within the target subspace, this freedom is only partial when the accessible input states and measurements are restricted.

In the main text, we compared the performance of two fitting models (Eq.(\ref{k_A_Z_comp}) and Eq.(\ref{new_leakageRB_formula})). The analysis was performed using the following rotated amplitude damping noise model:
\begin{equation}
    \mathcal{R}\big(\Lambda \big) = \Big( \mathcal{R}^{\intercal}\big( e^{-i\theta Y/2} \big) \mathcal{R}\big(\Lambda_{\gamma_1} \big) \mathcal{R}\big( e^{-i\theta Y/2} \big)  \Big) \otimes \mathcal{R}\big(\Lambda_{\gamma_2} \big) \; ,
\label{noise_model_leakage_app}
\end{equation}
where $\mathcal{R}\big( e^{-i\theta Y/2} \big)$ is the PTM representation of a unitary $Y-$rotation channel and $\mathcal{R}\big(\Lambda_{\gamma_i} \big)$ denotes the PTM representation of a single qubit amplitude damping channel with damping strength $\gamma_i$. Unless stated otherwise, all numerical results concerning the application of the gate-set shadow protocol to leakage RB, including those presented in Figs.~\ref{Comparison_Fits_LS1} and~\ref{Comparison_Fits_LS2}, were obtained using this noise model. Throughout the simulations, we fixed $\gamma_2=0.035$. The sequence lengths employed in the main text were $m\in \{2, 4, 6, 8, 10, 12, 16, 20, 24, 30, 40, 50, 60, 80, 100, 140, 200\}$.

\subsubsection{Computational fidelity} \label{comp_fid_est_app}
The computational fidelity requires accessing errors only within the computational subspace. This requires us to estimate the trace $\text{Tr} \big(\Pi_{\text{comp}} \; \Lambda \big)$, where $\Pi_{\text{comp}}$ denotes the projector, in the PTM representation, onto the subspace spanned by the vectorized operators supported entirely within the computational subspace~\cite{ClaRieWan2021}: 
\begin{equation}
    F_{\text{comp}} = \frac{\text{Tr} \big(\Pi_{\text{comp}} \; \Lambda \big) + d_1(1-L)}{d^2_1+d_1} \; .
\end{equation}
Any arbitrary operator $B$ supported only within the computational subspace can be expressed in the full Hilbert space as $|0 \rangle \langle0| \otimes B$. Hence, an orthonormal basis for these operators can be obtained by choosing the basis $\{E_j:=|0 \rangle \langle0| \otimes\sigma_j\}^3_{j=0}$. In the PTM representation, we may then express $\Pi_{\text{comp}}$ as follows 
\begin{equation}
\begin{split}
    \Pi_{\text{comp}} &= \sum^3_{j=0} |E_j \rangle\rangle \langle\langle E_j| 
    = \frac{1}{2} \sum^3_{j=0}  \Big( |\sigma_0 \otimes \sigma_j \rangle \rangle \langle \langle \sigma_0 \otimes \sigma_j| + |\sigma_3 \otimes \sigma_j \rangle \rangle \langle \langle \sigma_3 \otimes \sigma_j|\\ &+|\sigma_0 \otimes \sigma_j \rangle \rangle \langle \langle \sigma_3 \otimes \sigma_j| + |\sigma_3 \otimes \sigma_j \rangle \rangle \langle \langle \sigma_0 \otimes \sigma_j| \Big) \; .
\end{split} 
\label{comp_sub_proj_app}
\end{equation}
Comparing the components appearing in Eq.(\ref{comp_sub_proj_app}) with table~\ref{irreps_of_leakage_group_table}, we can conclude that there are four inequivalent irreducible representations contributing to the trace $\text{Tr} \big(\Pi_{\text{comp}} \; \Lambda \big)$: the trivial irrep (for $j=0$), the fourth non-trivial irrep (for $j=1$), the first non-trivial irrep (for $j=2$) and the fifth non-trivial irrep (for $j=3$). The fourth and fifth non-trivial irreps have no multiplicity, but the same is not true for the trivial irrep and the first non-trivial irrep. Defining $\mathbbm{P}_4$ and $\mathbbm{P}_5$ as the projectors to the fourth and fith irrep subspaces, respectively, we can show using table ~\ref{irreps_of_leakage_group_table} that
\begin{equation}
    \begin{split}
       & |v^{(4)}_x \rangle \rangle := \frac{1}{\sqrt{2}} \big(| \sigma_0 \otimes \sigma_1 \rangle \rangle + | \sigma_3 \otimes \sigma_1 \rangle \rangle \big) \; ,\; \mathbbm{P}_4 = |v^{(4)}_x \rangle \rangle \langle \langle v^{(4)}_x | \; , \\  
       & \text{Tr}\big( \mathcal{R}(\Lambda) \mathbbm{P}_4 \big)= \frac{1}{2} \Big( \langle \langle \sigma_0 \otimes \sigma_1 | \mathcal{R}(\Lambda) | \sigma_0 \otimes \sigma_1 \rangle \rangle + \langle \langle \sigma_0 \otimes \sigma_1 | \mathcal{R}(\Lambda) | \sigma_3 \otimes \sigma_1 \rangle \rangle \\ &+ \langle \langle \sigma_3 \otimes \sigma_1 | \mathcal{R}(\Lambda) | \sigma_0 \otimes \sigma_1 \rangle \rangle + \langle \langle \sigma_3 \otimes \sigma_1 | \mathcal{R}(\Lambda) | \sigma_3 \otimes \sigma_1 \rangle \rangle \Big) \; ,
        \end{split}
\end{equation}
and
\begin{equation}
    \begin{split}
       & |v^{(5)}_z \rangle \rangle := \frac{1}{\sqrt{2}} \big(| \sigma_0 \otimes \sigma_3 \rangle \rangle + | \sigma_3 \otimes \sigma_3 \rangle \rangle \big) \; ,\; \mathbbm{P}_5 = |v^{(5)}_z \rangle \rangle \langle \langle v^{(5)}_z | \; , \\
       & \text{Tr}\big( \mathcal{R}(\Lambda) \mathbbm{P}_5 \big)= \frac{1}{2} \Big( \langle \langle \sigma_0 \otimes \sigma_3 | \mathcal{R}(\Lambda) | \sigma_0 \otimes \sigma_3 \rangle \rangle + \langle \langle \sigma_0 \otimes \sigma_3 | \mathcal{R}(\Lambda) | \sigma_3 \otimes \sigma_3 \rangle \rangle \\ & + \langle \langle \sigma_3 \otimes \sigma_3 | \mathcal{R}(\Lambda) | \sigma_0 \otimes \sigma_3 \rangle \rangle + \langle \langle \sigma_3 \otimes \sigma_3 | \mathcal{R}(\Lambda) | \sigma_3 \otimes \sigma_3 \rangle \rangle \Big) \; .
    \end{split}
\end{equation}
Thus, the contributions of the fourth and fifth irreps to $\text{Tr} \big(\Pi_{\text{comp}} \; \Lambda \big)$ are directly determined by the decay parameters $\lambda_4=\text{Tr}\big( \mathcal{R}(\Lambda) \mathbbm{P}_4 \big)$ and $\lambda_5=\text{Tr}\big( \mathcal{R}(\Lambda) \mathbbm{P}_5 \big)$. Since both irreps are multiplicity free, their associated average sequence functions contain a single decay mode and there is no ambiguity in estimating these quantities.

For the trivial irrep contribution, evaluating $\text{Tr} \big(\Pi_{\text{comp}} \; \Lambda \big)$ requires knowledge of $\mathcal{R}(\Lambda)$ restricted to the subspace spanned by $|t_1 \rangle \rangle$ and $|t_2 \rangle \rangle$. More precisely, this contribution gives rise to the term $\frac{1}{2}\big( 1+\lambda +C\big)=1-L$. Consequently, estimating this contribution inherits the SPAM sensitivity associated with estimating the leakage parameter $L$.

The first non-trivial irrep also appears with multiplicity two. Unlike the trivial irrep, however, there is no trace preserving constraint that reduces the decay model to a single decay mode. As a result, estimating the corresponding contribution to $\text{Tr} \big(\Pi_{\text{comp}} \; \Lambda \big)$ requires estimating the matrix elements of $\Phi$ restricted to two copies of the first non-trivial irrep. To this end, we now construct the average sequence function associated with this irrep.

We start by re-labeling the vector states spanning the first non-trivial irrep as $|b_1 \rangle \rangle := |\sigma_0 \otimes \sigma_2 \rangle \rangle$ and $|b_2 \rangle \rangle := |\sigma_3 \otimes \sigma_2 \rangle \rangle$. Specializing Eq.(\ref{P_triv_general_app}) to this irrep yields 
\begin{equation}
  \mathlarger{\mathbb{P}}_{\text{triv}} = \sum^2_{i,j=1} |b_i \otimes b_j \rangle\rangle \langle \langle b_i \otimes b_j| = \sum^2_{i,j=1} | \mathbbm{P}^{(j,i)}_1 \rangle \rangle \langle \langle\mathbbm{P}^{(j,i)}_1| \; ,
\end{equation}
where we defined $\mathbbm{P}^{(j,i)}_1:= |b_j \rangle \rangle \langle \langle b_i|$. The resulting average sequence function reads
\begin{equation}
    k_A(m) = c^{(0)}_{f_A} \; \sum_x \sum^2_{i,j,k,l=1} \big[ \Phi^{m-1}\big]_{(j,i),(k,l)} \; \langle \langle b_l | \rho \rangle \rangle \langle \langle E_x| b_i \rangle \rangle \langle \langle b_k| \tilde{\rho} \rangle \rangle \langle \langle \tilde{E}_x| \mathcal{R}(\Lambda) | b_j \rangle \rangle \; ,
    \label{seq_fun_irrep1_LRB_app}
\end{equation}
with the matrix elements $\Phi_{(j,i),(k,l)}$ given by
\begin{equation}
    \Phi_{(j,i),(k,l)}= \langle \langle b_j| \mathcal{R}(\Lambda)|b_k \rangle \rangle \langle \langle b_l| A^{\intercal}|b_i \rangle \rangle \; .
\end{equation}
Recalling the definition of $\Phi$ provided in Eq.(\ref{mapping_to_state_space_model_app1}), namely $\Phi=\mathlarger{\mathbb{P}}_{\text{triv}} \big(A \otimes \mathcal{R}(\Lambda) \big)\mathlarger{\mathbb{P}}_{\text{triv}}$, only the restrictions of $A$ and $\mathcal{R}(\Lambda)$ to the subspace $\text{Span}\big\{|b_1 \rangle \rangle , |b_2 \rangle \rangle \big\}$ contribute to $\Phi$. We denote by $\mathcal{R}_{eff}(\Lambda)$ the $2\times2$ matrix with elements given by $\big( \mathcal{R}_{eff}(\Lambda) \big)_{i,j}:= \langle \langle b_i|\mathcal{R}(\Lambda)|b_j \rangle \rangle$. Likewise, we will refer to the restriction of $A$ to the subspace $\text{Span}\big\{|b_1 \rangle \rangle , |b_2 \rangle \rangle \big\}$ by $A_{eff}$.

It is convenient to rewrite the average sequence function in Eq.(\ref{seq_fun_irrep1_LRB_app}) as the bilinear form
\begin{equation}
\begin{split}
   & k_A(m) = c^{(0)}_{f_A} \; \vec{v}^{\intercal} \; \Phi^{m-1} \; \vec{u} \; , \\
   & \vec{v} = \begin{pmatrix}
                 \sum_x  \langle \langle \tilde{E}_x| \mathcal{R}(\Lambda) | b_1 \rangle \rangle \langle \langle E_x| b_1 \rangle \rangle \\
                 \sum_x  \langle \langle \tilde{E}_x| \mathcal{R}(\Lambda) | b_1 \rangle \rangle \langle \langle E_x| b_2 \rangle \rangle \\
                 \sum_x  \langle \langle \tilde{E}_x| \mathcal{R}(\Lambda) | b_2 \rangle \rangle \langle \langle E_x| b_1 \rangle \rangle \\
                 \sum_x  \langle \langle \tilde{E}_x| \mathcal{R}(\Lambda) | b_2 \rangle \rangle \langle \langle E_x| b_2 \rangle \rangle
               \end{pmatrix}
\;, \;              
\vec{u} =
\begin{pmatrix}
  \langle \langle b_1| \tilde{\rho} \rangle \rangle   \langle \langle b_1 | \rho \rangle \rangle \\
   \langle \langle b_1| \tilde{\rho} \rangle \rangle   \langle \langle b_2 | \rho \rangle \rangle \\
   \langle \langle b_2| \tilde{\rho} \rangle \rangle   \langle \langle b_1 | \rho \rangle \rangle \\
   \langle \langle b_2| \tilde{\rho} \rangle \rangle   \langle \langle b_2 | \rho \rangle \rangle \\
\end{pmatrix}
\end{split}
\; .
\end{equation}
Choosing then as ideal input state and POVM elements the set
\begin{equation}
    |\rho \rangle \rangle =  \frac{1}{2} \Big( |t_1 \rangle \rangle + |t_2 \rangle \rangle + |b_1 \rangle \rangle + |b_2 \rangle \rangle \Big) \; ,\; | E_0 \rangle \rangle = |t_1 \rangle \rangle + \frac{1}{2} \Big( |b_1 \rangle \rangle + |b_2 \rangle \rangle \Big) \; , \; | E_1 \rangle \rangle = |t_1 \rangle \rangle - \frac{1}{2} \Big(  |b_1 \rangle \rangle + |b_2 \rangle \rangle \Big) \; ,
    \label{inPOVMS_irrep1_LRB_app}
\end{equation}
and considering the limit of negligible SPAM noise, the vectors $\vec{u}$ and $\vec{v}$ simplify to
\begin{equation}
    \begin{split}
        \vec{u}  = \vec{u}_0 \otimes  \vec{u}_0 \;, \; 
        \vec{u}_0= \frac{1}{2}\begin{pmatrix} 
        1 \\
        1
        \end{pmatrix}
        \;, \;
        \vec{v} = \big(2\mathcal{R}^{\intercal}_{eff}(\Lambda)\vec{u}_0 \big) \otimes \vec{u}_0 \;.
    \end{split}
\end{equation}

Let us now choose the probe operator $A$ as
\begin{equation}
A=
\begin{pmatrix}
\bigzero_{2\times 2} & \bigzero_{2\times 2} & \bigzero_{2\times 12} \\
\rule{0pt}{1.5em}
\bigzero_{2\times 2} & \frac{1}{2} \begin{pmatrix}
    1 & 1\\
    1 & 1
\end{pmatrix} & \bigzero_{2\times 12} \\
\rule{0pt}{1.2em}
\bigzero_{12\times 2} & \bigzero_{12\times 2} & \bigzero_{12\times 12}
\end{pmatrix}
\;,
\end{equation}
where the only non-zero block is precisely the restriction of $A$ to the invariant subspace spanned by $\{|b_1 \rangle \rangle\ , |b_2 \rangle \rangle\}$, which we denote by
\begin{equation}
    A_{eff} : = \frac{1}{2} \begin{pmatrix}
    1 & 1\\
    1 & 1
\end{pmatrix}
\;.
\end{equation}
Since $A^2_{eff}=A_{eff}$, it follows that $A^n_{eff}=A_{eff}$ for all $n \geq 1$.

To then obtain a more compact expression for the average sequence function $k_A(m)$, we make use of the Cayley–Hamilton theorem, which states that any $n\times n$ matrix satisfies its own characteristic polynomial equation. More precisely, given a $n\times n$ matrix $M$ with characteristic polynomial $p_M(\lambda)=\text{det}(x \mathbbm{1}-M)=x^n + a_{n-1} x^{n-1}+\cdots+a_1 x+a_0$, the following matrix polynomial expression, $p_M(M)$, evaluates to the zero matrix: 
\begin{equation}
    p_M(M)  := M^n + a_{n-1} M^{n-1} + \cdots + a_1 M + a_0 \mathbbm{1}_n = \mathbf{0}_{n \times n} \;. 
\end{equation}
Since $\mathcal{R}_{eff}(\Lambda)$ is a $2 \times 2$ matrix, this theorem implies that all powers $\mathcal{R}^n_{eff}(\Lambda)$ can be expressed as a linear combination of $\mathbbm{1}_{2\times 2}$ and $\mathcal{R}_{eff}(\Lambda)$, with appropriate coefficients. These coefficients depend on the eigenvalues of $\mathcal{R}_{eff}(\Lambda)$. Hence, we can distinguish three cases for $k_A(m)$: 
\begin{enumerate}
    \item $\mathcal{R}_{eff}(\Lambda)$ is diagonalizable with two distinct eigenvalues, $\lambda_1$ and $\lambda_2$, which yields:
    \begin{gather}
        k_A(m) = c_1 \; \lambda^{m-1}_1 +  c_2 \;\lambda^{m-1}_2 \; , \\ c_1 = \frac{\lambda_1}{(\lambda_1-\lambda_2)} \Big( \frac{\Gamma}{2} - \lambda_2 \Big) \; \label{model1_app_LRB} , \; c_2 = \frac{\lambda_2}{(\lambda_1-\lambda_2)} \Big(  \lambda_1 -\frac{\Gamma}{2} \Big) \; , \\ \Gamma:= \big( \mathcal{R}_{eff}(\Lambda) \big)_{1,1} + \big( \mathcal{R}_{eff}(\Lambda) \big)_{1,2} + \big( \mathcal{R}_{eff}(\Lambda) \big)_{2,1} + \big( \mathcal{R}_{eff}(\Lambda) \big)_{2,2} \; .
    \end{gather}
    \item $\mathcal{R}_{eff}(\Lambda)$ is proportional to the identity, i.e. $\lambda_1=\lambda_2=\lambda$, in which case
    \begin{equation}
        k_A(m) = \lambda^{m} \; , \; \Gamma=2 \lambda \; .
        \label{model2_app_LRB}
    \end{equation}
    \item $\mathcal{R}_{eff}(\Lambda)$ is non-diagonalizable, resulting in
    \begin{equation}
         k_A(m) =  \lambda^{m-1} \Big( m \frac{\Gamma}{2} + (1-m) \lambda \Big) \; .
         \label{model3_app_LRB}
    \end{equation}
\end{enumerate}

\begin{figure}[h]
\centering
\includegraphics[width=0.8\linewidth]{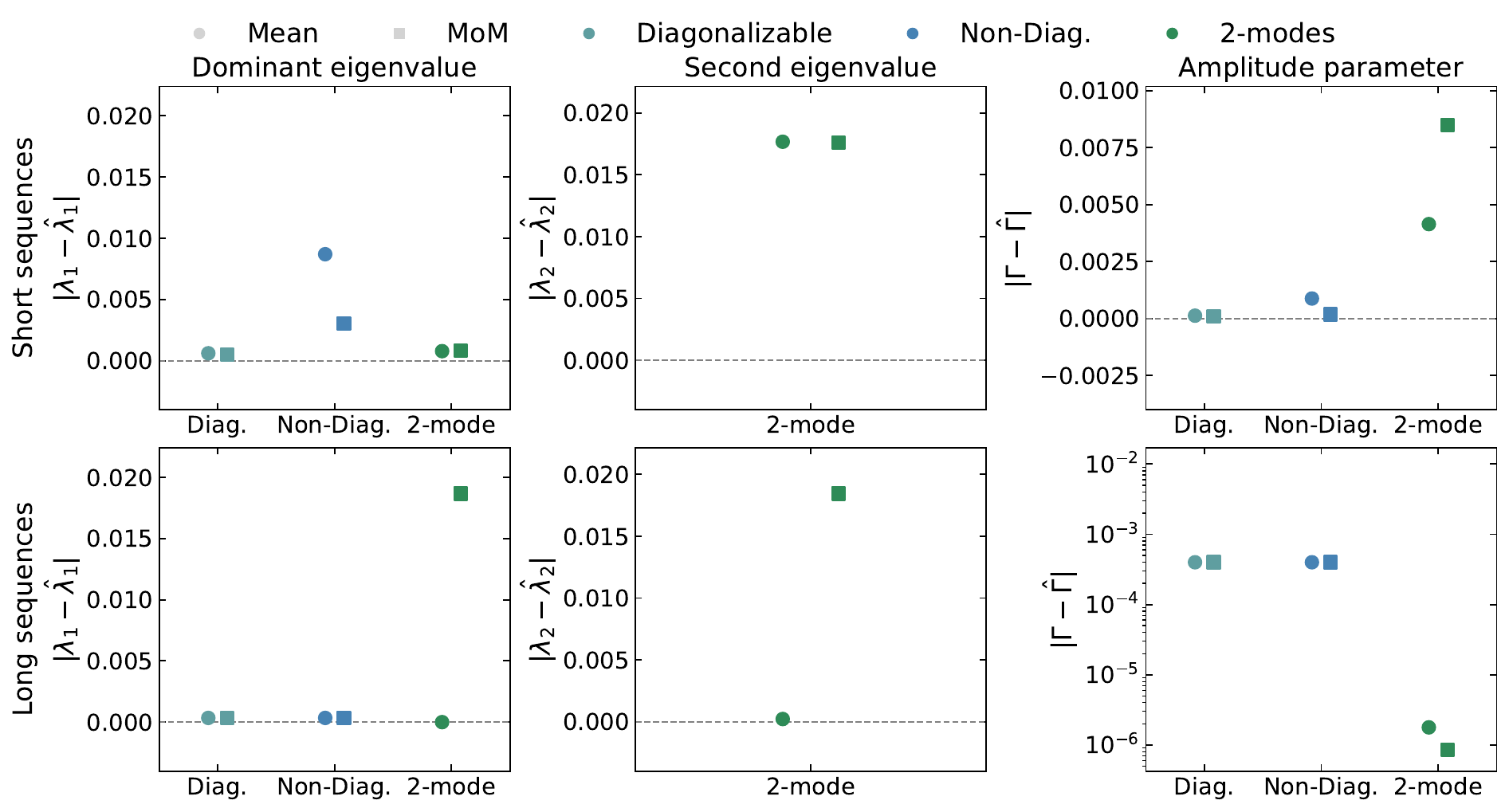}
\caption{Absolute error in the estimated eigenvalues and the parameter $\Gamma$ obtained using the three candidate decay models: the diagonalizable single-decay model (Diagonalizable), the non-diagonalizable single-decay model (Non-Diag.) and the two-decay model (2-modes). The upper row corresponds to the errors evaluated from the average sequence function constructed in the presence of stochastic measurement outcomes and using shorter sequence lengths, $m \in \{1,...,21\}.$ The lower row shows the errors obtained from the average sequence function that uses the exact Born probabilities and is evaluated by extending the probed sequence lengths to larger values, $m \in \{1,...,99\}$. For each model, we show the estimates obtained using both the mean and median-of-means (MoM).}
\label{short_vs_long_residuals_app_LRB}
\end{figure}

To estimate the computational fidelity, we need to be able to estimate the trace $\text{Tr} \big(\Pi_{\text{comp}} \; \Lambda \big)$. The contribution of the first non-trivial irrep to this trace is given by  
\begin{equation}
\text{Tr} \big(\Pi_{\text{comp}} \; \Lambda \big) \big|_{\mathrm{irrep}_1}= \frac{1}{2} \Big( \big( \mathcal{R}_{eff}(\Lambda) \big)_{1,1} + \big( \mathcal{R}_{eff}(\Lambda) \big)_{1,2} + \big( \mathcal{R}_{eff}(\Lambda) \big)_{2,1} + \big( \mathcal{R}_{eff}(\Lambda) \big)_{2,2} \Big) = \frac{\Gamma}{2} \; .  
\end{equation}
Thus, estimating the computational fidelity requires estimating the parameter $\Gamma$. Yet, as seen in Eqs.(\ref{model1_app_LRB})-(\ref{model3_app_LRB}), the relation between $\Gamma$ and the model parameters depends on the eigenvalue structure of $\mathcal{R}_{eff}(\Lambda)$, making its estimation model dependent. For instance, when $\Gamma$ is inferred from a two-decay model, its estimation depends on the amplitudes multiplying the decaying modes, namely $\Gamma=2(c_1+c_2)$, and is therefore expected to inherit their SPAM sensitivity, whereas in the diagonalizable degenerate eigenvalue case, $\Gamma$ can be directly inferred from the decay parameter and is hence immune to SPAM noise. Additionally, even in the absence of SPAM, accurately estimating $\Gamma$ depends on the information retained by the measured sequence data. As exemplified by Fig.~\ref{short_vs_long_residuals_app_LRB}, finite sampling and a limited range of sequence lengths may compromise our ability to clearly distinguish between the candidate decay models.

In Fig.~\ref{short_vs_long_residuals_app_LRB}, we illustrate the combined effect of finite sampling and limited range of sequence lengths for distinguishing between the candidate decay models. We employ the noise model introduced in Eq.(\ref{noise_model_leakage_app}), with $\gamma_1=0.025$ and $\gamma_2=0.035$. For this choice of parameters, $\mathcal{R}_{eff}(\Lambda)$ has two distinct eigenvalues, $\lambda_1 \approx 0.9823$ and $\lambda_2 \approx 0.9639$, so the ground-truth model corresponds to the two-decay model. However, for this particular noise model, $c_1>>c_2$, making the second mode weakly identifiable in the presence of finite sampling. Consequently, the single-decay models can provide similarly good descriptions of the average sequence data, despite corresponding to different algebraic and geometric multiplicity of the eigenvalues of $\mathcal{R}_{eff}(\Lambda)$. Although these models fit the average sequence function comparably well, they generally produce different estimates for $\lambda_1$ and the parameter $\Gamma$. For example, in the upper row of Fig.~\ref{short_vs_long_residuals_app_LRB}, the single decay models yield the smallest errors in the estimate of $\Gamma$, yet the non-diagonalizable model provides a less accurate estimate for $\lambda_1$. Within the sampling conditions considered here ($K=1000$ and stochastic measurement outcomes), extending the range of sequence lengths alone does not resolve this ambiguity. In contrast, replacing the stochastic measurement outcomes by the corresponding exact Born probabilities and simultaneously extending the sequence length range, shows that the tow-decay model provides substantially more accurate estimates of both decay modes and of $\Gamma$ (see Fig.~\ref{short_vs_long_residuals_app_LRB}, lower row). These observations suggest a pragmatic strategy in situations where the underlying decay model cannot be reliably identified from the estimated average sequence function. Whenever the goodness of fit does not single out a candidate model, and no additional information about the underlying noise process is available, a practical approach can be to fit all three models and use the spread in the inferred values of $\Gamma$ as an empirical diagnostic of model uncertainty.

The numerical example in this section illustrates that, depending on the underlying noise model and the information content of the measured sequence data (as determined, for example, by finite sampling and the range of sequence lengths probed), the appropriate decay model may not be uniquely identifiable. This ambiguity in model identification may in turn lead to different estimates of the target quantities, even when the competing models provide similarly good fits to the observed average sequence function. It therefore adds to the challenge of estimating the computational fidelity.

We emphasize that the decomposition of the trace $\text{Tr} \big(\Pi_{\text{comp}} \; \Lambda \big)$ into different irrep contributions, and consequently its expression in terms of decay parameters, is dependent on the choice of gate set. Different gate sets will in general induce different irrep decompositions, with different multiplicities. The gate-set shadow formalism is nevertheless general whenever the benchmarking gate set forms a group. Thus, the same procedure can be applied to other gate sets: one decomposes the projected computational trace into components belonging to different inequivalent irreps, identifies the average sequence functions associated with those components, and determines the corresponding expression for the computational fidelity. Following this procedure will also make explicit which components are identifiable from the resulting decay signals and which may remain non-identifiable.

\section{Numerical simulation details}~\label{app_num_det}
In this work, all numerical simulations pertaining the experimental stage of the gate-set shadow protocol were obtained using the PennyLane Python package~\cite{pennylane2018}. The code required to reproduce all numerical examples can be found in~\cite{SilGrep2025_git}, while the numerical data can be found in~\cite{SilGrep_zenodo}.

\subsection{Gate error model for Section~\ref{inter_G4_main}}~\label{inter_G4_main_app}
\begin{figure}[h]
\centering
\includegraphics[width=0.8\linewidth]{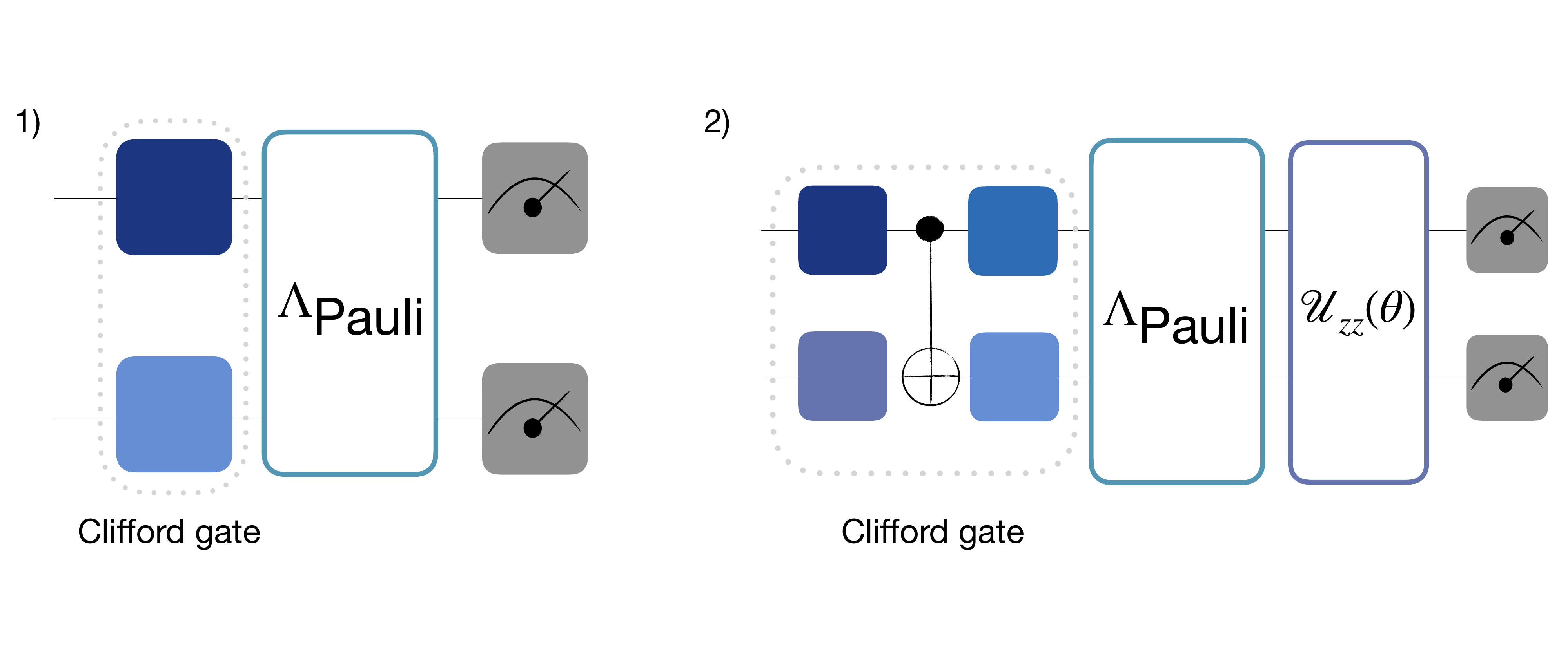}
\caption{Pictorial circuit-based representation of the gate errors used for simulating the results in Section~\ref{inter_G4_main}. The measurement symbols are shown only to indicate the final measurement performed at the end of each gate sequence.} \label{circuit_sim_interleaved}
\end{figure}

To obtain the numerical results in Fig.~\ref{fig_RB_cuves_interleaved}, we modeled gate errors as follows:
\begin{enumerate}
    \item Gate operations obtained from only applying single-qubit gates to the two-qubit system are perturbed by a Pauli noise channel~\cite{HarYuFla2021} (see Fig.~\ref{circuit_sim_interleaved}, panel 1):
    \begin{equation}
        \Lambda_{\text{Pauli}}(\rho) = \sum^{15}_{j=0} p_j  \;P_j \rho P_j \; , \text{with} \; P_j \in\{ \mathbbm{1}, X, Y, Z\}^{\otimes 2} \;.
    \end{equation}
    The set of coefficients $p_j$ form a probability distribution over the Pauli operators. 
    \item Gate operations that are produced by a composition of single-qubit gates with a single CNOT gate are perturbed by a composition of unitary errors, modeled by a coherent $zz-$rotation error (see Fig.~\ref{circuit_sim_interleaved}, panel 2), with Pauli noise:
    \begin{equation}
      \Lambda_1=\mathcal{U}_{zz}(\theta) \circ \Lambda_{\text{Pauli}} \; , \; \text{with} \;\mathcal{U}_{zz}(\theta)(\rho)=R_{zz}(\theta) \rho R^{\dagger}_{zz}(\theta) \;,  \; R_{zz}(\theta) = 
      \begin{pmatrix}
          e^{-i\theta/2} & 0 & 0 & 0 \\
          0 & e^{i\theta/2} & 0 & 0  \\
          0 & 0 & e^{i\theta/2} & 0  \\
          0 & 0 &      0        & e^{-i\theta/2}
      \end{pmatrix}
      \; .
    \end{equation}
    \item Gates that require two CNOT gates to implement are perturbed by applying the previous error channel twice and, likewise, gates that require three CNOT gates are perturbed by three consecutive applications of $ \Lambda_1$.
\end{enumerate}
We note that the $G(4)$ CNOT-dihedral group allows all of its elements to be built from gates requiring at most three CNOT gates~\cite{GarCros2020}. For the two-qubit Clifford group, we apply the same error model described above, following the class decomposition of the group discussed in the supplementary material of Ref.~\cite{CorGamChow2013}.

\section{Unital marginals reconstruction}\label{unital_rec_app}
Unital marginal channel reconstruction relies on the decomposition of the channel as a linear combination of elements of the local Clifford group, according to the formula~\cite{HelIoaKit2023}:
\begin{equation}
\begin{split}
    \mathcal{R}(\Lambda_i) =& \frac{|\mathbb{P}_i|^2}{|\mathbb{C}_i|} \sum_{C \in \mathbb{C}_i} \lambda(C^{\dagger},\Lambda) \; \mathbb{P}_i \mathcal{R}(C)  \mathbb{P}_i \; , \\ & \;\text{with }\;
\lambda(C^{\dagger},\Lambda) = \frac{\text{Tr} \big(\mathbb{P}_i \mathcal{R}(C^{\dagger}) \mathbb{P}_i \mathcal{R}(\Lambda) \big)}{|\mathbb{P}_i|} \;.
\end{split}
\label{marginals_recformula}
\end{equation}
To ensure clarity of the presentation, let us expand on the adopted notation in Eq.(\ref{marginals_recformula}). The element $C$ denotes a Clifford gate, while $\mathcal{R}(C)$ is its representative in the PTM formalism. Likewise, $\mathcal{R}(\Lambda)$ and $\mathcal{R}(\Lambda_i)$ are the PTM representations of the noise channel and its unital marginal channel, respectively. We denote the decay rate obtained with the probe $A=\mathcal{R}(C)$ by $\lambda(C^{\dagger},\Lambda)$, since $A^{\intercal}=\mathcal{R}(C^{\dagger})$ in the normalized Pauli basis. The symbol $\mathbb{C}_i$ denotes the set of local Clifford gates used in the reconstruction of the channel restricted to the $ith$ invariant subspace. For a two-qubit system, these sets are~\cite{HelIoaKit2023}
\begin{equation}
    \mathbb{C}_1 =\{C_j \otimes \mathbbm{1}  \}_j \; , \mathbb{C}_2 =\{\mathbbm{1}  \otimes C_j \}_j \; , \; \mathbb{C}_3 =\{C_j \otimes C_k \}_{(j,k)} \; .
\end{equation}
The sets can be identified from the conjugation action of the local Clifford gates on the local Pauli basis of each subspace defined in Tab.~\ref{tab_c1xc1_proj}\;. For example, on the subspace spanned by $\{ |\sigma_i \otimes \sigma_0 \rangle\rangle\}$, the action of a local Clifford gate $C_{a}\otimes C_b$ depends only on $C_a$, since $C_b$ acts trivially on $\sigma_0$. Therefore, all distinct operations on this subspace can be represented by the gates of the form $C_a \otimes \mathbbm{1}$. Analogously, all distinct operations on the subspace spanned by $\{ |\sigma_0 \otimes \sigma_i \rangle\rangle\}$ can be represented by gates of the form $\mathbbm{1}  \otimes C_j$. On the subspace spanned by $\{ |\sigma_i \otimes \sigma_j \rangle\rangle\}$, both Clifford gates generally act non-trivially, so that $\mathbb{C}_3$ is the local Clifford group itself.  We note that, in Eq.(\ref{marginals_recformula}), $\mathcal{R}(\Lambda_i)$ denotes the restriction of $\mathbb{P}_i \mathcal{R}(\Lambda)\mathbb{P}_i$ to the $ith$ invariant subspace. Consequently, $\mathcal{R}(\Lambda_1)$ and $\mathcal{R}(\Lambda_2)$ are $3\times 3$ matrices with support only on $W_1$ and $W_2$, respectively, while $\mathcal{R}(\Lambda_3)$ is a $9\times 9$ matrix with support only on $W_3$ (see Tab.~\ref{tab_c1xc1_proj}). The full block-diagonal unital marginal channel, $\mathcal{R}(\Lambda_F)$, can then be constructed as
\begin{equation}
    \mathcal{R}(\Lambda_F) = 1 \oplus \mathcal{R}(\Lambda_1) \oplus \mathcal{R}(\Lambda_2) \oplus \mathcal{R}(\Lambda_3) \; ,
\label{Lambda_full_app}    
\end{equation}

In Eq.(\ref{marginals_recformula}), $|\mathbb{C}_i|$ denotes the total number of elements in the set $\mathbb{C}_i$. The decomposition in Eq.(\ref{marginals_recformula}) is particularly convenient, because the expansion coefficients $\lambda(C^{\dagger},\Lambda)$ are precisely the figures of merit estimated in the gate-set shadow protocol. Indeed, each $\lambda(C^{\dagger},\Lambda)$ can be estimated by taking the average of the sequence correlation function $f_A$, with $A=\mathcal{R}(C)$, and fitting the resulting decay. We note that, in Eq.(\ref{unital_marginal_c1xc1_sec}) of the main text, we use $\lambda_{i,C}$ to denote the estimate of $\lambda(C^{\dagger},\Lambda)$.

\begin{figure}[t]
\centering
\includegraphics[width=0.8\linewidth]{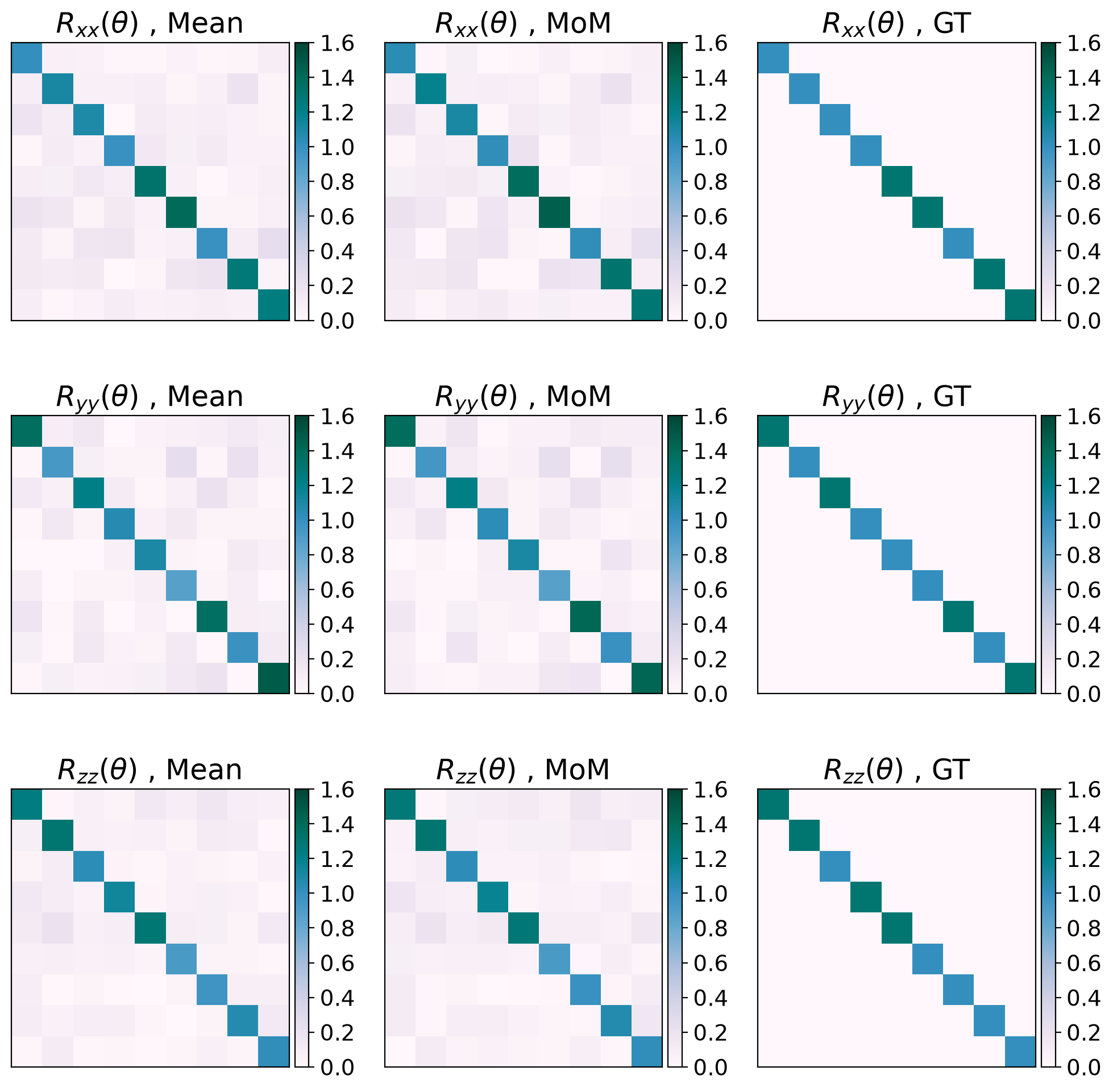}
\caption{$\delta^{(2)}_{CT}$ for coherent two-qubit gate errors, given by: $R_{xx}(\theta)$ (top panel), $R_{yy}(\theta)$ (middle panel) and $R_{zz}(\theta)$ (bottom panel). For all three cases, the angle $\theta$ was fixed at $\theta=0.5$. In each row, from left to right: the first two figures are obtained by simulating the gate-set shadow protocol, while the third figure corresponds to the actual value of $\delta^{(2)}_{CT}$, derived using full knowledge on the simulated error model (the ground truth (GT) value). The data was simulated using 15000 random sequences per sequence length.} \label{fig_correlated_L2_rot}
\end{figure}

\begin{figure}[h]
\centering
\includegraphics[width=0.8\linewidth]{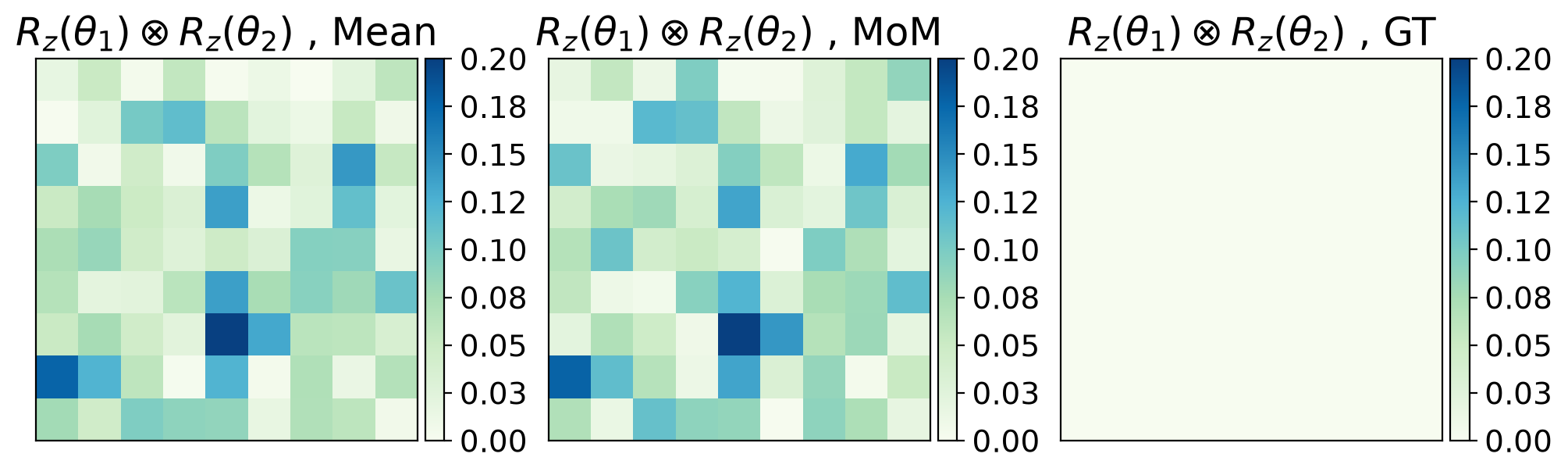}
\caption{$\delta^{(1)}_{CT}$ for the case of a gate error model given by the tensor product of two single-qubit rotations along the $z-$axis. The rotational angles were fixed to be $\theta_1=\theta_2=0.1$. From left to right: the first two figures are obtained by simulating the gate-set shadow protocol and reconstructing the unital marginals via Eq.(\ref{marginals_recformula}). The third figure corresponds to the actual value of $\delta^{(1)}_{CT}$ (ground truth (GT)), derived using full knowledge on the simulated error model. Consistent with the fact that the simulated gate noise is uncorrelated, the matrix elements are small in amplitude, indicating that the resulting noise is close to uncorrelated noise. The data was simulated using 15000 random sequences per sequence length.} \label{fig_correlated_L1_z1z2}
\end{figure}

\begin{figure}[h]
\centering
\includegraphics[width=0.8\linewidth]{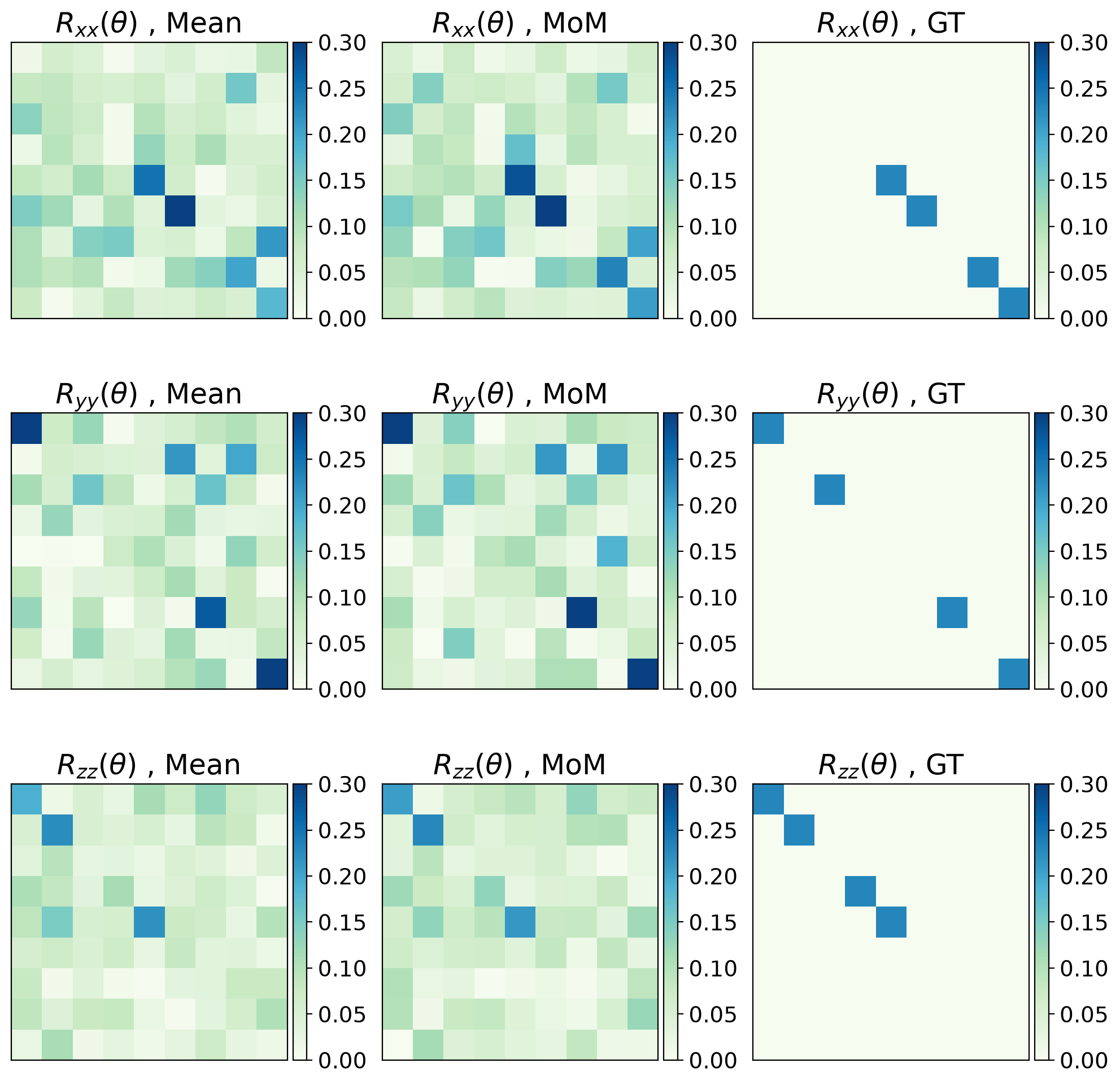}
\caption{$\delta^{(1)}_{CT}$ for the case where the gate error model corresponds to one of the following three two-qubit coherent gates: $R_{xx}(\theta)$ (top panel), $R_{yy}(\theta)$ (middle panel) and $R_{zz}(\theta)$ (bottom panel). For all three cases, the angle $\theta$ was fixed at $\theta=0.5$. In each row, from left to right: the first two figures are obtained by simulating the gate-set shadow protocol and reconstructing the unital marginals via Eq.(\ref{marginals_recformula}); the third figure corresponds to the actual value of $\delta^{(1)}_{CT}$ (ground truth (GT)), derived using full knowledge on the simulated error model. The data was simulated using 15000 random sequences per sequence length.} \label{fig_correlated_L1_rot}
\end{figure}

\begin{figure}[h]
\centering
\includegraphics[width=0.8\linewidth]{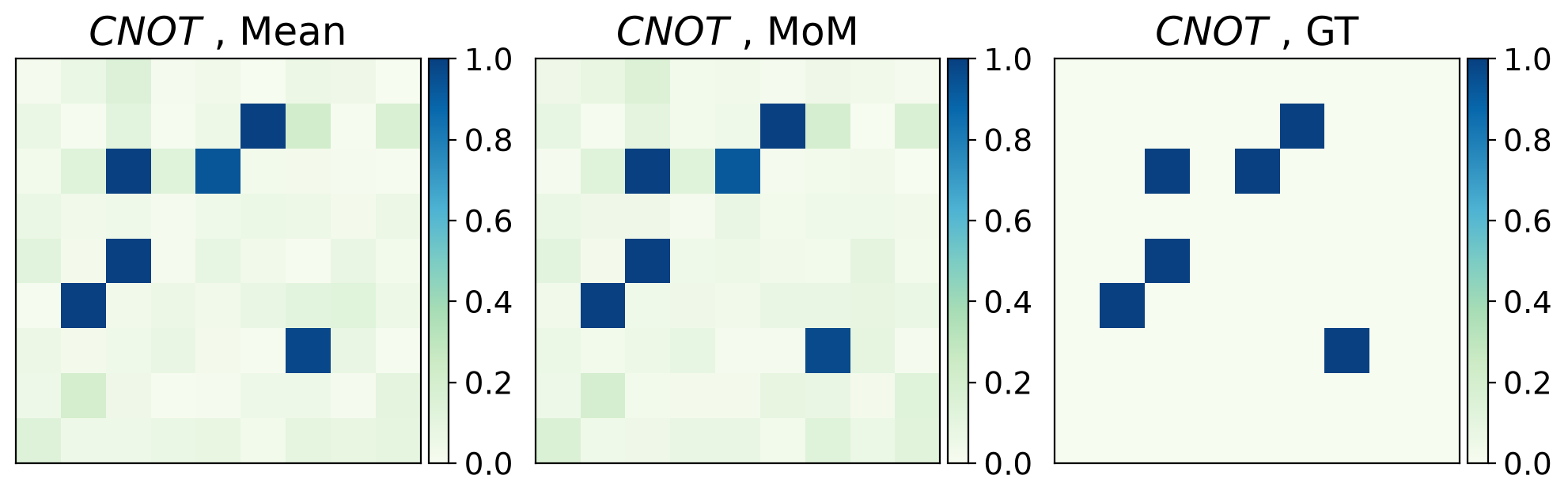}
\caption{$\delta^{(1)}_{CT}$ for the case where the gate error model corresponds to a CNOT operation. From left to right: the first two figures are obtained by simulating the gate-set shadow protocol and reconstructing the unital marginals via Eq.(\ref{marginals_recformula}); the third figure corresponds to the actual value of $\delta^{(1)}_{CT}$ (ground truth (GT)), derived using full knowledge on the simulated gate noise. The data was simulated using 15000 random sequences per sequence length.} \label{fig_correlated_L1_cnot}
\end{figure}

In Ref.~\cite{HelIoaKit2023}, there are two proposed crosstalk metrics:

\begin{equation}
    \delta^{(1)}_{CT} = |\mathcal{R}(\Lambda_3)-\mathcal{R}(\Lambda_1) \otimes \mathcal{R}(\Lambda_2)| \;\; , \;\;
    \delta^{(2)}_{CT} = |\big( \mathcal{R}(\Lambda_1) \otimes \mathcal{R}(\Lambda_2)\big)^{-1}\mathcal{R}(\Lambda_3)| \; ,
    \label{crosstalk_metrics_app}
\end{equation}
where $|.|$ denotes taking the element-wise absolute value of the resulting matrix. In the main text, we have highlighted $\delta^{(2)}_{CT}$, but note that, for the most part, $\delta^{(1)}_{CT}$ and $\delta^{(2)}_{CT}$ are related. We note, however, that $\delta^{(2)}_{CT}$ will be ill-defined if $\mathcal{R}(\Lambda_1) \otimes \mathcal{R}(\Lambda_2)$ is not an invertible matrix. In Figs.~\ref{fig_correlated_L1_z1z2}--\ref{fig_correlated_L1_cnot}, we illustrate the application of the metric $\delta^{(1)}_{CT}$ to quantify correlations in the system, for different simulated error channels. In Fig.~\ref{fig_correlated_L2_rot}, the metric $\delta^{(2)}_{CT}$ is applied to the entangling gate errors presented in the main text.

The pathological case of the CNOT error~\cite{GamCorMer2013} highlights that the noise marginals retain additional relevant information about the structure of the underlying gate noise. This can be understood as follows. Starting from the original gate noise, the construction of each unital marginal retains only its action within the corresponding subspace, while discarding couplings between different subspaces. If only the trace of each unital marginal is subsequently retained, the off-diagonal couplings within that subspace are also discarded, and the diagonal information is reduced to a single scalar. Taking this last trace retrieves the type of information accessible through protocols such as simultaneous RB~\cite{GamCorMer2013}.


For all error models besides the CNOT case, the POVM elements were fixed to be the measurements in the z-computational basis, i.e. $\{E_x\}_x=\{|00\rangle \langle00|, |01\rangle \langle01|, |10\rangle \langle10|, |11\rangle \langle11| \}$. However, if we use the same set of measurements for the case where the gate error is described by a CNOT channel, the resulting dataset cannot accurately retrieve structured information on the gate noise within all invariant subspaces. We may use the freedom in choosing the set of POVM operators and initial state to bypass this issue. Indeed, the numerical results in Fig.~\ref{fig_correlated_L1_cnot} employ a different choice for initial state and POVM elements, namely they make use of the initial state and POVM elements described in Eq.(\ref{POVM_and_input_G4}). The CNOT case, although arguably an unrealistic example, highlights the interplay between the performance of the gate-set shadow protocol and the choice of input and measurement basis.

In the main text, we also presented the conversion of the unital marginal channels into Pauli channels. Obtaining the Pauli channel can be done by computing the average of $\mathcal{R}(\Lambda_F)$ over the set of all 2-qubit Pauli operations, in postprocessing. In the PTM formalism, the average of a given quantum operation $\mathcal{E}$, over all $n$-qubit Pauli gates, is given by:
\begin{equation}
    \mathcal{R}(\mathcal{E}_P) = \frac{1}{4^n} \sum^{4^n-1}_{i=0} \mathcal{R}(\mathcal{P}_i) \; \mathcal{R}(\mathcal{E}) \; \mathcal{R}(\mathcal{P}_i) \;,
    \label{pauli_twirl}
\end{equation}
where $\mathcal{P}_i$ denotes the Pauli conjugation channel, $\mathcal{P}_i(\rho)=P_i \rho P_i$. Hence, the PTM $\mathcal{R}(\mathcal{P}_i)$ is diagonal, with matrix elements $\big(\mathcal{R}(\mathcal{P}_i) \big)_{a,c}= \delta_{a,c}(-1)^{\langle P_c, P_i \rangle}$, where $\langle P_c, P_i \rangle=0$ if the Pauli operators $P_c$ and $P_i$ commute, and $\langle P_c, P_i \rangle=1$ if they anti-commute.
In Eq.(\ref{pauli_twirl}), the resulting Pauli average channel is a diagonal channel, with diagonal elements 
\begin{equation}
 \big(\mathcal{R}(\mathcal{E}_P) \big)_{b,b}=\big(\mathcal{R}(\mathcal{E}) \big)_{b,b} \;.   
\end{equation}
In the PTM formalism, any Pauli channel $\Lambda_P(\rho)=\sum_b p_b \;P_b\rho P_b$ is represented by a diagonal matrix, with matrix elements $\sum^{4^n-1}_{c=0}p_c (-1)^{\langle P_c, P_b \rangle}$, where $\{p_c\}$ is the set of Pauli error rates. By identifying $\mathcal{R}(\mathcal{E}_P)$ with the PTM representation of a Pauli channel, the corresponding distribution $\{p_a\}$ can be obtained through the relation~\cite{MckCroWoo2020}:
\begin{equation}
    p_a = \frac{1}{4^n} \sum^{4^n-1}_{b=0}  (-1)^{\langle P_a, P_b \rangle} \; \big(\mathcal{R}(\mathcal{E}_P) \big)_{b,b} \; .
    \label{ptm_to_kraus}
\end{equation}

The motivation behind this transformation is as follows. Each unital margianl already corresponds to the restriction of the gate noise to a particular subspace. Pauli averaging then retains the diagonal part of these marginals and expresses it as a distribution of Pauli error rates. Since each Pauli error has a definite support and weight, this distribution can reveal which Pauli errors dominate the effective error channel.

Imprecision in the reconstruction of the unital channel marginals can cause some of the estimated error rates $p_a$ to be negative, particularly if their \emph{true} value is zero. To reinforce the constraint that the amplitudes $p_a$ should correspond to a probability distribution, we perform a final Euclidean projection onto the probability simplex~\cite{WanCar2013}. After this final step, the resulting set of amplitudes $\{p_a\}_a$ is guaranteed to satisfy the constraint $\sum_a p_a=1 \;, \; p_a \geq0$, although inaccuracies in the channel marginal reconstruction may still lead to a final probability distribution that differs from the ground truth one. A similar imprecision is present for the correlated RB coefficients. In particular, for the noise models considered here, the ground truth $\varepsilon$ parameters are non-negative. Finite inaccuracies in the estimates for the $\lambda$ parameters can cause small negative amplitudes for some of the estimated $\{ \hat{\varepsilon} \}$ set, particularly when the corresponding ground truth amplitude is zero. In simulations, we first evaluate $\hat{\varepsilon}_2$. Since the amplitudes $\{ \hat{\varepsilon} \}$ should be non-negative for a CPTP fixed-subspace-weight parametrized channel~\cite{MckCroWoo2020}, we have set a heuristic threshold value, where if $\hat{\varepsilon}_2$ is negative but has a small amplitude, i.e. $|\hat{\varepsilon}_2|<0.1$, we fix it to be zero, and proceed similarly for the remaining $\hat{\varepsilon}$ parameters. This threshold value explains why not all amplitudes in Tab.~\ref{tab:epsilon} have a lower error estimate. In Tab.~\ref{tab:epsilon}, the error estimates were computed by considering how the set $\{\hat{\varepsilon}\}$ changed when the simultaneous RB decay rates increased or decreased, based on their respective error estimates. The simultaneous RB error estimates were obtained from the covariance matrix returned by the fitting procedure, with the standard error estimated from the square root of the corresponding diagonal element. We then recomputed each $\hat{\varepsilon}_i$ by varying each decay rate by $\pm$ its estimated standard error, evaluated all combinations of these shifts, and used the largest and smallest resulting values to define the upper and lower uncertainty bounds for $\hat{\varepsilon}_i$.

\twocolumngrid

\bibliographystyle{JHEP}
\bibliography{biblio.bib}

@article{SarProRud2020,
     author = {Sarovar, M. and Proctor, T. and Rudinger, K. and Young, K. and Nielsen, E. and Blume-Kohout, R.},
     title = "{Detecting crosstalk errors in quantum information processors}",
     doi = "https://doi.org/10.22331/q-2020-09-11-321",
     journal = "Quantum",
     volume = "4",
     pages = "321",
     year = "2020"}

@article{MckCroWoo2020,
     author = {McKay, D. C. and Cross, A. W. and Wood, C. J. and Gambetta, J. M.},
     title = "{Correlated Randomized Benchmarking}",
     eprint = "2003.02354",
     archivePrefix = "arXiv",
     doi = "https://doi.org/10.48550/arXiv.2003.02354",
     year = "2020"}

@article{HelIoaKit2023,
     author = {Helsen, J. and Ioannou, M. and Kitzinger, J. and Onorati, E. and Werner, A. H. and Eisert, J. and Roth, I.},
     title = "{Shadow estimation of gate-set properties from random sequences}",
     doi = "https://doi.org/10.1038/s41467-023-39382-9",
     journal = "Nat. Commun.",
     volume = "14",
     pages = "5039",
     year = "2023"}

@article{HelXueVan2019,
     author = {Helsen, J. and Xue, X. and Vandersypen, L.M.K. and Wehner, S.},
     title = "{A new class of efficient randomized benchmarking protocol}",
     doi = "https://doi.org/10.1038/s41534-019-0182-7",
     journal = "npj Quantum Inf.",
     volume = "5",
     pages = "71",
     year = "2019"}

@article{XueWatHel2019,
  title = {Benchmarking Gate Fidelities in a $\mathrm{Si}/\mathrm{SiGe}$ Two-Qubit Device},
  author = {Xue, X. and Watson, T. F. and Helsen, J. and Ward, D. R. and Savage, D. E. and Lagally, M. G. and Coppersmith, S. N. and Eriksson, M. A. and Wehner, S. and Vandersypen, L. M. K.},
  journal = {Phys. Rev. X},
  volume = {9},
  pages = {021011},
  year = {2019},
  doi = {https://doi.org/10.1103/PhysRevX.9.021011}
}

@article{ClaRieWan2021,
  title = {Character {R}andomized {B}enchmarking for {N}on-{M}ultiplicity-{F}ree {G}roups {W}ith {A}pplications to {S}ubspace, {L}eakage, and {M}atchgate {R}andomized {B}enchmarking},
  author = {Claes, J. and Rieffel, E. and Wang, Z.},
  journal = {PRX Quantum},
  volume = {2},
  pages = {010351},
  year = {2021},
  doi = {https://doi.org/10.1103/PRXQuantum.2.010351}
}

@article{WanCar2013,
     title = "{Projection onto the probability simplex: {A}n efficient algorithm with a simple proof, and an application}",
    author = {Wang, W. and Carreira-Perpinán, M.A.},
     eprint = "1309.1541",
     archivePrefix = "arXiv",
     doi = "https://doi.org/10.48550/arXiv.1309.1541",
     year = "2013"}

@article{WooGam2018,
  title = {Quantification and characterization of leakage errors},
  author = {Wood, C. J. and Gambetta, J. M.},
  journal = {Phys. Rev. A},
  volume = {97},
  issue = {3},
  pages = {032306},
  year = {2018},
  doi = {https://doi.org/10.1103/PhysRevA.97.032306}
}

@article{MagGamJoh2012,
  title = {Efficient {M}easurement of {Q}uantum {G}ate {E}rror by {I}nterleaved {R}andomized {B}enchmarktrg},
  author = {Magesan, E. and Gambetta, J. M. and Johnson, B. R. and Ryan, C. A. and Chow, J. M. and Merkel, S. T. and da Silva, M. P. and Keefe, G. A. and Rothwell, M. B. and Ohki, T. A. and Ketchen, M. B. and Steffen, M.},
  journal = {Phys. Rev. Lett.},
  volume = {109},
  issue = {8},
  pages = {080505},
  year = {2012},
  doi = {https://doi.org/10.1103/PhysRevLett.109.080505}
}

@article{CarWalEm2019,
     author = "Carignan-Dugas, A. and Wallman, J. J. and Emerson, J.",
     title = "{Bounding the average gate fidelity of composite channels using the unitarity}",
     journal = {New J. Phys.},
     volume = {21},
     pages = {053016},
     year = {2019},
     doi = {https://doi.org/10.1088/1367-2630/ab1800}
}

@article{KimSilRya2014,
  title = {Robust {E}xtraction of {T}omographic {I}nformation via {R}andomized {B}enchmarking},
  author = {Kimmel, S. and da Silva, M. P. and Ryan, C. A. and Johnson, B. R. and Ohki, T.},
  journal = {Phys. Rev. X},
  volume = {4},
  issue = {1},
  pages = {011050},
  year = {2014},
  doi = {https://doi.org/10.1103/PhysRevX.4.011050}
}

@article{GarCros2020,
  title = {Synthesis of {CNOT}-{D}ihedral circuits with optimal number of two qubit gates},
  author = {Garion, S. and Cross, A. W.},
  journal = {Quantum},
  volume = {4},
  pages = {369},
  year = {2020},
  doi = {https://doi.org/10.22331/q-2020-12-07-369}
}

@article{CrosMagBis2016,
  title = {Scalable randomised benchmarking of non-{C}lifford gates},
  author = {Cross, A. W. and Magesan, E. and Bishop, L. and Smolin, J. A. and Gambetta, J. M.},
  journal = {npj Quantum Inf.},
  volume = {2},
  pages = {16012},
  year = {2016},
  doi = {https://doi.org/10.1038/npjqi.2016.12}
}

@article{DirHelWeh2019,
  title = {Efficient unitarity randomized benchmarking of few-qubit {C}lifford gates},
  author = {Dirkse, B. and Helsen, J. and Wehner, S.},
  journal = {Phys. Rev. A},
  volume = {99},
  pages = {012315},
  year = {2019},
  doi = {https://doi.org/10.1103/PhysRevA.99.012315}
}

@article{HuaKuePres2020,
  title = {Predicting many properties of a quantum system from very few measurements},
  author = {Huang, H. Y. and Kueng, R. and Preskill, J.},
  journal = {Nat. Phys.},
  volume = {16},
  pages = {1050},
  year = {2020},
  doi = {https://doi.org/10.1038/s41567-020-0932-7}
}

@article{LugMen2019,
  title = {Mean {E}stimation and {R}egression {U}nder {H}eavy-{T}ailed {D}istributions: {A} {S}urvey},
  author = {Lugosi, G. and Mendelson, S.},
  journal = {Found. Comput. Math},
  volume = {19},
  pages = {1145},
  year = {2019},
  doi = {https://doi.org/10.1007/s10208-019-09427-x}
}

@book{Eftib1993,
    author ={Efron, B. and Tibshirani, R. J.} ,
    title = {An {I}ntroduction to the {B}ootstrap} ,
    publisher = {Monographs on statistics and applied probability} ,
   volume = {57},
    year = {1993}
}

@article{Green2015,
     author = {Greenbaum, D.},
     title = "{Introduction to {Q}uantum {G}ate {S}et {T}omography}",
     eprint = "1509.02921",
     archivePrefix = "arXiv",
     doi = "https://arxiv.org/abs/1509.02921",
     year = "2015"}

@book{NielChua2010,
    author ={Nielsen, M. A. and Chuang, I. L.} ,
    title = {Quantum {C}omputation and {Q}uantum {I}nformation} ,
    publisher = {Cambridge University Press, Cambridge, England} ,
    year = {2010}
}

@article{MagGamEme2011,
  title = {Scalable and Robust Randomized Benchmarking of Quantum Processes},
  author = {Magesan, E. and Gambetta, J. M. and Emerson, J.},
  journal = {Phys. Rev. Lett.},
  volume = {106},
  pages = {180504},
  year = {2011},
  doi = {https://doi.org/10.1103/PhysRevLett.106.180504}
}

@article{MagGamEme2012,
  title = {Characterizing quantum gates via randomized benchmarking},
  author = {Magesan, E. and Gambetta, J. M. and Emerson, J.},
  journal = {Phys. Rev. A},
  volume = {85},
  pages = {042311},
  year = {2012},
  doi = {https://doi.org/10.1103/PhysRevA.85.042311}
}

@article{SilGrep2025,
  title = {Hands-on introduction to randomized benchmarking},
  author = {Silva, A. and Greplova, E.},
  journal = {SciPost Phys. Lect. Notes},
  volume = {97},
  year = {2025},
  doi = {https://doi.org/10.21468/SciPostPhysLectNotes.97}
}

@article{YanChuGuo2024,
  title = {Coupler-Assisted Leakage Reduction for Scalable Quantum Error Correction with Superconducting Qubits},
  author = {Yang, X. and Chu, J. and Guo, Z. and Huang, W. and Liang, Y. and Liu, J. and Qiu, J. and Sun, X. and Tao, Z. and Zhang, J. and Zhang, J. and Zhang, L. and Zhou, Y. and Guo, W. and Hu, L. and Jiang, J. and Liu, Y. and Linpeng, X. and Chen, T. and Chen, Y. and Niu, J. and Liu, S. and Zhong, Y. and Yu, D.},
  journal = {Phys. Rev. Lett.},
  volume = {133},
  pages = {170601},
  year = {2024},
  doi = {https://doi.org/10.1103/PhysRevLett.133.170601}
}

@article{Fow2013,
  title = {Coping with qubit leakage in topological codes},
  author = {Fowler, Austin G.},
  journal = {Phys. Rev. A},
  volume = {88},
  pages = {042308},
  year = {2013},
  doi = {https://doi.org/10.1103/PhysRevA.88.042308}
}

@article{ZhaYanWan2018,
  title = {Leakage and sweet spots in triple-quantum-dot spin qubits: {A} molecular-orbital study},
  author = {Zhang, C. and Yang, X.-C. and Wang, X.},
  journal = {Phys. Rev. A},
  volume = {97},
  pages = {042326},
  year = {2018},
  doi = {https://doi.org/10.1103/PhysRevA.97.042326}
}

@article{KleFran2005,
  title = {Quantum {E}rror {C}orrection in {S}patially {C}orrelated {Q}uantum {N}oise},
  author = {Klesse, R. and Frank, S.},
  journal = {Phys. Rev. Lett.},
  volume = {95},
  pages = {230503},
  year = {2005},
  doi = {https://doi.org/10.1103/PhysRevLett.95.230503}
}

@article{Goo2025,
  title = {Quantum error correction below the surface code threshold},
  author = {Acharya, R. and Abanin,  D. A. and Aghababaie-Beni, L. and Aleiner, I. and Andersen, T. I. and Ansmann,  M. and Arute, F. and Arya, K. and Asfaw, A. and Astrakhantsev, N. and Atalaya, J. and Babbush, R. and Bacon, D. and Ballard, B. and Bardin, J. C. and et al.},
  journal = {Nature},
  volume = {638},
  pages = {920},
  year = {2025},
  doi = {https://doi.org/10.1038/s41586-024-08449-y}
}

@article{RyaBohLee2021,
  title = {Realization of Real-Time Fault-Tolerant Quantum Error Correction},
  author = {Ryan-Anderson, C. and Bohnet, J. G. and Lee, K. and Gresh, D. and Hankin, A. and Gaebler, J. P. and Francois, D. and Chernoguzov, A. and Lucchetti, D. and Brown, N. C. and Gatterman, T. M. and Halit, S. K. and Gilmore, K. and Gerber, J. A. and Neyenhuis, B. and Hayes, D. and Stutz, R. P.},
  journal = {Phys. Rev. X},
  volume = {11},
  pages = {041058},
  year = {2021},
  doi = {https://doi.org/10.1103/PhysRevX.11.041058}
}

@article{DevMunNem2013,
  title = {Quantum error correction for beginners},
  author = {Devitt, S. J. and Munro, W. J. and Nemoto, K.},
  journal = {Rep. Prog. Phys.},
  volume = {76},
  pages = {076001},
  year = {2013},
  doi = {https://doi.org/10.1088/0034-4885/76/7/076001}
}

@article{kniLafZur1998,
  title = {Resilient quantum computation: error models and thresholds},
  author = {Knill, E. and Laflamme, R. and Zurek, W. H.},
  journal = {Proc. R. Soc. Lond. A},
  volume = {454},
  pages = {365},
  year = {1998},
  doi = {https://doi.org/10.1098/rspa.1998.0166}
}

@article{Pres1997,
     author = {Preskill, J.},
     title = "{Fault-tolerant quantum computation}",
     eprint = "quant-ph/9605011",
     archivePrefix = "arXiv",
     doi = "https://doi.org/10.48550/arXiv.quant-ph/9712048",
     year = "1997"}

@article{TorWooAcha2023,
  title = {Quantum process tomography with unsupervised learning and tensor networks},
  author = {Torlai, G and Wood, C.J. and Acharya, A. and Carleo, G. and Carrasquilla, J. and Aolita, L.},
  journal = {Nat. Commun.},
  volume = {14},
  pages = {2858},
  year = {2023},
  doi = {https://doi.org/10.1038/s41467-023-38332-9}
}

@article{AriPre2001,
  title = {Quantum {T}omography for {M}easuring {E}xperimentally the {M}atrix {E}lements of an {A}rbitrary {Q}uantum {O}peration},
  author = {D'Ariano, G. M. and Lo Presti, P.},
  journal = {Phys. Rev. Lett.},
  volume = {86},
  pages = {4195},
  year = {2001},
  doi = {https://doi.org/10.1103/PhysRevLett.86.4195}
}

@article{ChuNiel1997,
  title = {Prescription for experimental determination of the dynamics of a quantum black box},
  author = {Chuang, I. L. and Nielsen, M. A.},
  journal = {J. Mod. Opt.},
  volume = {44},
  pages = {2455},
  year = {1997},
  doi = {https://doi.org/10.1080/09500349708231894}
}

@article{BalKalDeu2014,
  title = {Quantum process tomography of unitary and near-unitary maps},
  author = {Baldwin, C. H. and Kalev, A. and Deutsch, I. H.},
  journal = {Phys. Rev. A},
  volume = {90},
  pages = {012110},
  year = {2014},
  doi = {https://doi.org/10.1103/PhysRevA.90.012110}
}

@article{HelRotOno2022,
  title = {General {F}ramework for {R}andomized {B}enchmarking},
  author = {Helsen, J. and Roth, I. and Onorati, E. and Werner, A.H. and Eisert, J.},
  journal = {PRX Quantum},
  volume = {3},
  pages = {020357},
  year = {2022},
  doi = {https://doi.org/10.1103/PRXQuantum.3.020357}
}

@article{WalGraHar2015,
  title = {Estimating the coherence of noise},
  author = {Wallman, J. and Granade, C. and Harper, R. and Flammia, S. T.},
  journal = {New J. Phys.},
  volume = {17},
  pages = {113020},
  year = {2015},
  doi = {https://doi.org/10.1088/1367-2630/17/11/113020}
}

@article{HarFla2017,
  title = {Estimating the fidelity of {T} gates using standard interleaved randomized benchmarking},
  author = {Harper, R. and Flammia, S. T.},
  journal = {Quantum Sci. Technol.},
  volume = {2},
  pages = {015008},
  year = {2017},
  doi = {https://doi.org/10.1088/2058-9565/aa5f8d}
}

@book{FulHar2013,
    author ={Fulton, W. and Harris, J.} ,
    title = {Representation theory: a first course},
    publisher = {Springer Science and Business Media},
   volume = {129},
    year = {2013}
}

@article{GroAudEis2007,
  title = {Evenly distributed unitaries: {O}n the structure of unitary designs},
  author = {Gross, D. and Audenaert, K. and Eisert, J.},
  journal = {J. Math. Phys.},
  volume = {48},
  pages = {052104},
  year = {2007},
  doi = {https://doi.org/10.1063/1.2716992}
}

@book{Tinkham1964,
    author ={Tinkham, M.} ,
    title = {Group theory and quantum mechanics} ,
    publisher = {Dover, New York} ,
    year = {1964}
}

@article{WoBiCo2015, 
author = {Wood, C. J. and Biamonte, J. D. and Cory, D. G.}, 
title = {Tensor networks and graphical calculus for open quantum systems}, 
journal = {Quant. Inf. Comp.},
year = {2015},  
volume = {15},  
pages = {759–811}
}

@article{pennylane2018,
  title={Pennylane: Automatic differentiation of hybrid quantum-classical computations},
  author={Bergholm, Ville and Izaac, Josh and Schuld, Maria and Gogolin, Christian and Ahmed, Shahnawaz and Ajith, Vishnu and Alam, M Sohaib and Alonso-Linaje, Guillermo and AkashNarayanan, B and Asadi, Ali and others},
  eprint = "1811.04968",
  archivePrefix = "arXiv",
  doi = "https://doi.org/10.48550/arXiv.1811.04968",
  year={2018}
}

@article{HarYuFla2021,
  title = {Fast {E}stimation of {S}parse {Q}uantum {N}oise},
  author = {Harper, R. and Yu, W. and Flammia, S. T.},
  journal = {PRX Quantum},
  volume = {2},
  pages = {010322},
  year = {2021},
  doi = {https://doi.org/10.1103/PRXQuantum.2.010322}
}

@article{Chow2012,
  title = {Universal {Q}uantum {G}ate {S}et {A}pproaching {F}ault-{T}olerant {T}hresholds with {S}uperconducting {Q}ubits},
  author = {Chow, J. M. and Gambetta, J. M. and C\'orcoles, A. D. and Merkel, S. T. and Smolin, J. A. and Rigetti, C. and Poletto, S. and Keefe, G. A. and Rothwell, M. B. and Rozen, J. R. and Ketchen, M. B. and Steffen, M.},
  journal = {Phys. Rev. Lett.},
  volume = {109},
  pages = {060501},
  year = {2012},
  doi = {https://doi.org/10.1103/PhysRevLett.109.060501}
}

@article{Mele2024,
  title={Introduction to {H}aar {M}easure {T}ools in {Q}uantum {I}nformation: {A} {B}eginner's {T}utorial},
  author={Mele, A. A.},
  journal = {Quantum},
  volume = {8},
  pages = {1340},
  year = {2024},
  doi = {https://doi.org/10.22331/q-2024-05-08-1340}
}

@article{CorGamChow2013,
  title = {Process verification of two-qubit quantum gates by randomized benchmarking},
  author = {C\'orcoles, A. D. and Gambetta, J. M. and Chow, J. M. and Smolin, J. A. and Ware, M. and Strand, J. and Plourde, B. L. T. and Steffen, M.},
  journal = {Phys. Rev. A},
  volume = {87},
  pages = {030301},
  year = {2013},
  doi = {10.1103/PhysRevA.87.030301}
}

@misc{SilGrep2025_git,
  author = {Silva, A. and Greplova, E.},
  title = {Every {B}enchmark {A}ll at {O}nce},
  howpublished = {Accompanying code can be found here: \url{https://gitlab.com/QMAI/papers/everybenchmarkallatonce}}
}

@article{PauHog2025,
     author = {H\o{}gsgaard, M.M. and Paudice, A.},
     title = "{Uniform {M}ean {E}stimation for {H}eavy-{T}ailed {D}istributions via {M}edian-of-{M}eans}",
     eprint = "2506.14673",
     archivePrefix = "arXiv",
     primaryClass = {stat.ML},
     doi = "https://doi.org/10.48550/arXiv.2506.14673",
     year = "2025"}

@article{ZhaSar2015,
  title = {Identification of open quantum systems from observable time traces},
  author = {Zhang, Jun and Sarovar, Mohan},
  journal = {Phys. Rev. A},
  volume = {91},
  issue = {5},
  pages = {052121},
  numpages = {12},
  year = {2015},
  month = {May},
  publisher = {American Physical Society},
  doi = {10.1103/PhysRevA.91.052121},
  url = {https://link.aps.org/doi/10.1103/PhysRevA.91.052121}
}

@article{BenRyaLou2019,
doi = {10.1088/1367-2630/ab3598},
url = {https://doi.org/10.1088/1367-2630/ab3598},
year = {2019},
month = {aug},
publisher = {IOP Publishing},
volume = {21},
number = {8},
pages = {083013},
author = {Bennink, Ryan S and Lougovski, Pavel},
title = {Quantum process identification: a method for characterizing non-markovian quantum dynamics},
journal = {New Journal of Physics}
}

@article{YaZhuChai2023,
  title={Benchmarking universal quantum gates via channel spectrum},
  author={G., Yanwu and Zhuang, W.-F. and Chai, X. and Liu, D. E.},
  journal={Nature Communications},
  volume={14},
  number={1},
  pages={5880},
  year={2023},
  publisher={Nature Publishing Group UK London}
}

@article{HelBatTer2019,
  title={Spectral quantum tomography},
  author={Helsen, J. and Battistel, F. and Terhal, B. M.},
  journal={npj Quantum Information},
  volume={5},
  number={1},
  pages={74},
  year={2019},
  publisher={Nature Publishing Group UK London}
}

@article{HarFlaWall2020,
  title={Efficient learning of quantum noise},
  author={Harper, R. and Flammia, S.T. and Wallman, J.J.},
  journal={Nat. Phys.},
  volume={16},
  pages={1184-1188},
  year={2020},
  doi = {10.1038/s41567-020-0992-8}
}

@article{NieGamRud2021,
  title={Gate Set Tomography},
  author={Nielsen, E. and Gamble, J.K. and Rudinger, K. and Scholten, T. and Young, K. and Blume-Kohout, R.},
  journal={Quantum},
  volume={5},
  pages={557},
  year={2021},
  doi = {10.22331/q-2021-10-05-557}
}

@article{BluGamNie2013,
     author = {Blume-Kohout, R. and Gamble, J. K. and Nielsen, E. and Mizrahi, J. and Sterk, J. D. and Maunz, P.},
     title = "{Robust, self-consistent, closed-form tomography of quantum logic gates on a trapped ion qubit}",
     eprint = "1310.4492",
     archivePrefix = "arXiv",
     primaryClass = {quant-ph},
     doi = "https://doi.org/10.48550/arXiv.1310.4492",
     year = "2013"}

@article{ProRudYou2017,
  title = {What Randomized Benchmarking Actually Measures},
  author = {Proctor, T. and Rudinger, K. and Young, K. and Sarovar, M. and Blume-Kohout, R.},
  journal = {Phys. Rev. Lett.},
  volume = {119},
  issue = {13},
  pages = {130502},
  numpages = {6},
  year = {2017},
  month = {Sep},
  publisher = {American Physical Society},
  doi = {10.1103/PhysRevLett.119.130502},
  url = {https://link.aps.org/doi/10.1103/PhysRevLett.119.130502}
}

@article{CheBal2025,
  title = {Randomized benchmarking with leakage errors},
  author = {Chen, Yi-Hsiang and Baldwin, Charles H.},
  journal = {Phys. Rev. Res.},
  volume = {7},
  issue = {4},
  pages = {043065},
  numpages = {18},
  year = {2025},
  month = {Oct},
  publisher = {American Physical Society},
  doi = {10.1103/j8sz-6ws9},
  url = {https://link.aps.org/doi/10.1103/j8sz-6ws9}
}

@article{AndJonRee2019,
  title={Quantifying error and leakage in an encoded Si/SiGe triple-dot qubit},
  author={Andrews, R.W. and Jones, C. and Reed, M.D. and Jones, A. M. and Ha, S. D. and Jura, M. P. and Kerckhoﬀ, J. and Levendorf, M. and Meenehan, S. and Merkel, S. T. and Smith, A. and Sun, B. and Weinstein, A. J. and Rakher, M. T. and Ladd, T. D. and Borselli, M. G.},
  journal={Nat. Nanotechnol.},
  volume={14},
  pages={747–750},
  year={2019},
  doi = {10.1038/s41565-019-0500-4}
}

@article{ErhWallPos2019,
  title={Characterizing large-scale quantum computers via cycle benchmarking},
  author={Erhard, A. and Wallman, J.J. and Postler, L. and Meth, M. and Stricker, R. and Martinez, E. A. and Schindler, P. and Monz, T. and Emerson, J. and Blatt, R.},
  journal={Nat. Commun.},
  volume={10},
  pages={5347},
  year={2019},
  doi = {10.1038/s41467-019-13068-7}
}

@article{GarKanLan2021,
  title = {Experimental implementation of non-Clifford interleaved randomized benchmarking with a controlled-$S$ gate},
  author = {Garion, S. and Kanazawa, N. and Landa, H. and McKay, D. C. and Sheldon, S. and Cross, A. W. and Wood, C. J.},
  journal = {Phys. Rev. Res.},
  volume = {3},
  issue = {1},
  pages = {013204},
  numpages = {9},
  year = {2021},
  month = {Mar},
  publisher = {American Physical Society},
  doi = {10.1103/PhysRevResearch.3.013204},
  url = {https://link.aps.org/doi/10.1103/PhysRevResearch.3.013204}
}

@article{CheDinHua2022,
  title = {Randomized Benchmarking beyond Groups},
  author = {Chen, J. and Ding, D. and Huang, C.},
  journal = {PRX Quantum},
  volume = {3},
  issue = {3},
  pages = {030320},
  numpages = {24},
  year = {2022},
  month = {Aug},
  publisher = {American Physical Society},
  doi = {10.1103/PRXQuantum.3.030320},
  url = {https://link.aps.org/doi/10.1103/PRXQuantum.3.030320}
}

@article{HinHotBlu2024,
  title = {Fully Scalable Randomized Benchmarking Without Motion Reversal},
  author = {Hines, J. and Hothem, D. and Blume-Kohout, R. and Whaley, B. and Proctor, T.},
  journal = {PRX Quantum},
  volume = {5},
  issue = {3},
  pages = {030334},
  numpages = {19},
  year = {2024},
  month = {Aug},
  publisher = {American Physical Society},
  doi = {10.1103/PRXQuantum.5.030334},
  url = {https://link.aps.org/doi/10.1103/PRXQuantum.5.030334}
}

@article{ProSerRud2022,
  title = {Scalable Randomized Benchmarking of Quantum Computers Using Mirror Circuits},
  author = {Proctor, T. and Seritan, S. and Rudinger, K. and Nielsen, E. and Blume-Kohout, R. and Young, K.},
  journal = {Phys. Rev. Lett.},
  volume = {129},
  issue = {15},
  pages = {150502},
  numpages = {7},
  year = {2022},
  month = {Oct},
  publisher = {American Physical Society},
  doi = {10.1103/PhysRevLett.129.150502},
  url = {https://link.aps.org/doi/10.1103/PhysRevLett.129.150502}
}

@article{ChaWil2015,
  title = {Complete randomized benchmarking protocol accounting for leakage errors},
  author = {Chasseur, T. and Wilhelm, F. K.},
  journal = {Phys. Rev. A},
  volume = {92},
  issue = {4},
  pages = {042333},
  numpages = {7},
  year = {2015},
  month = {Oct},
  publisher = {American Physical Society},
  doi = {10.1103/PhysRevA.92.042333},
  url = {https://link.aps.org/doi/10.1103/PhysRevA.92.042333}
}

@article{EpsCrosMag2014,
  title = {Investigating the limits of randomized benchmarking protocols},
  author = {Epstein, J. M. and Cross, A. W. and Magesan, E. and Gambetta, J. M.},
  journal = {Phys. Rev. A},
  volume = {89},
  issue = {6},
  pages = {062321},
  numpages = {12},
  year = {2014},
  month = {Jun},
  publisher = {American Physical Society},
  doi = {10.1103/PhysRevA.89.062321},
  url = {https://link.aps.org/doi/10.1103/PhysRevA.89.062321}
}

@article{WalBarEme2016,
  title={Robust characterization of leakage errors},
  author={Wallman, J. J. and Barnhill, M. and Emerson, J.},
  journal={New J. Phys.},
  volume={18},
  pages={043021},
  year={2016},
  doi = {10.1088/1367-2630/18/4/043021}
}

@article{WuWanYu2024,
  title={Leakage Benchmarking for Universal Gate Sets},
  author={Wu, B. and Wang, X. and Yuan, X. and Huang, C. and Chen, J.},
  journal={Entropy},
  volume={26},
  pages={71},
  year={2024},
  doi = {10.3390/e26010071}
}

@INPROCEEDINGS{GueSinMaj2021,
  author={Guého, D. and Singla, P. and Majji, M. and Juang, J.-N.},
  booktitle={2021 60th IEEE Conference on Decision and Control (CDC)}, 
  title={Advances in System Identification: Theory and Applications}, 
  year={2021},
  volume={},
  number={},
  pages={22-30},
  doi={10.1109/CDC45484.2021.9683394}
}

@book{kees2011,
  title={System identification: an introduction},
  author={Keesman, K. J.},
  year={2011},
  publisher={Springer Science \& Business Media}
}

@article{Wall2018,
  title = {Randomized benchmarking with gate-dependent noise},
  author = {Wallman, J. J.},
  journal = {Quantum},
  volume = {2},
  pages = {47},
  year = {2018},
  doi = {https://doi.org/10.22331/q-2018-01-29-47}
}

@article{CheKelJul2016,
  title = {Measuring and Suppressing Quantum State Leakage in a Superconducting Qubit},
  author = {Chen, Zijun and Kelly, Julian and Quintana, Chris and Barends, R. and Campbell, B. and Chen, Yu and Chiaro, B. and Dunsworth, A. and Fowler, A. G. and Lucero, E. and Jeffrey, E. and Megrant, A. and Mutus, J. and Neeley, M. and Neill, C. and O'Malley, P. J. J. and Roushan, P. and Sank, D. and Vainsencher, A. and Wenner, J. and White, T. C. and Korotkov, A. N. and Martinis, John M.},
  journal = {Phys. Rev. Lett.},
  volume = {116},
  issue = {2},
  pages = {020501},
  numpages = {5},
  year = {2016},
  month = {Jan},
  publisher = {American Physical Society},
  doi = {10.1103/PhysRevLett.116.020501},
  url = {https://link.aps.org/doi/10.1103/PhysRevLett.116.020501}
}

@article{MunStoSta2025,
  title = {High-Fidelity Universal Gates in the ${}^{171}$$\mathrm{Yb}$ Ground-State Nuclear-Spin Qubit},
  author = {Muniz, J. A. and Stone, M. and Stack, D. T. and Jaffe, M. and Kindem, J. M. and Wadleigh, L. and Zalys-Geller, E. and Zhang, X. and Chen, C.-A. and Norcia, M. A. and Epstein, J. and Halperin, E. and Hummel, F. and Wilkason, T. and Li, M. and Barnes, K. and Battaglino, P. and Bohdanowicz, T. C. and Booth, G. and Brown, A. and Brown, M. O. and Cairncross, W. B. and Cassella, K. and Coxe, R. and Crow, D. and Feldkamp, M. and Griger, C. and Heinz, A. and Jones, A. M. W. and Kim, H. and King, J. and Kotru, K. and Lauigan, J. and Marjanovic, J. and Megidish, E. and Meredith, M. and McDonald, M. and Morshead, R. and Narayanaswami, S. and Nishiguchi, C. and Paule, T. and Pawlak, K. A. and Pudenz, K. L. and P\'erez, D. Rodr\'{\i}guez and Ryou, A. and Simon, J. and Smull, A. and Urbanek, M. and van de Veerdonk, R. J. M. and Vendeiro, Z. and Wu, T.-Y. and Xie, X. and Bloom, B. J.},
  journal = {PRX Quantum},
  volume = {6},
  issue = {2},
  pages = {020334},
  numpages = {18},
  year = {2025},
  month = {May},
  publisher = {American Physical Society},
  doi = {10.1103/PRXQuantum.6.020334},
  url = {https://link.aps.org/doi/10.1103/PRXQuantum.6.020334}
}

@article{MillOstCor2026,
     author = {Miller, Ashe and Ostrove, Corey and Hines, Jordan and Siekierski, Noah and Young, Kevin and Blume-Kohout, Robin and Proctor, Timothy},
     title = "{Scalable linearized gate set tomography}",
     eprint = "2605.11158",
     archivePrefix = "arXiv",
     primaryClass = {quant-ph},
     doi = "https://doi.org/10.48550/arXiv.2605.11158",
     year = "2026"}

@article{JarEngBert2004,
  title = {Minimal qubit tomography},
  author = {\ifmmode \check{R}\else \v{R}\fi{}eh\'a\ifmmode \check{c}\else \v{c}\fi{}ek, Jaroslav and Englert, Berthold-Georg and Kaszlikowski, Dagomir},
  journal = {Phys. Rev. A},
  volume = {70},
  issue = {5},
  pages = {052321},
  numpages = {13},
  year = {2004},
  month = {Nov},
  publisher = {American Physical Society},
  doi = {10.1103/PhysRevA.70.052321},
  url = {https://link.aps.org/doi/10.1103/PhysRevA.70.052321}
}

@article{WonZenLi2026,
  title={Non-Markovian exceptional points by interpolating quantum channels},
  author={Wong, W.C. and Zeng, B. and Li, J.},
  journal={npj Quantum Inf.},
  volume={12},
  pages={63},
  year={2026},
  doi = {10.1038/s41534-026-01205-2}
}

@article{GirCirJen2022,
  title = {Estimation of correlations and nonseparability in quantum channels via unitarity benchmarking},
  author = {Girling, Matthew and C\^{\i}rstoiu, Cristina and Jennings, David},
  journal = {Phys. Rev. Res.},
  volume = {4},
  issue = {2},
  pages = {023041},
  numpages = {29},
  year = {2022},
  month = {Apr},
  publisher = {American Physical Society},
  doi = {10.1103/PhysRevResearch.4.023041},
  url = {https://link.aps.org/doi/10.1103/PhysRevResearch.4.023041}
}

@book{Sim1996,
    author ={Simon, B.} ,
    title = {Representations of Finite and Compact Groups},
    publisher = {American Mathematical Society},
   volume = {10},
   year = {1996}
}

@article{HelWalWeh2018,
  title = {Representations of the multi-qubit {C}lifford group},
  author = {Helsen, J. and Wallman, J. J. and Wehner, S.},
  journal = {J. Math. Phys.},
  volume = {59},
  pages = {072201},
  year = {2018},
  doi = {https://doi.org/10.1063/1.4997688}
}

@article{JuaPap1985,
  title={An eigensystem realization algorithm for modal parameter identification and model reduction},
  author={Juang, J.-N. and Pappa, R. S.},
  journal={J. Guid. Control Dyn.},
  volume={8},
  number={5},
  pages={620-627},
  year={1985}
}

@misc{Wolf2012,
  author       = {Wolf, M.},
  title        = {Quantum channels and operations guided tour},
  year         = {2012},
  howpublished = {\url{https://mediatum.ub.tum.de/doc/1701036/document.pdf}}
}

@article{WolGar2010,
     author = {Wolf, M. M. and Perez-Garcia, D.},
     title = "{The Inverse Eigenvalue Problem for Quantum Channels}",
     eprint = "1005.4545",
     archivePrefix = "arXiv",
     doi = "https://doi.org/10.48550/arXiv.1005.4545",
     year = "2010"}

@article{RudPucZyc2018,
  title={Gauge invariant information concerning quantum channels},
  author={Rudnicki, \l{L}. and Pucha\l{l}a, Z. and Zyczkowski, K.},
  journal={Quantum},
  volume={2},
  number={60},
  year={2018}
}

@misc{DHPC2024,
author = {{D}elft {H}igh {P}erformance {C}omputing {C}entre ({DHPC})},
title = {{D}elft{B}lue {S}upercomputer ({P}hase 2)},
year = {2024},
howpublished = {\url{https://www.tudelft.nl/dhpc/ark:/44463/DelftBluePhase2}},
ark = {ark:/44463/DelftBluePhase2}
}

@misc{SilGrep_zenodo,
  author       = {Silva, Ana and Greplova, Eliska},
  title        = {Accompanying materials for: Every Benchmark All at Once},
  year         = {2026},
  doi          = {\href{https://doi.org/10.5281/zenodo.23233958}
                        {10.5281/zenodo.23233958}}}

\end{document}